\documentclass[preprints,article,accept,moreauthors,pdftex]{Definitions/mdpi} 
\setitemize{parsep=6pt,itemsep=0pt,leftmargin=*,labelsep=5.5mm}
\setenumerate{parsep=6pt,itemsep=0pt,leftmargin=*,labelsep=5.5mm}
\setlist[description]{itemsep=0mm}   

\firstpage{1} 
\pubvolume{xx}
\issuenum{1}
\articlenumber{5}
\pubyear{2020}
\copyrightyear{2020}
\history{Received: 17 June 2020; Accepted: 14 August 2020; Published: 18 August 2020}
\updates{yes}

\Title{Symmetry, Transactions, and the Mechanism of Wave Function Collapse}

\Author{John Gleason Cramer $^{1,}$* and Carver Andress Mead $^{2}$}

\AuthorNames{John G.~Cramer and Carver A.~Mead}

\address{%
$^{1}$ \quad Department of Physics, University of Washington, Seattle, WA 98195, USA\\
$^{2}$ \quad California Institute of Technology, Pasadena, CA 91125, USA; carver@caltech.edu\\}

\corres{Correspondence: jcramer@uw.edu}

\abstract{The Transactional Interpretation of quantum mechanics exploits the intrinsic time-symmetry of wave mechanics to interpret the $\psi$ and $\psi$* wave functions present in all wave mechanics calculations as representing retarded and advanced waves moving in opposite time directions that form a quantum ``handshake'' or transaction.  This handshake is a 4D standing-wave that builds up across space-time to transfer the conserved quantities of energy, momentum, and angular momentum in an interaction.  Here, we derive a two-atom quantum formalism describing a transaction.  We show that the bi-directional electromagnetic coupling between atoms can be factored into a matched pair of vector potential Green's functions:  one retarded and one advanced, and that this combination uniquely enforces the conservation of energy in a transaction.  Thus factored, the single-electron wave functions of electromagnetically-coupled atoms can be analyzed using Schr\"odinger's original wave mechanics.  The technique generalizes to any number of electromagnetically coupled single-electron states---no higher-dimensional space is needed.  Using this technique, we show a worked example of the transfer of energy from a hydrogen atom in an excited state to a nearby hydrogen atom in its ground state.  It is seen that the initial exchange creates a dynamically unstable situation that avalanches to the completed transaction, demonstrating that wave function collapse, considered mysterious in the literature, can be implemented with solutions of Schr\"odinger's original wave mechanics, coupled by this unique combination of retarded/advanced vector potentials, without the introduction of any additional mechanism or formalism.  We also analyze a simplified version of the photon-splitting and Freedman--Clauser three-electron experiments and show that their results can be predicted by this formalism.}


\begin{document}

\setcounter{section}{0} 

\section{Introduction}

Quantum mechanics (QM) was never properly finished.  Instead, it was left in an exceedingly unsatisfactory state by its founders.
Many attempts by highly qualified individuals to improve the situation have failed to produce any consensus
about either (a) the precise nature of the problem, or (b) what a better form of QM might look like.

At the most basic level, a simple observation illustrates the central conceptual problem:

An excited atom somewhere in the universe transfers {\it all} of its excitation energy to another single atom,
independent of the presence of the vast number of alternative atoms that could have received all or part of the energy.
The obvious ``photon-as-particle'' interpretation of this situation has a one-way symmetry:  The excited source atom is depicted as emitting a particle, a {\it photon}
of electromagnetic energy that is somehow oscillating with angular frequency $\omega$ while moving in a particular direction.  The photon is depicted as carrying a quantum of energy $\hbar\omega$, a momentum $\hbar\omega/c$, and an angular momentum $\hbar$ through space,
until it is later absorbed by some unexcited atom.  The emission and absorption are treated as independent isolated events without internal structure.  It is insisted that the only real and meaningful quantities describing this process are {\it probabilities}, since these are measurable.  The necessarily abrupt change in the quantum wave function of the system when the photon arrives (and an observer potentially gains information) is called ``wave function collapse'' and is considered to be a mysterious process that the founders of QM found it necessary to ``put in by hand'' without providing any mechanism.  [The missing mechanism behind wave function collapse is sometimes called ``the measurement problem'', particularly by acolytes of Heisenberg's knowledge interpretation.  In our view, measurement requires wave function collapse {\it but does not cause it.}]  [Side comments will be put in square brackets]

Referring to statistical quantum theory, which is reputed to apply only to {\em ensembles} of similar systems, Albert Einstein \cite{Einstein_1949} had this to say:
\begin{quote}{\it
``I do not believe that this fundamental concept will provide a useful
basis for the whole of physics.''

``I am, in fact, firmly convinced that the
essentially statistical character
of contemporary quantum theory is solely to be ascribed to the fact that this [theory]
operates with an incomplete description of physical systems.''

``One arrives at very implausible theoretical conceptions, if one attempts
to maintain the thesis that the statistical quantum theory is in principle capable of
producing a complete description of an individual physical system ...''

``Roughly stated, the conclusion is this:  Within the framework of
statistical quantum theory, there is no such thing as a complete description of the
individual system.  More cautiously, it might be put as follows:  The attempt to
conceive the quantum-theoretical description as the complete description of the
individual systems leads to unnatural theoretical interpretations, which become
immediately unnecessary if one accepts the interpretation that the description refers
to ensembles of systems and not to individual systems.  In that case, the
whole 'egg-walking' performed in order to avoid the 'physically real' becomes
superfluous.  There exists, however, a simple psychological reason for the fact that
this most nearly obvious interpretation is being shunned---for, if the statistical
quantum theory does not pretend to describe the individual system (and its
development in time) completely, it~appears unavoidable to look elsewhere for a
complete description of the individual system.  In doing so, it would be clear from
the very beginning that the elements of such a description are not contained within
the conceptual scheme of the statistical quantum theory.  With this. one would admit
that, in~principle, this scheme could not serve as the basis of theoretical physics.
Assuming the success of efforts to accomplish a complete physical description, the
statistical quantum theory would, within the framework of future physics, take an
approximately analogous position to the statistical mechanics within the framework of
classical mechanics.  I am rather firmly convinced that the development of
theoretical physics will be of this type, but the path will be lengthy and difficult.''

``If it should be possible to move forward to a complete description, it is
likely that the laws would represent relations among all the conceptual elements of
this description which, per se,  have nothing to do with statistics.''}
\end{quote}

In what follows we put forth a simple approach to {\it describing the individual system (and its development in time),} which Einstein
believed was missing from statistical quantum theory and which must be present before any theory of physics could be considered to be complete.

The way forward was suggested by the phenomenon
of {\it entanglement}.  Over the past few decades, many increasingly exquisite Einstein--Podolsky--Rosen \cite{Einstein_1935} (EPR) experiments \cite{Bell_1964,Bell_1966,Camhy-Val_1970,FreedmanClauser_1972,Fry_1976,Clauser_1976,Aspect_1981,Aspect_1982,Aspect_1982a} have demonstrated that multi-body quantum systems with separated  components that are subject to conservation laws exhibit a property called ``quantum entanglement'' \cite{Cramer_2016}: Their component wave functions are inextricably locked together, and they display a nonlocal correlated behavior enforced over an arbitrary interval of space-time without any hint of an underlying mechanism or any show of respect for our cherished classical ``arrow of time.''  Entanglement is the most mysterious of the many so-called ``quantum mysteries.''

It has thus become clear that the quantum transfer of energy must have quite a different symmetry from that implied by this simple ``photon-as-particle'' interpretation.  Within the framework of statistical QM, the intrinsic symmetry of the energy transfer and the mechanisms behind wave function collapse and entanglement have been greatly clarified by the {\bf Transactional Interpretation of quantum mechanics} (TI), developed over several decades by one of us and recently described in some detail in the book {\textbf {\textit {The Quantum Handshake}}} \cite{Cramer_2016}.  [We note that Ruth Kastner has extended her ``probabilist'' variant of the TI, which embraces the Heisenberg/probability view and characterizes transactions as events in many-dimensional Hilbert space, into the quantum-relativistic domain \cite{Kastner_2012, Kastner_2018} and has used it to extend and enhance the ``decoherence'' approach to quantum interpretation \cite{Kastner_2020}].

This paper begins with a tutorial review of the TI approach to a credible photon mechanism developed in the book
 {\textbf {\textit {Collective Electrodynamics}}} \cite{Mead_2000}, followed by a deeper dive into the electrodynamics of the quantum handshake,
 and finally includes descriptions of several historic experiments that have excluded entire classes of theories.  We conclude that the
 approach described here has {\it not} been excluded by any experiment to date.

\subsection{Wheeler--Feynman Electrodynamics}

The Transactional Interpretation was inspired by classical time-symmetric
Wheeler--Feynman electrodynamics \cite{Wheeler_1945, Wheeler_1949} (WFE), sometimes called ``absorber theory.''  Basically, WFE assumes that electrodynamics must be time-symmetric, with equally valid retarded waves (that arrive {\em after} they are emitted) and advanced waves (that arrive {\em before} they are emitted).  WFE describes a ``handshake'' process accounting for emission recoil in which the emission of a retarded wave stimulates a future absorber to produce an advanced wave that arrives back at the emitter at the instant of emission.  WFE~is based on electrodynamic time symmetry and has been shown to be completely interchangeable with conventional classical electrodynamics in its predictions.\

WFE asserts that the breaking of the intrinsic  time-symmetry to produce the electromagnetic arrow of time, i.e., the observed dominance of retarded radiation and absence of advanced radiation in the universe, arises from the presence of more
absorption in the future than in the past.  In an expanding universe, that assertion is questionable.  One of us has suggested an alternative cosmological explanation \cite{Cramer_1983},
which employs advanced-wave termination and reflection from the singularity of the Big Bang.

\subsection{The Transactional Interpretation of Quantum Mechanics}

The Transactional Interpretation of quantum mechanics \cite{Cramer_2016} takes the concept of a WFE handshake from the classical regime into the quantum realm of photons and massive particles.  The retarded and advanced waves of WFE become the quantum wave functions $\psi$ and $\psi$*.  Note that the complex conjugation of $\psi$* is in effect the application of the Wigner time-reversal operator, thus representing an advanced wave function that carries negative energy from the present to the past.

Let us here clarify what an {\it interpretation} of quantum mechanics actually is.  An interpretation serves the function of explaining and clarifying the formalism and procedures of its theory.  In our view, the mathematics is (and should be) exclusively contained in the formalism itself.  The interpretation should not introduce additional variant formalism.  [We note, however, that this principle is violated by the Bohm--de Broglie ``interpretation'' with its ``quantum potentials'' and uncertainty-principle-violating trajectories, by the Ghirardi--Rimini--Weber ``interpretation'' with its nonlinear stochastic term, and by many other so-called interpretations that take the questionable liberty of modifying the standard QM formalism.  In that sense, these are alternative variant quantum theories, {\em not} interpretations at all.]\

The present work is a calculation describing the formation of a transaction that was inspired by the Transactional Interpretation but has not previously been a part of it.  In Section \ref{Relevance to TI} below, we discuss how the TI is impacted by this work.
We use Schr\"odinger's original wave mechanics formalism {\it with the inclusion of retarded and advanced electromagnetic four-potentials}
to describe and illuminate the processes of transaction formation and the collapse of the wave function.
We  show that this approach can provide a detailed mathematical description of a ``quantum-jump'' in which what seems to be a photon is
emitted by one hydrogen atom in an excited state and excites another hydrogen atom initially
in its ground state.  Thus, the mysterious process of wave function collapse becomes just a phenomenon involving an exchange of
advanced/retarded electromagnetic waves between atomic wave functions described by the Schr\"odinger formalism.

\begin{figure}[H]
\centering
\includegraphics[width=14 cm]{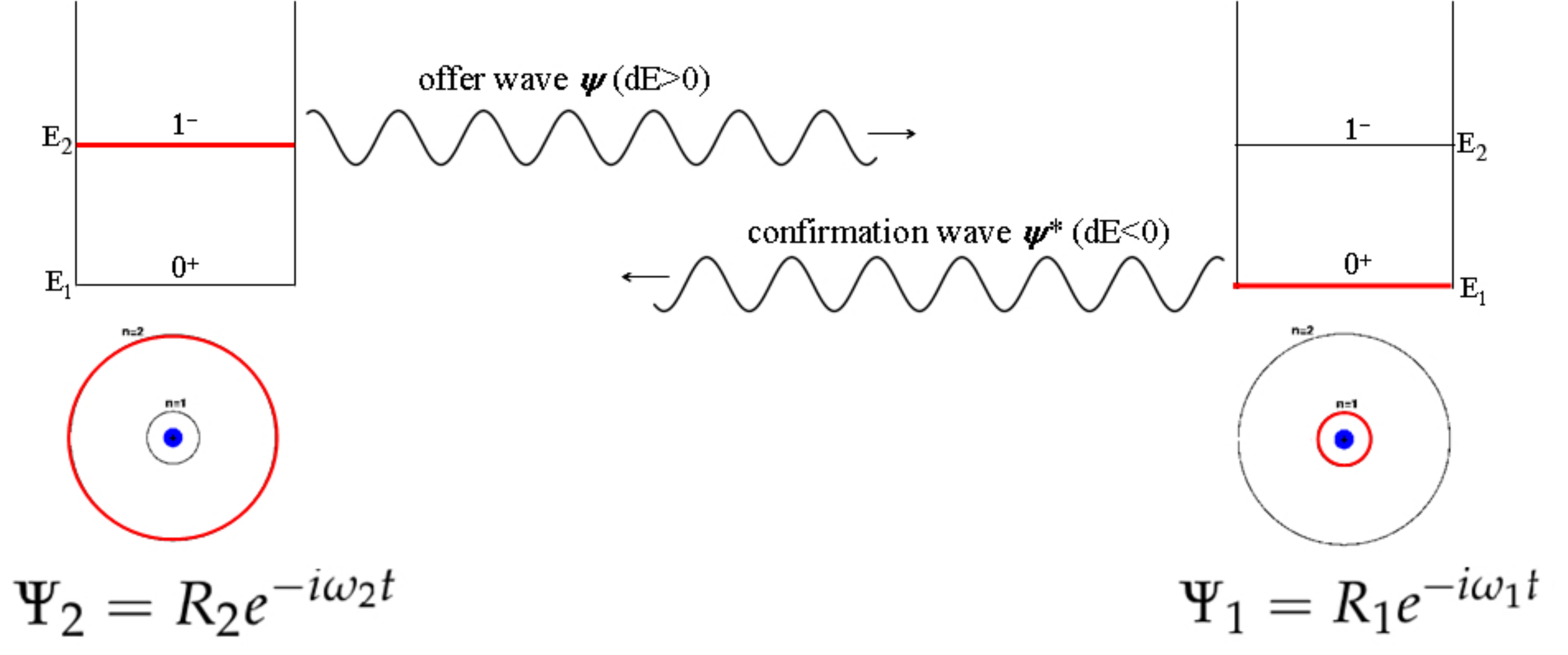}
\caption{Model of transaction formation: An emitter atom $2$ in a space-antisymmetric excited state
of energy $E_2$ and an absorber atom $1$ in a space-symmetric ground state of energy $E_1$ both have slight admixtures of the other state, giving both atoms dipole moments that oscillate with the same difference-frequency $\omega_0 = \omega_2 - \omega_1$.  If the relative phase of the initial small offer wave $\psi$ and confirmation wave $\psi*$  is optimal, this condition initiates energy transfer, which avalanches to complete transfer of one photon-worth of energy $\hbar\omega_0$.}
\label{twoatom}
\end{figure}

As illustrated schematically in Figure~\ref{twoatom}, the process described involves the initial existence in each atom of a very small admixture of the wave function for the opposite state, thereby forming two-component states in both atoms.  This causes them to become weak dipole radiators oscillating at the same difference-frequency $\omega_0$.  The interaction that follows, characterized by a retarded-advanced exchange of 4-vector potentials, leads to an exponential build-up of a transaction, resulting in the complete transfer of one photon worth of energy
$\hbar\omega_0$ from one atom to the other.  This process is described in more detail below.\

\
\vspace{-14pt}

\section{Physical Mechanism of the Transfer}
\label{Physical Mechanism}

The standard formalism of QM consists of a set of arbitrary rules, conventionally viewed as dealing only with probabilities.
When illuminated by the TI, that formalism hints at an underlying physical mechanism that might be understood, 
in the usual sense of the concept {\it understood}.
The~first glimpse of such an understanding, and of the physical nature of the transactional symmetry, \mbox{was
suggested} by Gilbert N. Lewis in 1926 \cite{Lewis_1926,Lewis_1926a}, the same year he gave electromagnetic energy transfer the name~``photon'':

\begin{quote}{\it
``It is generally assumed that a radiating body emits light in every direction, quite regardless of whether there are near or distant objects which may ultimately absorb that light; in other words that it radiates 'into space'...''

``I am going to make the contrary assumption that an atom never emits light except to another atom...''

``I propose to eliminate the idea of mere emission of light and substitute the idea of {\it transmission},
or~a process of exchange of energy between two definite atoms...
Both atoms must play coordinate and symmetrical parts in the process of exchange...''

``In a pure geometry it would surprise us to find that a true theorem becomes false when the page
upon which the figure is drawn is turned upside down.  A dissymmetry alien to the pure geometry
of relativity has been introduced by our notion of causality.''}
\end{quote}

In what follows, we demonstrate that the pair of coupled Schr\"odinger equations describing the two atoms,
as coupled by a relativistically correct description of the electromagnetic field, exhibit a unique solution.
This solution has exactly the symmetry of the TI and thus provides a {\it physically understandable} mechanism
for the experimentally observed behavior:
Both atoms, in the words of Lewis, ``{\it play coordinate and symmetrical parts in the process of exchange.}''\

The solution gives a smooth transition in each of the atomic wave functions, brought to abrupt closure by the highly nonlinear
increase in coupling as the transition proceeds.
The origin of statistical behavior and ``quantum randomness'' can be understood in terms of the random distribution
of wave-function amplitudes and phases provided by the perturbations of the many other potential recipient atoms; 
no ``hidden variables'' are required.  Although much remains to be done,
these~findings might be viewed as a next step towards a physical understanding of the process of quantum energy~transfer.

We will close by indicating the deep, fundamental questions that we have not addressed, and that must be understood
before anything like a complete physical understanding of QM is in hand.

\section{Quantum States}
\label{Quantum States}

In 1926, Schr\"odinger, seeking a wave-equation description of a quantum system with mass, adopted Planck's notion that energy was somehow proportional to frequency together with  deBroglie's idea that momentum was the propagation vector of a wave
and crafted his wave equation for the time evolution of the wave function $\Psi$ \cite{Schrodinger_1926a}:
\begin{equation}
-\frac{\hbar}{2m\,i}\nabla^2\Psi +\frac{q\,V}{i\,\hbar} \Psi = \frac{\partial\Psi}{\partial t}.
\label{SchEqwithV}
\end{equation}
Here, $V$ is the electrical potential, $m$ is the electron mass, and $q$ is the (negative) charge on the electron. 
Thus, what is the meaning of the wave function $\Psi$ that is being characterized? 
In modern treatments, $\Psi$ is called a ``probability amplitude'', which has only a probabilistic interpretation.
In what follows, however, we return to Schr\"odinger's original vision, which
provides a detailed physical picture of the wave function and how it interacts with other charges:

\begin{quote}{\it
``The hypothesis that we have to admit is very simple, namely that the square of the absolute value of $\Psi$
is proportional to an electric density, which causes emission of light according to the laws of ordinary electrodynamics.''}
\end{quote}

That vision has inspired generations of talented conceptual thinkers to invent solutions to technical problems using Schr\"odinger's approach. 
Foremost among these was Ed Jaynes who, with a number of students and collaborators,
attacked a host of quantum problems in this manner \cite{Jaynes_1958,Jaynes_1963,Lamb_1969,Crisp_1969,Stroud_1970,Leiter_1970,Jaynes_1970,Jaynes_1973MW}. 
A great deal of physical understanding was obtained, in particular concerning
lasers and the coherent optics made possible by them.  The theory evolved rapidly and had an enabling role in the explosive progress
of that field.  Indeed, the continued rapid technical progress into the present is due, in no small part, to the understanding gained through
application of the Jaynes way of thinking.  A detailed review of the progress up to 1972 was reported in a
conference that year \cite{Jaynes_1973MW}.  By then this class of theory was called {\bf neoclassical} (NCT)
because of its use of Maxwell's equations.  While there was no question about the utility of NCT in the conceptualization and
technical realization of amazing quantum-optics devices and their application, there was a deep concern about whether it could possibly be
correct {\it at the fundamental level}---maybe it was just a clever bunch of hacks.  The tension over this concern was a major focus
of the 1972 {\em Third Rochester Conference on Coherence and Quantum Optics} \cite{Mandel_1973}, and several experiments testing NCT predictions were discussed there by Jaynes \cite{Jaynes_1973MW}.  
\vskip10pt
He ended his presentation this way:
\begin{quote}{\it
``We have not commented on the beautiful experiment reported
here by Clauser \cite{Clauser_1973MW}, which opens up an entirely new area of fundamental
importance to the issues facing us...'' 

``What it seems to boil down to is this: a perfectly harmless
looking experimental fact (nonoccurence of coincidences at $90^\circ$),
which amounts to determining a single experimental point---and with
a statistical measurement of unimpressive statistical accuracy---can, at a single stroke, throw out a whole infinite class of alternative
theories of electrodynamics, namely all local causal theories.''

``...if the experimental result is confirmed by
others, then this will surely go down as one of the most incredible
intellectual achievements in the history of science, and my own work will lie in ruins.''}
\end{quote}

The experiment he was alluding to was that of Freedman and Clauser \cite{FreedmanClauser_1972}, and in particular to their observaton of an essentially zero coincidence rate with crossed polarizers. 
The Freedman--Clauser experiment (see Subsection \ref{F-C Expt} below), with its use of entangled photon pairs, was the vanguard of an entire new direction in quantum physics that now goes under the rubric of {\bf Tests of Bell's Inequality} \cite{Bell_1964,Bell_1966} and/or {\bf EPR experiments} \cite{FreedmanClauser_1972,Fry_1976,Clauser_1976,Aspect_1981,Aspect_1982,Aspect_1982a}. 
Both the historic EPR experiment and its analysis have been repeated many times with ever-increasing precision, and always with the same outcome: a difinitive violaton of Bell's inequalities.  Local causal theories were dead!  [Much of the literature on violations of Bell's inequalities in EPR experiments has unfortunately emphasized the refutation of {\em local hidden-variable theories}.  In our view, this is a regrettable historical accident attributable to Bell.  {\it Nonlocal} hidden-variable theories have been shown to be compatible with EPR results.  It is {\it locality} that has been refuted.  Entangled systems exhibit correlations that can only be accomodated by quantum nonlocality.  The TI supplies the mechanism for that nonlocality.]

In fact, it was the manifest quantum nonlocality evident in the early EPR experiments of the 1970s that led to the synthesis of the transactional interpretation in the 1980s \cite{Cramer_1980,Cramer_1983,Cramer_1986}, designed to compactly explain entanglement and nonlocality.  This in turn led to
the search for an underlying transaction mechanism, as reported in 2000 in {\textbf {\textit {Collective Electrodynamics}}} \cite{Mead_2000}.  As we detail below, the quantum handshake, as mediated by advanced/retarded electromagnetic four-potentials, provides the effective non-locality so evident in modern versions of these EPR experiments. 
In Section \ref{Historic Tests}, we analyze the Freedman--Clauser experiment in detail and show that their result is a natural outcome of our approach. 
Jaynes' work does not lie in ruins---all that it needed for survival was the non-local quantum handshake!  What follows is an extension and modification of NCT
using a different non-Maxwellian form of E\&M~\cite{Mead_2000} and including our non-local Transactional approach.  We illustrate the approach with the simplest possible physical arrangements, 
described with the major goal of conceptual understanding rather than exhaustion.  Obviously, much more work needs to be done, which we 
point out where~appropriate.

\subsection{Atoms}

We will begin by visualizing the electron as Schr\"odinger and Jaynes did: as having a smooth charge distribution in three-dimensional space,
whose density is given by $\Psi^* \Psi$. 
There is no need for statistics and probabilities at any point in these calculations, and none of the quantities have statistical meaning. 
The probabilistic outcome of quantum experiments has the same origin as it does in all other experiments---random perturbations
beyond the control of the experimenter.  We return to the topic of probability after we have established the nature of the transaction.

For a local region of positive potential $V$, for example near a positive proton, the negative electron's wave function has a local potential energy ($q V$) minimum in which the electron's wave function can form local {\bf bound states}.  The spatial shape of the wave function
amplitude is a trade-off between getting close to the proton, which lowers its potential energy, and bunching together too much, which
increases its $\nabla^2$ ``kinetic energy.''  Equation~\eqref{SchEqwithV} is simply a mathematical expression of this trade-off, a statement of the physical relation between mass, energy, and momentum in the form of a wave equation.

A discrete set of states called {\bf eigenstates} are standing-wave solutions of Equation~\eqref{SchEqwithV} and have the form
$\Psi=R e^{-i\omega t}$, where $R$ and $V$ are functions of only the spatial coordinates,
and the angular frequency $\omega$ is itself independent of time.
For the hydrogen atom, the potential $V=\epsilon_0 q_p/r$, where~$q_p$ is the positive charge on the nucleus, equal in magnitude to the electron charge $q$.
Two of the lowest-energy solutions to Equation~\eqref{SchEqwithV} with this potential are:
\begin{equation}
\begin{aligned}
\Psi_{100} &= \frac{e^{-r}}{\sqrt{\pi}}\,e^{-i\omega_1 t}\qquad\qquad
\Psi_{210} = \frac{r\, e^{-r/2}\cos{\!(\theta)}}{4\sqrt{6\pi}}\,e^{-i\omega_2 t},
\end{aligned}
\label{SchEqSols}
\end{equation}
where the dimensionless radial coordinate $r$ is the radial distance divided by the {\bf Bohr radius} $a_0$:
\begin{equation}
\begin{aligned}
a_0\equiv\frac{4\pi\epsilon_0\hbar^2}{m q^2}=0.0529~{\rm nm,} 
\end{aligned}
\label{BohrRad}
\end{equation}
and $\theta$ is the azimuthal angle from the North Pole ($+z$ axis) of the spherical coordinate system.

\begin{figure}[H]
\centering
\includegraphics[height=6cm]{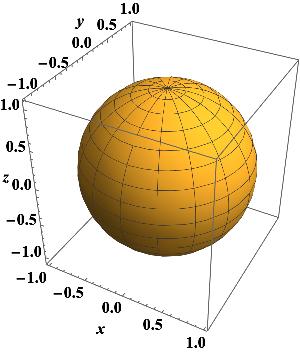}\hskip2.5cm\includegraphics[height=7.5cm]{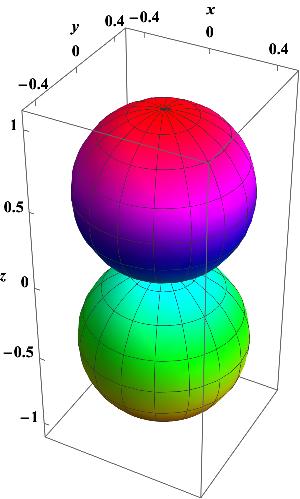}
\caption{Angular dependence of the spatial wave function amplitudes for the lowest (100, \textbf{left}) and next higher (210, \textbf{right})
states of the hydrogen atom, plotted as unit radius in spherical coordinates from Equation~\eqref{SchEqSols}.
The 100 wave function has spherical symmetry: positive in all directions. \mbox{The 210}~wave~function is antisymmetric
along the $z$-axis, as indicated by the color change.  In practice, the direction of the $z$-axis will be established by an external 
electromagnetic field, as we shall analyze~shortly.}
\label{H10H21wf}
\end{figure}

\begin{figure}[H]
\centering
\includegraphics[width=\linewidth/2]{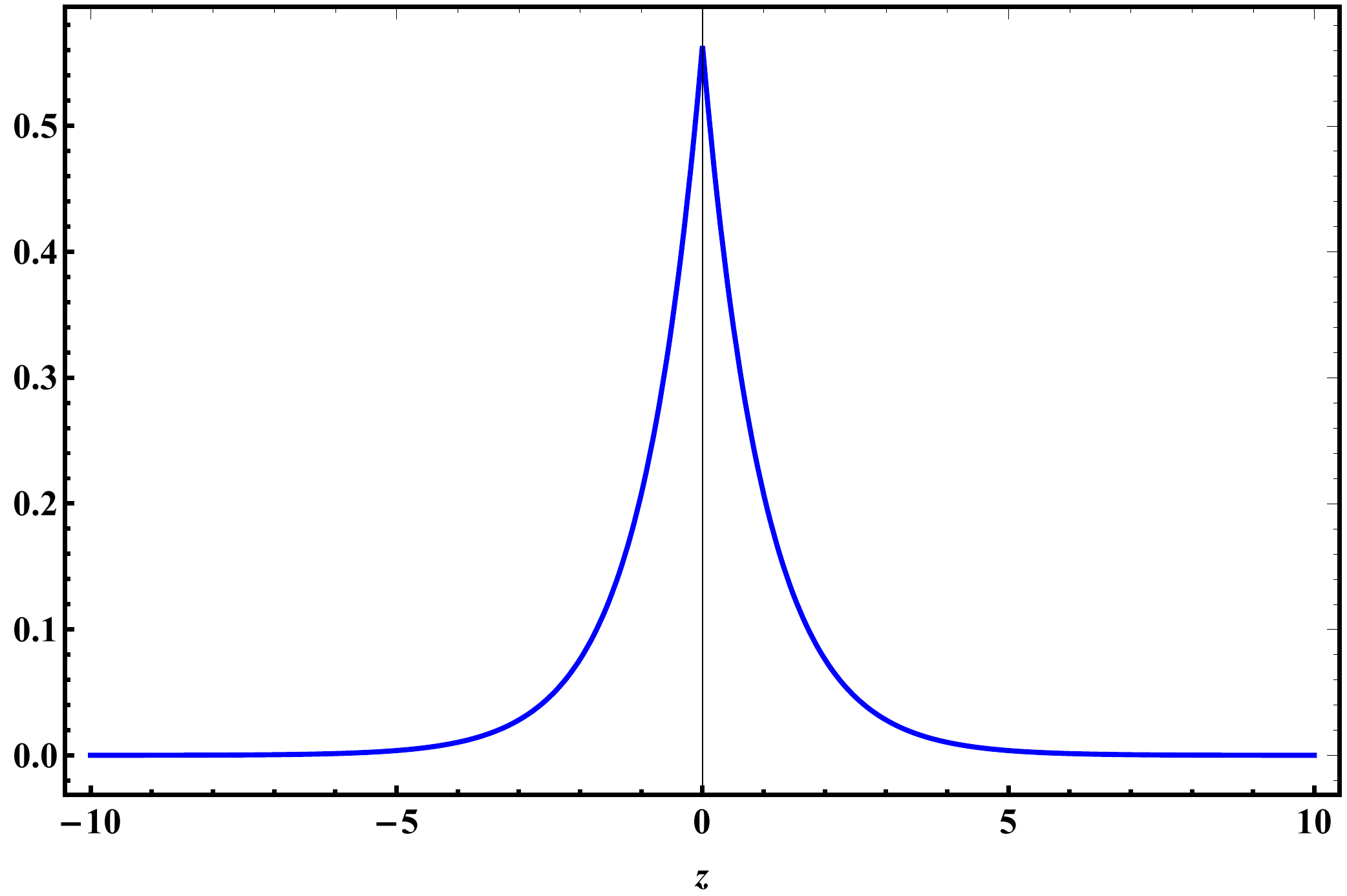}\includegraphics[width=\linewidth/2]{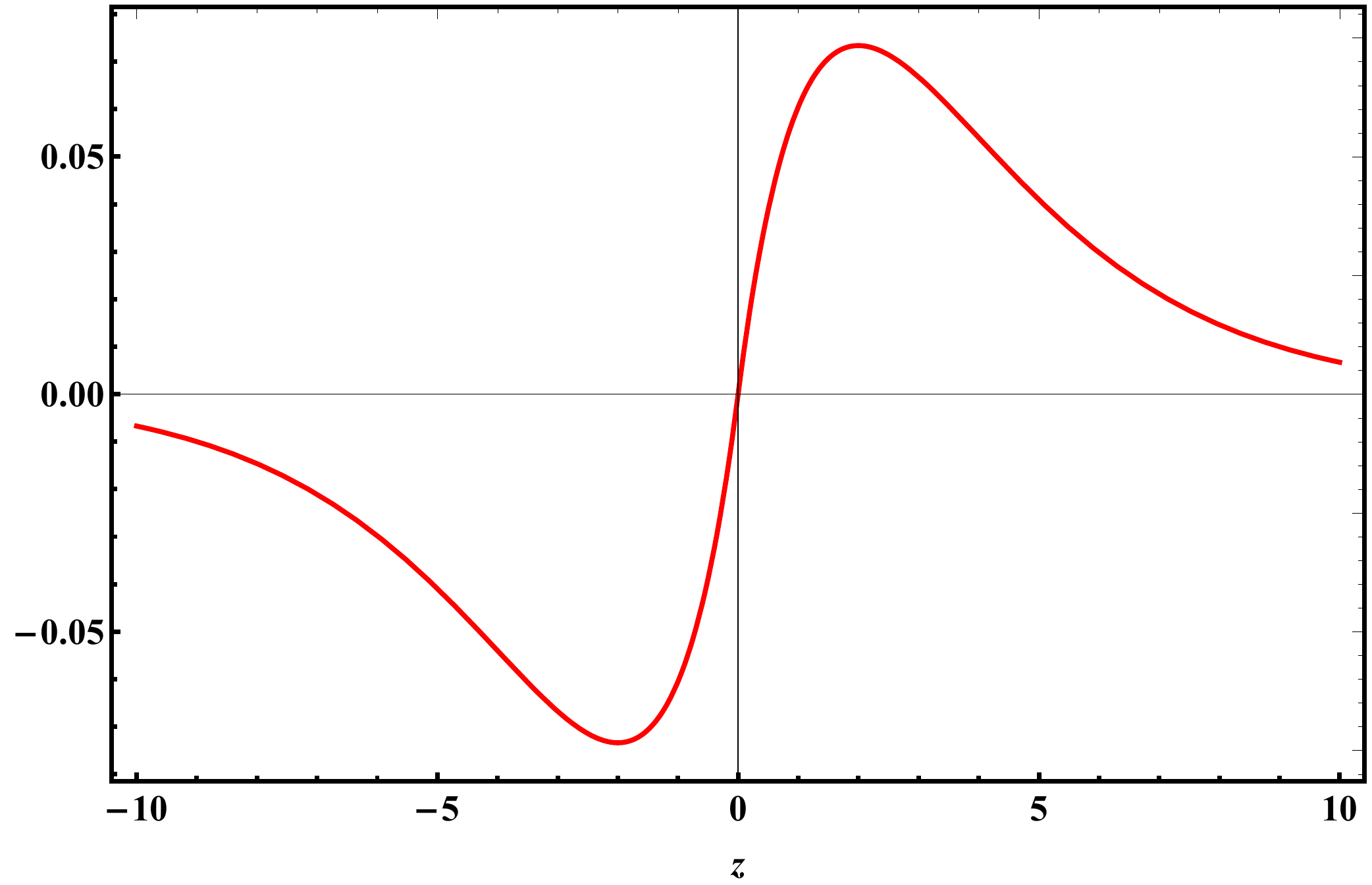}
\caption{Wave function amplitudes $\Psi$ for the 100 and 210 states, along the $z$-axis of the hydrogen atom. 
The horizontal axis in all plots is the position along the $z$-axis in units of the Bohr radius.
\label{H10H21amp}}
\end{figure}

\begin{figure}[H]
\centering
\includegraphics[width=\linewidth/2]{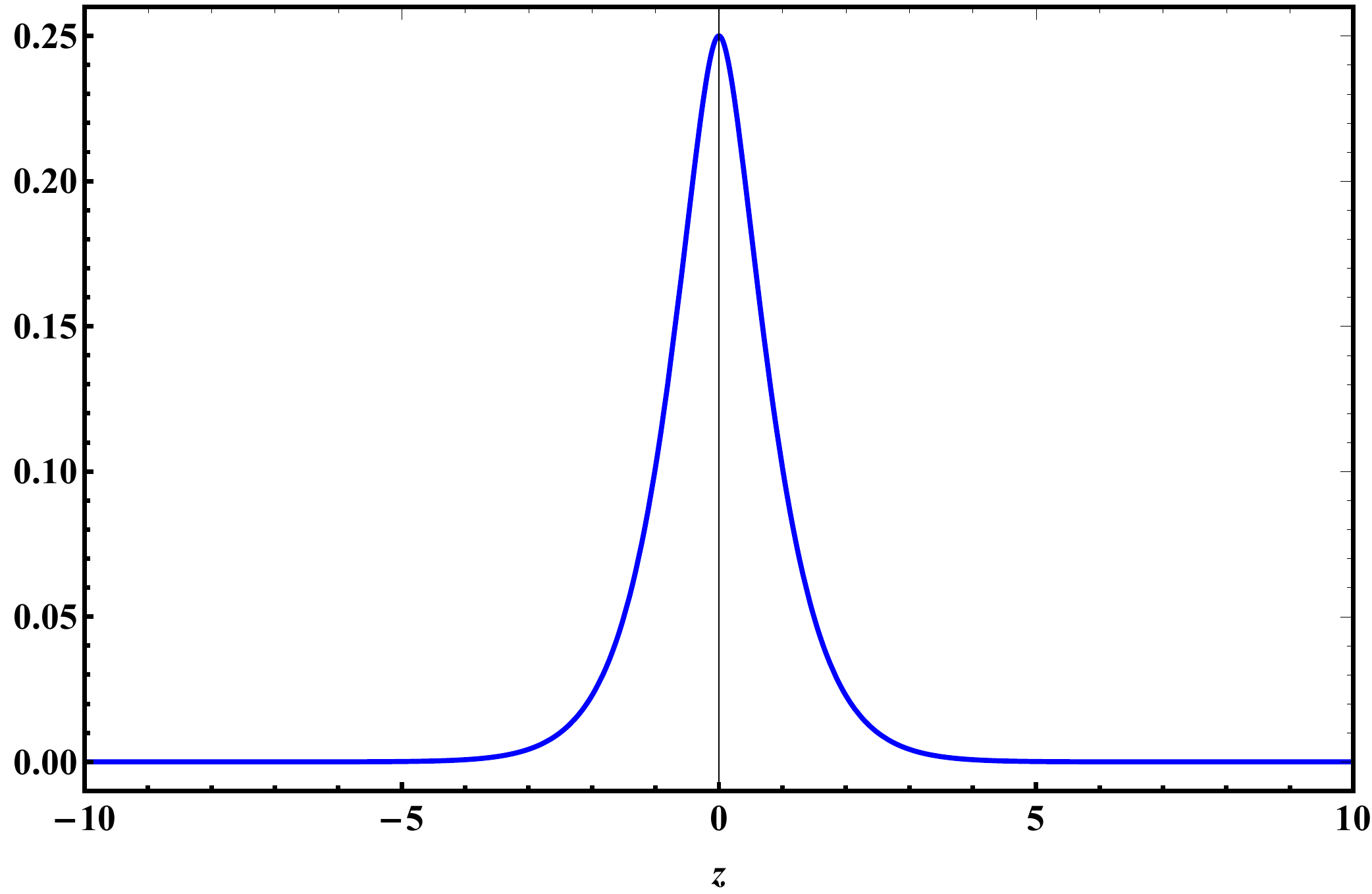}\includegraphics[width=\linewidth/2]{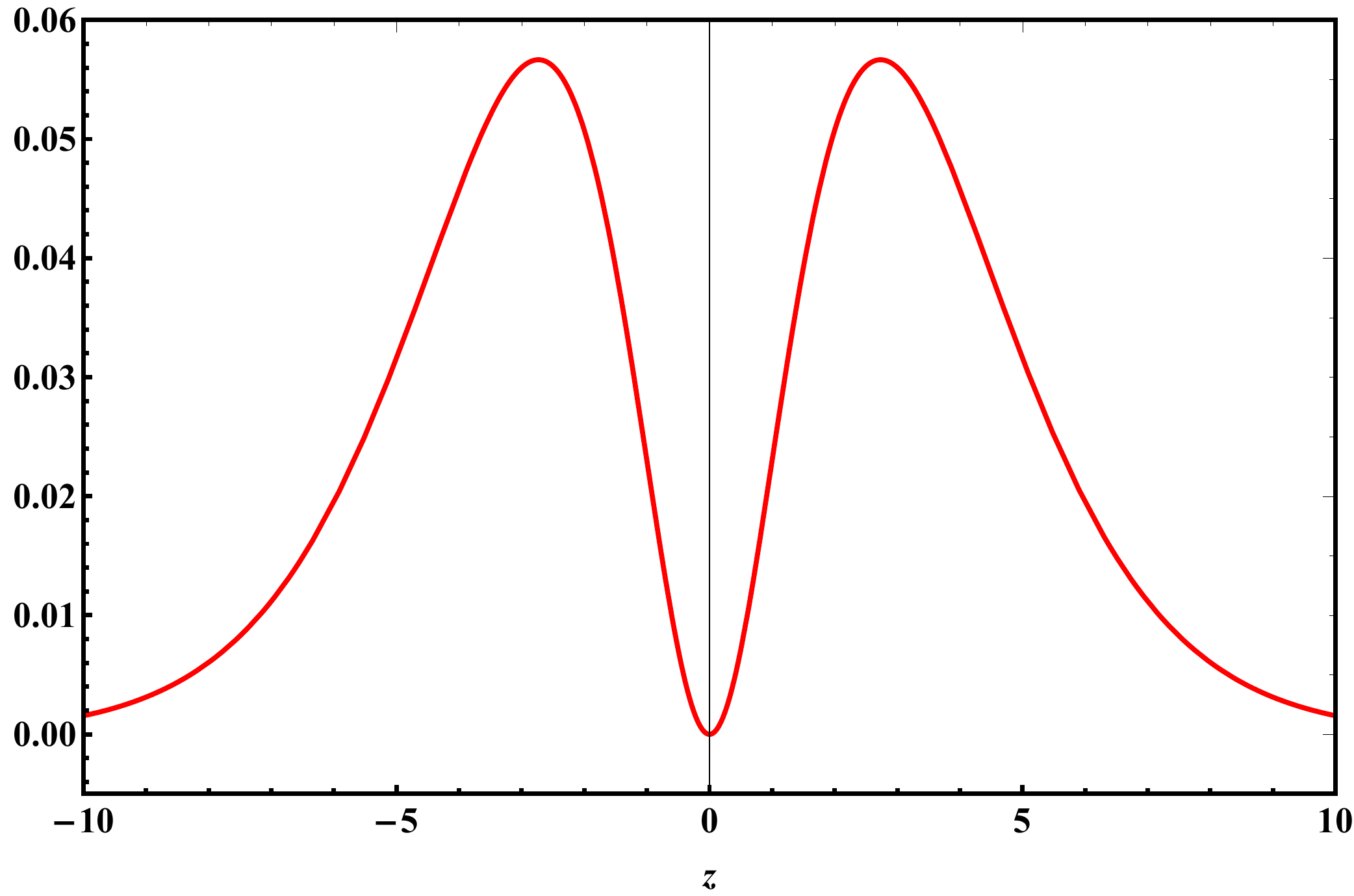}
\caption{Contribution of $x-y$ ``slices'' at position $z$ of wave function density $\Psi^* \Psi$ to the total charge or mass
of the 100 and 210 states of the hydrogen atom. Both curves integrate to 1.
\label{H10H21rho}}
\end{figure}

The spatial ``shape'' of the two lowest energy eigenstates for the hydrogen atom is shown in Figure~\ref{H10H21wf}.  Here, we focus on the excited-state wave function $\Psi_{210}$ that has no angular momentum projection on the $z$-axis.  For the moment, we set aside the wave functions  $\Psi_{21\pm1}$, which have $+1$ and $-1$ angular momentum $z$-axis projections. 
Because, for any single eigenstate, the electron density is $\Psi^* \Psi=R e^{i\omega t}R e^{-i\omega t}=R^2$, the charge density
is not a function of time, so none of the other properties of the wave function change with time. 
The individual eigenstates are thus {\bf stationary states}. 
The~lowest energy bound eigenstate for a given form of potential minimum is called its {\bf ground state},
shown on the left in Figure \ref{H10H21amp}.  The corresponding charge densities are shown in Figure \ref{H10H21rho}.

In 1926, Schr\"odinger had already derived the energies and wave functions for the stationary solutions of his equation for the hydrogen atom.  His physical insight that the absolute square $\Psi^*\Psi$ of the wave function was the {\em electron density} had enabled him to work out the energy shifts of these levels caused by external applied electric and magnetic fields, the expected strengths of the transitions between pairs of energy levels, and the polarization of light from certain transitions.   

These predictions could be compared directly with experimental data, which had been previously observed but not understood. 
He reported that these calculations were:
\begin{quote}{\it
``...not at all difficult, but {\it very} tedious.  In spite of their tediousness, it is rather fascinating to see all the well-known but not
understood ``rules'' come out one after the other as the result of familiar elementary and absolutely cogent analysis, like {\it e.g.}
the fact that $\int_0^{2\pi} \cos{m\phi}\ \cos{n\phi}\ d\phi$ vanishes unless $n=m$.  Once the hypothesis about $\Psi^*\Psi$ has been made,
no accessory hypothesis is needed or is possible;  none could help us if the ``rules'' did not come out correctly. 
However, fortunately they do~\cite{Schrodinger_1928a, Schrodinger_1926a}.''}
\end{quote}

The Schr\"odinger/Jaynes approach enables us to describe continuous quantum transitions in an intuitively appealing way: 
We extend the electromagnetic coupling described in Collective Electrodynamics \cite{Mead_2000} to the wave function of a single electron,
and require only the most rudimentary techniques of Schr\"odinger's original quantum theory.

\section{The Two-State System}
\label{Two-State System}

The first two eigenstates of the Hydrogen atom, from Equation~\eqref{SchEqSols}, form an ideal two-state system.\\
We refer to the 100 ground state as State~1, with wave function $\Psi_1$ and energy $E_1$,  
and to the 210 excited state as State~2, with wave function $\Psi_2$ and energy $E_2>E_1$:
\begin{equation}
\Psi_1 = R_1 e^{-i\omega_1 t} 
\qquad \qquad \Psi_2 = R_2 e^{-i\omega_2 t},
\label{wavefuncs}
\end{equation}
where $\omega_1=E_1/\hbar$, $\omega_2=E_2/\hbar$, and $R_1$ and $R_2$ are real,
time-independent functions of the space coordinates. 
The wave functions represent totally continuous matter
waves, and all of the usual operations involving the wave function are
methods for computing properties of this continuous distribution.  The only
particularly quantal assumption involved is that the wave function obeys a {\bf normalization condition}:
\begin{equation}
\int\Psi^*  \Psi \ d{\rm vol}= 1,
\label{norm}
\end{equation}
where the integral is taken over the entire three-dimensional volume where $\Psi$ is
non-vanishing.  [Envelope
functions like $R_1$ and $R_2$ generally die out exponentially with distance 
sufficiently far from the ``region of interest,'' such as an atom.  
Integrals like this one and those that follow in principle extend to infinity but in practice are taken out far enough that the part
being neglected is within the tolerance of the calculation.].

Equation~\eqref{norm} ensures that the total charge will be a single electronic charge, and
the total mass will be a single electronic mass.

By construction, the eigenstates of the Schr\"odinger equation are real and orthogonal: 
\begin{equation}
\begin{aligned}
\int R_1R_2\ d{\rm vol}=0.
\end{aligned}
\label{Rproperties}
\end{equation}
The first moment $\left<z\right>$ of the electron distribution along the atom's $z$-axis is:
\begin{equation}
\left<z\right>\equiv\int\Psi^* z\, \Psi\ d{\rm vol},
\label{moment}
\end{equation}

{In statistical treatments, $\left< z \right>$ would be called the
"expectation value of $z$'', whereas for our continuous distribution it is called the ``average value of $z$'' or the ``first moment of 
z.''~~The electron wave function
is a {\em wave packet} and is subject to all the Fourier properties of one, as treated at some length in Ref. \cite{Cramer_2016}. 
Statistical QM insisted that electrons were ``point particles'', so one was no longer able to visualize how they could exhibit interference or other wave properties, so a set of rules was constructed to make the outcomes of statistical experiments come out right.  Among these was the
{\bf uncertainty principle}, which simply restated the Fourier properties of an object described by waves in a statistical context.
No statistical attributes are attached to any properties of the wave function in this treatment.}

Equation~\eqref{moment} gives the position of the center of negative charge of the electron wave function
relative to the positive charge on the nucleus.  When multiplied by the electronic charge $q$, it is called {\it the 
electric dipole~moment}\index{dipole!moment} $ q\left<z\right>$ of the charge distribution of the atom:\index{charge distribution}  
\begin{equation}
q\left<z\right>\ = q\int\Psi^* z\, \Psi\ d{\rm vol.}
\label{dipolemoment}
\end{equation}
From  Equations~\eqref{moment} and~\eqref{wavefuncs}, the dipole moment for the $i$th \index{eigenstate}eigenstate~is:
\begin{equation} q\left<z_i\right>\ = q\int\Psi_i^* z\, \Psi_i\ d{\rm vol} 
= q\int R_i^* z\, R_i\ d{\rm vol} = q\int R_i^2\, z\ d{\rm vol.}
\label{eigenmoment}
\end{equation}

Pure eigenstates of the system will have a definite parity, i.e., they will have wave functions 
with either even symmetry [$\Psi(z)=\Psi(-z)$], or odd symmetry [$\Psi(z)=-\Psi(-z)$].  For either symmetry,
the integral over $R^{2}z$ vanishes, and the dipole moment is zero.  
We note that, even if the wave function did not have even or odd symmetry, the dipole moment, and all
higher moments as well, would be independent of time.  By their very nature, eigenstates are
\index{stationary!state}stationary states and can be visualized as standing-waves---none of their
physical spatial attributes can be functions of time.  In order to radiate electromagnetic energy,
the charge distribution {\it must change with time}.

 The notion of {\it stationarity} is the quantum answer to the original question about atoms depicted as electrons orbiting a central nucleus like a tiny Solar System:\\
\hbox to 22pt{}{\it  Why doesn't the electron orbiting the nucleus radiate its energy away?}\\
In his 1917 book, {\textbf {\textit {The Electron}}},  R.A. Millikan \cite{Millikan_1917} 
anticipates the solution in his comment about the
\begin{quote}
{\it''\dots apparent contradiction involved in the non-radiating electronic orbit---a
contradiction which would disappear, however, if the negative electron itself, when inside the
atom, were a ring of some sort, capable of expanding to various radii, and capable, only when freed
from the atom, of assuming essentially the properties of a point charge.''}
\end{quote}

Millikan was the first researcher to directly observe and measure the quantized charge on the electron with
his famous oil-drop experiment, for which he later received the Nobel prize.  Ten years before the statistical quantum theory
was put in place, he had clearly seen
that a continuous, symmetric electronic charge distribution would not radiate, and that the
real problem was the assumption of a point charge.  


\section{Transitions}
\label{Transitions}

The eigenstates of the system form a complete basis set, so any behavior of the system can be expressed by
forming a linear combination (superposition) of its eigenstates.

The general form of such a superposition of our two chosen eigenstates is:
\begin{equation}
\Psi=ae^{i\phi_a}R_1e^{-i\omega_1 t}+be^{i\phi_b}R_2e^{-i\omega_2 t},
\label{mixedphicomplex}
\end{equation}
where $a$ and $b$ are real amplitudes, and $\phi_a$ and $\phi_b$ are real constants that determine the phases of the oscillations $\omega_1$ and $\omega_2$.\\

\begin{figure}[H]
\centering
\includegraphics[width=\linewidth/2]{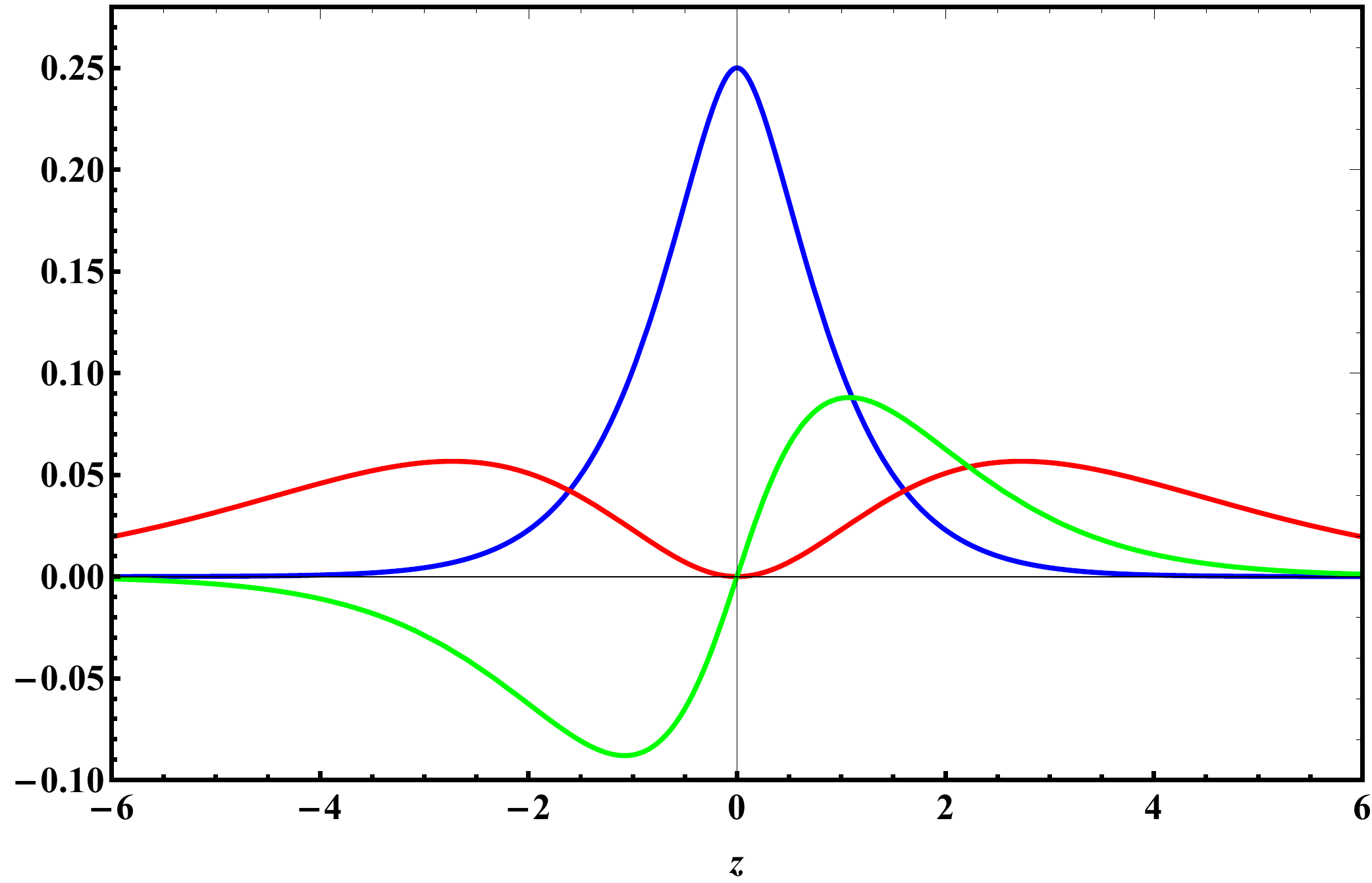}\includegraphics[width=\linewidth/2]{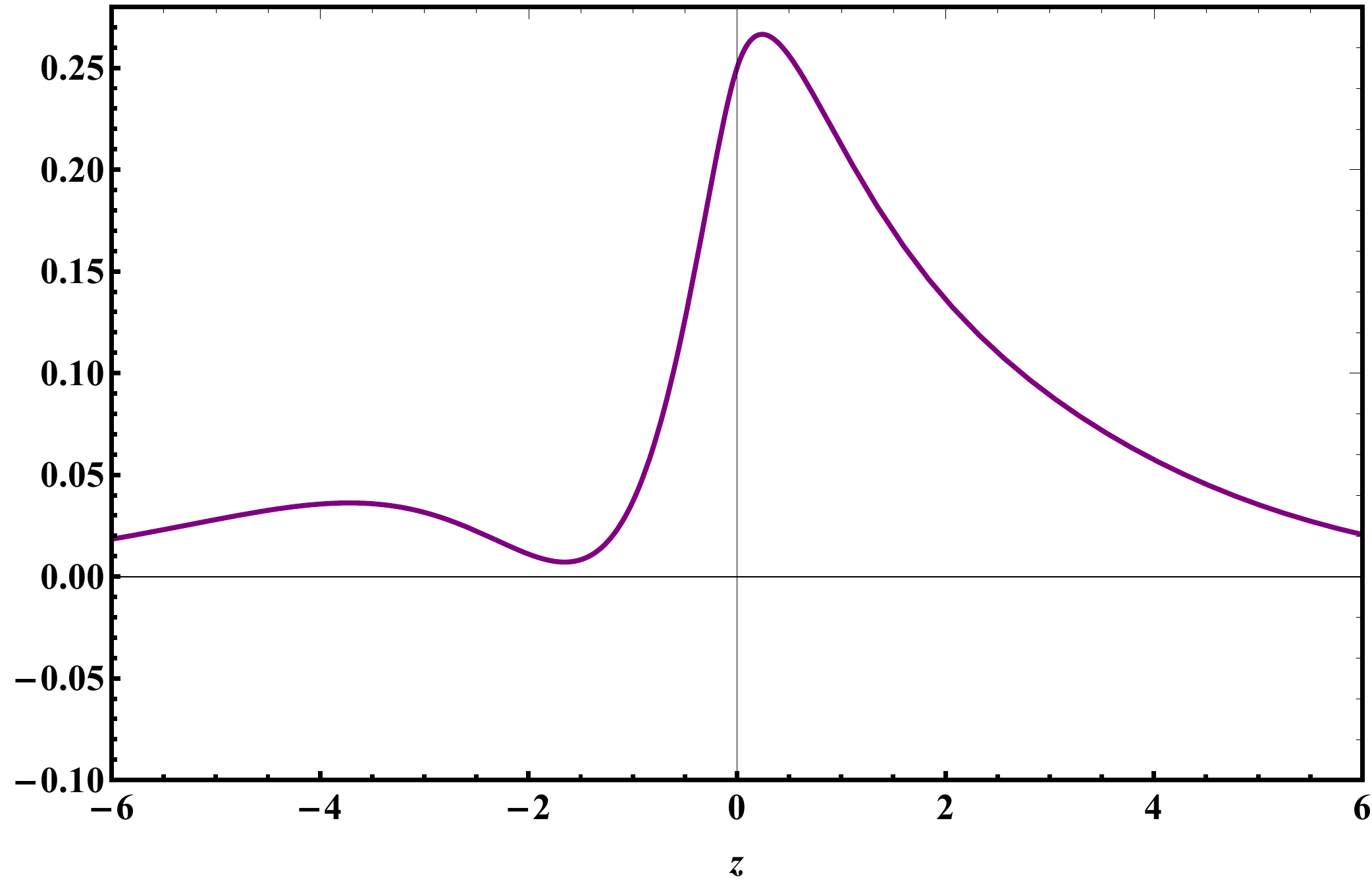}
\caption{{\textbf{Left}:} Plot of the three terms in the wave-function density in  Equation~\eqref{rhomixwf} for an equal $\left(a=b=1/\sqrt{2}\right)$
superposition of the ground state ($R^2_1$, blue) and first excited state ($R^2_2$, red) of the hydrogen atom.  The green curve
is a snapshot of the time-dependent $R_1R_2$ product term, which oscillates at difference frequency $\omega_0$. {\textbf{Right}:} Snapshot of the total charge density, which is the sum
of the three curves in the left plot. The magnitudes plotted are
the contribution to the total charge in an $x-y$ ``slice'' of $\Psi^*\Psi$ at the indicated $z$~coordinate.
All plots are shown for the time such that  $\cos{\!(\omega_0 t+\phi)}=1$. 
The horizontal axis in each plot is the spatial coordinate along the $z$-axis of the atom,
given in units of the Bohr radius $a_0$.  \href{run:Mov2aCM.mov}{Animation here} \cite{AnimLink}
\label{H10H21mix}}
\end{figure}

Using $\omega_0=\omega_2-\omega_1$ and $ \phi=\phi_a-\phi_b$, the charge density $\rho$ of the two-component-state wave function is:
\begin{equation}
\begin{aligned}
\rho&=q\Psi^*\,\Psi \cr
\frac{\rho}{q}&=\left(ae^{-i\phi_a}R_1e^{i\omega_1 t}+be^{-i\phi_b}R_2e^{i\omega_2 t}\right)
\left(ae^{i\phi_a}R_1e^{-i\omega_1 t}+be^{i\phi_b} R_2e^{-i\omega_2 t}\right)\cr
&= a^2 R_1^2+b^2 R_2^2
+\left(ae^{-i\phi_a}be^{i\phi_b}\,e^{-i\omega_0 t}+be^{-i\phi_b}ae^{i\phi_a}\,e^{i\omega_0 t}\right)R_1R_2 \cr
&=a^2 R_1^2+b^2 R_2^2
+ab\left(e^{i(\phi_b-\phi_a)}\,e^{-i\omega_0 t}+e^{i(\phi_a-\phi_b)}\,e^{i\omega_0 t}\right)R_1R_2\cr
&=a^2 R_1^2+b^2 R_2^2+ab\left(e^{-i(\omega_0 t-\phi)}+e^{i(\omega_0 t+\phi)}\right)R_1R_2\cr
&=a^2 R_1^2+b^2 R_2^2+2abR_1R_2\cos{\!(\omega_0 t+\phi)}.
\end{aligned}
\label{rhomixwf}
\end{equation}
Thus, the charge density of the two-component wave function is made up of the charge densities of the two separate wave functions,
shown in Figure~\ref{H10H21rho},
plus a term proportional to the product of the two wave function amplitudes.  It reduces to the individual charge density
of the ground state when $a=1,b=0$ and to that of the excited state when $a=0,b=1$.  The product term, shown in green in
 Figure~\ref{H10H21mix}, is the only non-stationary term;  it oscillates at the transition frequency $\omega_0$.  The integral
 of the total density shown in the right-hand plot is equal to 1 for any phase of the cosine term, since there is only one electron
 in this two-component state.

All the $\Psi^*\Psi$ plots represent the density of negative charge of the electron.  The atom as a whole is neutral
because of the equal positive charge on the nucleus.  The dipole is formed when the center of charge of the
electron wave function is displaced from the central positive charge of the nucleus.

The two-component wave function must be normalized, since it is the state~of~one~electron: 
\begin{equation}
\begin{aligned}
&\int\Psi^*\,\Psi \ d{\rm vol}=1\cr
&=\int\left(ae^{-i\phi_a}R_1e^{i\omega_1 t}+be^{-i\phi_b}R_2e^{i\omega_2 t}\right)\cr
&\quad\quad\left(ae^{i\phi_a}R_1e^{-i\omega_1 t}+be^{i\phi_b} R_2e^{-i\omega_2 t}\right) \ d{\rm vol}\cr
&=a^2\int R_1^2\,d{\rm vol}+b^2\int R_2^2d{\rm vol}\cr
&+\left(ae^{-i\phi_a}be^{i\phi_b}\,e^{-i\omega_0 t}+be^{-i\phi_b}ae^{i\phi_a}\,e^{i\omega_0 t}\right)\int R_1R_2d{\rm vol}.
\end{aligned}
\label{abSchsolnorm}
\end{equation}
Recognizing from Equations~\eqref{norm} and~\eqref{Rproperties} that the individual eigenfunctions
are normalized and~orthogonal:
\begin{equation}
\begin{aligned}
\int R_1^2\, d{\rm vol}=1 \qquad\int R_2^2\, d{\rm vol}=1 \qquad \int R_1R_2\ d{\rm vol}=0.
\end{aligned}
\label{Rproperties1}
\end{equation}
Equation~\eqref{abSchsolnorm} becomes
\begin{equation}
\begin{aligned}
&\int\Psi^*\,\Psi \ d{\rm vol}=1=a^2+b^2.
\end{aligned}
\label{abnormcomplex}
\end{equation}
Thus, $a^2$ represents the fraction of the two-component wave function made up of the lower state $\Psi_1$,
and~$b^2$ represents the fraction made up of the upper state $\Psi_2$. 
The total energy $E$ of a system whose wave function is a superposition of two eigenstates is:
\begin{equation}
E = a^2 E_1 + b^2 E_2.
\label{energy}
\end{equation}
Using the normalization condition $a^2+b^2=1$ and solving Equation~\eqref{energy} for $b^2$, we obtain:
\begin{equation}
b^2 = \frac{{E - E_1}}{{E_2 - E_1}}.
\label{Bsq}
\end{equation}
In other words, $b^2$ is just the energy of the wave function, normalized to the
transition energy, and~using $E_1$ as its reference energy.  Taking $E_1$ as our zero of energy
and $E_0=E_2 - E_1$, Equation~\eqref{Bsq}~becomes:
\begin{equation}
E= {E_0}b^2 \quad\Rightarrow\quad \frac{\partial E}{\partial t}=E_0\frac{\partial \left(b^2\right)}{\partial t}
\label{BsqE}.
\end{equation}
Defining: 
\begin{equation}
d_{12} \equiv 2q\int R_1 R_2\,z\,d{\rm vol} = 2q\left<z\right>_{\rm max},
\label{d12eq}
\end{equation}
the dipole moment $q\left<z\right>$ of such a superposition\index{superposition} can,
from Equation~\eqref{rhomixwf}, be written:

\begin{equation}\begin{aligned}
q\left<z\right>=d_{12}ab\,\cos{\!(\omega_0 t +\phi)}.
\end{aligned}
\label{abmixdipole}
\end{equation}
The factor $d_{12}$ we call the {\bf dipole strength} for the transition.
When one $R$ is an even function of $z$ and the other is an odd
function of $z$, as in the case of the 100 and 210 states of the hydrogen atom,
then $d_{12}$ is nonzero, and the transition is said to be {\bf electric dipole allowed}.
When both $R_1$ and $R_2$ are either even or odd functions
of $z$, $d_{12}=0$, and the transition is said to be {\bf electric dipole forbidden}.

Even in this case, some other moment of the distribution generally will
be nonzero, and the transition can proceed by magnetic dipole, magnetic quadrupole, or other
higher-order moments. 

For now, we will concentrate on transitions that are electric dipole allowed.

We have the time dependence of the electron dipole moment $q\left<z\right>$ from Equation~\eqref{abmixdipole},
from which we can derive the velocity and acceleration of the charge:
\begin{equation}
\begin{aligned}
q\left<z\right>&=d_{12}ab\,\cos{\!(\omega_0 t +\phi)}\cr
q\frac{\partial\left<z\right>}{\partial t}&=-\omega_0d_{12}ab\sin{\!(\omega_0 t+\phi)}
+d_{12}\,\cos{\!(\omega_0 t +\phi)}\frac{\partial (ab)}{\partial t}\cr
&\approx-\omega_0d_{12}ab\sin{\!(\omega_0 t+\phi)}\cr
q\frac{\partial^2\left<z\right>}{\partial t^2}&\approx-\omega_0^2d_{12}ab\cos{\!(\omega_0 t+\phi)},
\end{aligned}
\label{velocity}
\end{equation}
where the approximation arises because we will only consider situations where the coefficients $a$ and $b$ change
slowly with time over a large number of cycles of the transition frequency:
$\left(\frac{\partial (ab)}{\partial t}\ll ab\,\omega_0\right)$.

The motion of the electron mass density endows the electron with a momentum $\vec p$:
\begin{equation}
\begin{aligned}
\vec p = m \vec v \quad\Rightarrow\quad
p_z= m \frac{\partial\left<z\right>}{\partial t}
\approx-\frac{m}{q}\omega_0d_{12}ab\sin{\!(\omega_0 t+\phi)}.
\end{aligned}
\label{momentum}
\end{equation}

\section{Atom in an Applied Field}
\label{Atom in an Applied Field}

Schr\"odinger had a detailed physical picture of the wave function, and he gave an elegant derivation of
the process underlying the change of atomic state mediated by electromagnetic
coupling.  [{The original derivation in Ref. \cite{Schrodinger_1928} p.~137, 
is not nearly as readable as that in Schr\"odinger's second and third 1928 lectures \cite{Schrodinger_1928b}, 
where the state transition is described in Section \ref{Two Atoms at a Distance} starting at the bottom of page~31, for which the second lecture is preparatory.
There he solved the problem more generally, including the effect of a slight detuning of the field frequency from the atom's transition frequency.}]
Instead of directly tackling the transfer of energy between two atoms, he considered the response of a
single atom to a small externally applied vector potential field $\vec A$.
He found that the immediate effect of an applied vector
potential is to change the momentum $p$ of the electron wave function:
\begin{equation}
\begin{aligned}
p_z&= m \frac{\partial\left<z\right>}{\partial t}-q A_z\cr
\frac{\partial p_z}{\partial t}&= m \frac{\partial^2\left<z\right>}{\partial t^2}-q \frac{\partial A_z}{\partial t}.
\end{aligned}
\label{momentum}
\end{equation}
Thus, the quantity $-q \frac{\partial A_z}{\partial t}$ acts as a {\it force}, causing an {\it acceleration} of the electron wave function. 

This is the physical reason that $-\frac{\partial A_z}{\partial t}$ can be treated as an {\it electric field} ${\cal E}_z$.  [{At a large distance
from an overall charge-neutral charge distribution like an atom, the longitudinal gradient of the scalar potential just cancels the longitudinal
component of ${\partial \vec A}/{\partial t}$, so what is left is $\vec{\cal E}={-\partial A_\perp}/{\partial t}$, which is purely transverse.}].
\begin{equation}
{\cal E}_z=-\frac{\partial A_z}{\partial t}.
\label{EfromA}
\end{equation}
In the simplest case, the $q A_z$ term makes only a tiny perturbation to the momentum over a single cycle
of the $\omega_0$ oscillation, so its effects will only be appreciable over many cycles.

We consider an additional simplification, where the
frequency of the applied field is exactly equal to the transition frequency $\omega_0$ of the atom:
\begin{equation}
{A}_z={A}\cos{(\omega_0 t)}\quad\Rightarrow\quad-\frac{\partial A_z}{\partial t}={\cal E}_z={\omega_0 A}\sin{(\omega_0 t)}.
\label{OscA}
\end{equation}
In such evaluations, we need to be very careful to identify {\it exactly which energy} we are calculating: 

The electric field is merely a bookkeeping device to keep track of
the energy that an electron in one atom exchanges with another
electron in another atom, in such a way that the total energy is conserved.  We will evaluate how much energy a given electron
gains from or loses to the field, recognizing that the negative of that energy represents work done by the electron on
the source electron responsible for the field. 
The force on the electron is just $q{\cal E}_z$. 
Because ${\cal E}_z={\omega_0 A}\sin{(\omega_0 t)}$, for a stationary charge, 
the~force is in the $+z$ direction as much as it is in the $-z$ direction, and, averaged over
one cycle of the electric field, the work averages to zero.  
However, if the charge itself oscillates with the electric field,
it~will gain energy $\Delta E$ from the work done by the field on the electron over one cycle:
\begin{equation}
\begin{aligned}
\frac{\Delta E}{\rm cycle}&=\int q {\cal E}_z\, dz=\int_0^{2\pi/\omega_0} q {\cal E}_z \frac{\partial \left<z\right>}{\partial t}\,dt,
\end{aligned}
\label{intfdx}
\end{equation}
where $\left<z\right>$ is the $z$ position of the electron center of charge from Equation~\eqref{velocity}.

When the electron is ``coasting downhill'' {\it with} the electric field, it gains energy and $\Delta E$ is positive. 
When~the electron is moving ``uphill'' {\it against} the electric field, the electron loses energy
and $\Delta E$~is~negative. 
 
As long as the energy gained or lost in each cycle is small compared with $E_0$, we can define a
continuous {\bf power} (rate of change of electron energy), which is understood to be an average over many cycles.  
The time required for one cycle is $2\pi/\omega_0$, so Equation~\eqref{intfdx} becomes: 
\begin{equation}
\begin{aligned}
\frac{\partial E}{\partial t}&=\frac{\omega_0\Delta E}{2\pi}
=\frac{\omega_0}{2\pi}\int_0^{2\pi/\omega_0} q {\cal E}_z \frac{\partial \left<z\right>}{\partial t}\,dt
=\frac{1}{2\pi}\int_0^{2\pi} q {\cal E}_z \frac{\partial \left<z\right>}{\partial t}\,d(\omega_0 t).
\end{aligned}
\label{dEdtav}
\end{equation}

 \section{Electromagnetic Coupling}
 \label{Electromagnetic Coupling}
 
 Because our use of electromagnetism is conceptually quite different from that in traditional Maxwell treatments (including Jaynes' NCT), we review here the foundations of that discipline from the present perspective.  [{A more detailed discussion from the present viewpoint is given in Mead, {\textbf {\textit {Collective Electrodynamics}}} \cite{Mead_2000}. The standard treatment is given in Jackson,  {\textbf {\textit {Classical Electrodynamics, 3rd Edition}}}, Chapter 8 \cite{Jackson}.}]
It is shown in Ref. \cite{Mead_2000} that electromagnetism is of totally quantum origin.  We saw in Equation~\eqref{momentum} that
it is the vector potential $\vec A$ that appears as part of the momentum of the wave function, signifying the coupling of
one wave function to one or more other wave functions.  Thus,  to stay in a totally quantum context, we must work with
electromagnetic relations based on the vector potential and related quantities.
The entire content of electromagnetism is contained in the relativistically-correct Riemann--Sommerfeld second-order differential equation:
\begin{equation}
\left(\nabla^2 -\frac{\partial^2}{{\partial t^2}}\right){\bf A} = -{\mu_0}\ {\bf J},
\label{d4A}
\end{equation}
where ${\bf A}=[\vec A, V/c]$ is the four-potential and ${\bf J}=[\vec J, c\rho]$ is the four-current, $\vec A$ is the vector potential,
$V$~is the scalar potential, $\vec J$ is the physical current density (no displacement current), and $\rho$ is the physical charge density, all expressed in the same inertial frame.

Connection with the usual electric and magnetic field quantities $\vec {\cal E}$ and $\vec B$ is as follows:
\begin{equation}
\vec{\cal E}=-\nabla {V}-\frac{\partial {\vec A}}{\partial t}\qquad\qquad \vec B=\nabla\times\vec A.
\label{pot2fld}
\end{equation}
Thus, once we have the four-potential $\bf A$, we can derive any electromagnetic relations we wish.

Equation~\eqref{d4A} has a completely general Green's Function solution
for the four-potential ${\bf A}(t)$ at a point in space,
from four-current density ${\bf J}(r,t)$ in volume elements $d$vol at at distance $r$ from that~point:
\begin{equation}
\begin{aligned}
{\bf A}(t) &=  \frac{\mu_0}{4\pi }\int \left.\frac{{\bf J}(r,t')}{r}\right|^{t'= t\pm\frac{r}{c}} \ d{\rm vol},
\end{aligned}
\label{A4int}
\end{equation}
where $r$ is the distance from element $d$vol to the point where $\bf A$ is evaluated, assumed large compared to the size of
the atomic wave functions, and $c$ is the speed of light.  

Equation~\eqref{A4int} is the first fundamental equation of electromagnetic coupling:  The vector potential,
which will appear as part of an electron's momentum, is simply the sum of all current elements on that electron's light cone,
each weighted inversely with its distance from that electron. 
The second-order nature of derivatives in Equation~\eqref{d4A} does not favor any particular sign of space or time. 
Thus, the four-potential from a current element on the {\it past} light cone of the electron ($t-r/c$) will be ``felt'' by the electron at later time $t$,
and is termed a {\bf retarded} field.  Conversely, the four-potential from a current element on the {\it future} light cone of the electron ($t+r/c$)
will be ``felt'' by the electron at earlier time $t$, and~is termed an {\bf advanced} field.  Historically, with rare exception, advanced fields
have been discarded as non-physical because evidence for them has been explained in other ways.  We~shall see that modern quantum
experiments provide overwhelming evidence for their active role in \mbox{{\bf quantum entanglement.}}

Equation~\eqref{A4int} can be expressed in terms of more familiar E\&M quantities: 
\begin{equation}
\begin{aligned}
{\vec A}(t) &=  \frac{\mu_0}{4\pi }\int \left.\frac{{\vec J}(r,t')}{r}\right|^{t'= t\pm\frac{r}{c}} \! d{\rm vol}\qquad \qquad \qquad 
{V}(t) =  \frac{\mu_0c^2}{4\pi }\int \left.\frac{{\rho}(r,t')}{r}\right|^{t'= t\pm\frac{r}{c}} \! d{\rm vol}\cr
\end{aligned}
\label{AV3int}
\end{equation}

If the current density $\vec J$ is due to the movement of a small, unified ``cloud'' of charge, as is the case for the wave function of an atomic
electron, and the motion of the wave function is non-relativistic, the $\vec J$ integral just becomes the movement of the center of charge
relative to its average position at the~nucleus:
\begin{equation}
\begin{aligned}
{\vec A}(t) &\approx  \frac{\mu_0}{4\pi }\int \left.\frac{{\rho\vec v}(r,t')}{r}\right|^{t'= t\pm\frac{r}{c}} \! d{\rm vol}
\approx  \frac{\mu_0}{4\pi }\left.\frac{{q\,\vec v(r,t')}}{r}\right|^{t'= t\pm\frac{r}{c}}
\end{aligned}
\label{Awf3int}
\end{equation}
If, as we have chosen previously, the motion is in the $z$ direction, 
\begin{equation}
\begin{aligned}
A_z(t)\approx  \frac{q\mu_0}{4\pi r}\left.\frac{\partial\!\left<z(r,t')\right>}{\partial t'}\right|^{t'= t\pm\frac{r}{c}} 
\end{aligned}
\label{AfromI1}
\end{equation}
If we use the current element as our origin of time, the signs are reversed:
\begin{equation}
\begin{aligned}
\left.A_z(t')\right|^{t'= t\mp\frac{r}{c}} \approx  \frac{q\mu_0}{4\pi r}\frac{\partial\!\left<z(r,t)\right>}{\partial t}
\end{aligned}
\label{AfromI2}
\end{equation}
In this case, $t+r/c$ represents the retarded field and $t-r/c$ represents the advanced field.

We shall use these two forms for the simple examples presented below.

An important difference between standard Maxwell E\&M practice and our use of the four-potential to couple atomic wave functions is highlighted by Wheeler and Feynman \cite{Wheeler_1945}:
\begin{quote}{\it
``There is no such concept as 'the' field,
an independent entity with degrees of freedom of its own.''}
\end{quote}

The field has no degrees of freedom of its own.  It is simply a convenient bookkeeping device for keeping track of the total effect of an arbitrary number of charges
on a particular charge distribution in some region of space.  The general form of interaction energy $E$ is given by:
\begin{equation}
\begin{aligned}
E&=
\int ({\vec A}\cdot{\vec J}+\rho V)\ d{\rm vol}
\end{aligned}
\label{AJint}
\end{equation}
Equation~\eqref{AJint} is the second  fundamental equation of electromagnetic coupling. 
The interaction described in Section \ref{Atom in an Applied Field} is a simplified case of this relation
for a single atom.  The energy minimum created by the positive nucleus is the $V$, and the $\rho$ is the charge distribution of the electron
wave function.  The $\vec A$ from a second atom is very small, and is assumed to not change the eigenfunctions. 
 
From Equations~\eqref{AJint} and~\eqref{A4int}, we see that, when applied to two atoms, what is being described is a way to factor a bi-directional
connection between them so that each can be analyzed separately by Schr\"odinger's equation as a one-electron problem, 
using the vector potential from the other as part of its energy.

The fact that the four-potential field from a charge is defined everywhere on its light cone does not imply that
it is ``radiating into space'', carrying energy with it. 
Energy is only transferred at the position of another charge.  Since all charges are the finite charge
densities of wave functions, {\bf there are no self-energy infinities in this formulation}.

One widely-held viewpoint treats the ``quantum vacuum'' as being made up of an infinite number
of quantum harmonic oscillators.  The problem with this view is that each such oscillator would have a zero-point energy that would contribute to the energy density of space in any gravitational treatment of cosmology. 
Even when the energies of the oscillators are cut off at some high value, the contribution of this ``dark energy'' is
120 orders of magnitude larger than that needed to agree with astrophysical observations. 
Such a disagreement between theory and observation (called the ``cosmological constant problem''), 
even after numerous attempts to reduce it, is {\em many orders of magnitude worse than any other theory-vs-observation discrepancy in the history of science}!  However, somehow this viewpoint remains a central part of the standard model of particle physics and standard practice in QM.

Our approach does not suffer from this serious defect, since its vacuum has no degrees of freedom of its own. 
Where, then, is radiated energy going if an atom's excitation decays and does not interact locally?  The obvious candidate is the enormous continuum of states of matter in the early universe, source of the {\bf cosmic microwave background}, to which atoms here and now are coupled by the quantum handshake.  
For independent discussions from the two of us, see Ref. \cite{Mead_2000} p.94 and Ref. \cite{Cramer_1983}.

 \section{Two Coupled Atoms}
 \label{Two Coupled Atoms}
 
  The central point of this paper is to understand the {\it photon} mechanism by which energy is transferred from
 a {\it single} excited atom (atom $\alpha$) to another {\it single} atom (atom $\beta$) initially in its ground~state.
 
 We proceed with the simplest and most idealized case of two identical atoms, where:
 \begin{itemize}[leftmargin=8mm,labelsep=3.8mm]
\item[(1)] Excited atom $\alpha$ will start in a state where $b\approx 1$ and $a$ is very small, but never zero because of its
ever-present random statistical interactions with a vast number of other atoms in the universe,~and
 \item[(2)] Likewise, atom $\beta$ will start in a state where $a\approx 1$ and $b$ is very small, but never zero for the same~reason.\
 \end{itemize}
 
Thus, each atom starts in a two-component state that has an oscillating electrical current described by Equation~\eqref{velocity}:
\begin{equation}
\begin{aligned}
q\frac{\partial\left<z_\alpha\right>}{\partial t}&\approx-\omega_0d_{12}a_\alpha b_\alpha\,\sin{\!(\omega_0 t)}\cr
q\frac{\partial\left<z_\beta\right>}{\partial t}&\approx-\omega_0d_{12}a_\beta b_\beta\,\sin{\!(\omega_0 t+\phi)},
\end{aligned}
\label{2dmoments}
\end{equation}
where we have taken excited atom $\alpha$ as our reference for the phase of the oscillations ($\phi_{\alpha}=0$),\\
and the approximation assumes that $a$ and $b$ are changing slowly on the scale of $\omega_0$.\
            
Although that random starting point will contain small excitations of a wide range of phases,\\
 we simplify the problem by assuming the following:
 \vspace{6pt}

\noindent {\it All} of the vector potential ${A}_\beta$ at atom $\alpha$ is supplied by atom $\beta$,\\
\noindent{\it All} of the vector potential ${A}_\alpha$ at atom $\beta$ is supplied by atom $\alpha$,\\
\noindent The dipole moments of both atoms are in the $z$ direction, \\
\noindent The atoms are separated by a distance $r$ in a direction orthogonal to $z$, 
 \vspace{6pt}

The vector potential at distance $r$ from a small charge distribution oscillating in the $z$-direction is from Equation~\eqref{AfromI1}:
\begin{equation}
{A_z(t)} = \frac{q\mu_0}{4\pi r}\, \left.\frac{\partial\left<z(r,t')\right>}{\partial t'}\right|^{t'=t\pm \frac{r}{c}}.
\label{A3I2}
\end{equation}\\
Since all electron motions and fields are in the $z$ direction, we can henceforth omit the $z$ subscript.

When the distance $r$ is small compared with the wavelength, i.e., $r\ll 2\pi c/\omega_0$, the delay $r/c$
can be neglected.  Since atomic dimensions are of the order of $10^{-10}$m
and the wavelength is of the order of $10^{-7}$m,
this case can be accommodated.  We shall find that the results we arrive at here are directly adaptable to
the centrally important case in which the atoms are separated by an arbitrarily distance, which will be analyzed
in Section \ref{Two Atoms at a Distance}.  Using Equation~\eqref{EfromA} and~\eqref{AfromI1}, the vector potentials, and~hence the electric fields,
from the two atoms become:
 \begin{equation}
\begin{aligned}
A_{\alpha} \approx\frac{q\mu_0}{4\pi r}\, \frac{\partial\left<z_\alpha\right>}{\partial t}\quad\Rightarrow\quad
{\cal E}_\alpha=-\frac{\partial A_{\alpha}}{\partial t}\approx
 -\frac{q\mu_0}{4\pi r}\frac{\partial^2\left<z_\alpha\right>}{\partial t^2}\cr
A_{\beta} \approx \frac{q\mu_0}{4\pi r}\, \frac{\partial\left<z_\beta\right>}{\partial t}\quad\Rightarrow\quad
{\cal E}_\beta=-\frac{\partial A_{\beta}}{\partial t}\approx -\frac{q\mu_0}{4\pi r}\frac{\partial^2\left<z_\beta\right>}{\partial t^2}.
\end{aligned}
\label{AandEfromI}
\end{equation}
When atom $\alpha$ is subject to electric field ${\cal E}_{\beta}$ and atom $\beta$ is subject to electric field ${\cal E}_{\alpha}$,
the energy of both atoms will change with time in such a way that the total energy is conserved. 
Thus, the superposition amplitudes $a$ and $b$ of both atoms change with time, as described by Equation~\eqref{BsqE} and~\eqref{dEdtav}, from~which:
\begin{equation}
\begin{aligned}
\frac{\partial E_\alpha}{\partial t}
&\approx\frac{1}{2\pi}\int_0^{2\pi} q {\cal E}_\beta \frac{\partial \left<z_\alpha\right>}{\partial t}\,d(\omega_0 t)
=- \frac{q^2\mu_0}{8\pi^2 r}\int_0^{2\pi} \frac{\partial^2\left<z_\beta\right>}{\partial t^2}
 \frac{\partial \left<z_\alpha\right>}{\partial t}\,d(\omega_0 t)\cr
\frac{\partial E_\beta}{\partial t}
&\approx\frac{1}{2\pi}\int_0^{2\pi} q {\cal E}_\alpha \frac{\partial \left<z_\beta\right>}{\partial t}\,d(\omega_0 t)
=- \frac{q^2\mu_0}{8\pi^2 r}\int_0^{2\pi} \frac{\partial^2\left<z_\alpha\right>}{\partial t^2}
 \frac{\partial \left<z_\beta\right>}{\partial t}\,d(\omega_0 t).\cr
\end{aligned}
\label{dEdtav2}
\end{equation}
From Equation~\eqref{dEdtav2}, using the $\left<z\right>$ derivatives from Equation~\eqref{velocity}:
\begin{equation}
\begin{aligned}
\frac{\partial E_\alpha}{\partial t}
&\approx- \frac{\mu_0}{8\pi^2 r}\int_0^{2\pi} \big(-\omega_0^2d_{12}a_\beta b_\beta \cos{\!(\omega_0 t+\phi)}\big)\,
\big(-\omega_0d_{12} a_\alpha b_\alpha\sin{\!(\omega_0 t)}\,d(\omega_0 t)\big)\cr
&\approx- \frac{\mu_0\omega_0^3d_{12}^2a_\beta b_\beta a_\alpha b_\alpha}{8\pi^2 r}
\int_0^{2\pi} \cos{\!(\omega_0 t+\phi)}\sin{\!(\omega_0 t)}\,d(\omega_0 t)
= \frac{\mu_0\omega_0^3d_{12}^2a_\beta b_\beta a_\alpha b_\alpha}{8\pi r}\sin{\!(\phi)}\cr
\frac{\partial E_\beta}{\partial t}
&\approx -\frac{\mu_0}{8\pi^2 r}\int_0^{2\pi} \big(-\omega_0^2d_{12}a_\alpha b_\alpha \cos{\!(\omega_0 t)}\big)\,
\big(-\omega_0d_{12} a_\beta b_\beta\sin{\!(\omega_0 t+\phi)}\,d(\omega_0 t)\big)\cr
&\approx -\frac{\mu_0\omega_0^3d_{12}^2a_\alpha b_\alpha a_\beta b_\beta }{8\pi^2 r}
\int_0^{2\pi} \cos{\!(\omega_0 t)}\sin{\!(\omega_0 t+\phi)}\,d(\omega_0 t)
= -\frac{\mu_0\omega_0^3d_{12}^2 a_\alpha b_\alpha a_\beta b_\beta}{8\pi r}\sin{\!(\phi)}.\cr
\end{aligned}
\label{dEdtav3}
\end{equation}

These equations describe energy transfer between the two atoms in either direction, depending on the sign of $\sin{\!(\phi)}$. 
For transfer from atom~$\alpha$ to atom~$\beta$, ${\partial E_\alpha}/{\partial t}$ is negative.
Since this transaction dominated all competing potential transfers, its amplitude will be maximum, so $\sin{\!(\phi)}=-1$. 

\begin{figure}[H]
\centering
\includegraphics[height=5.5cm]{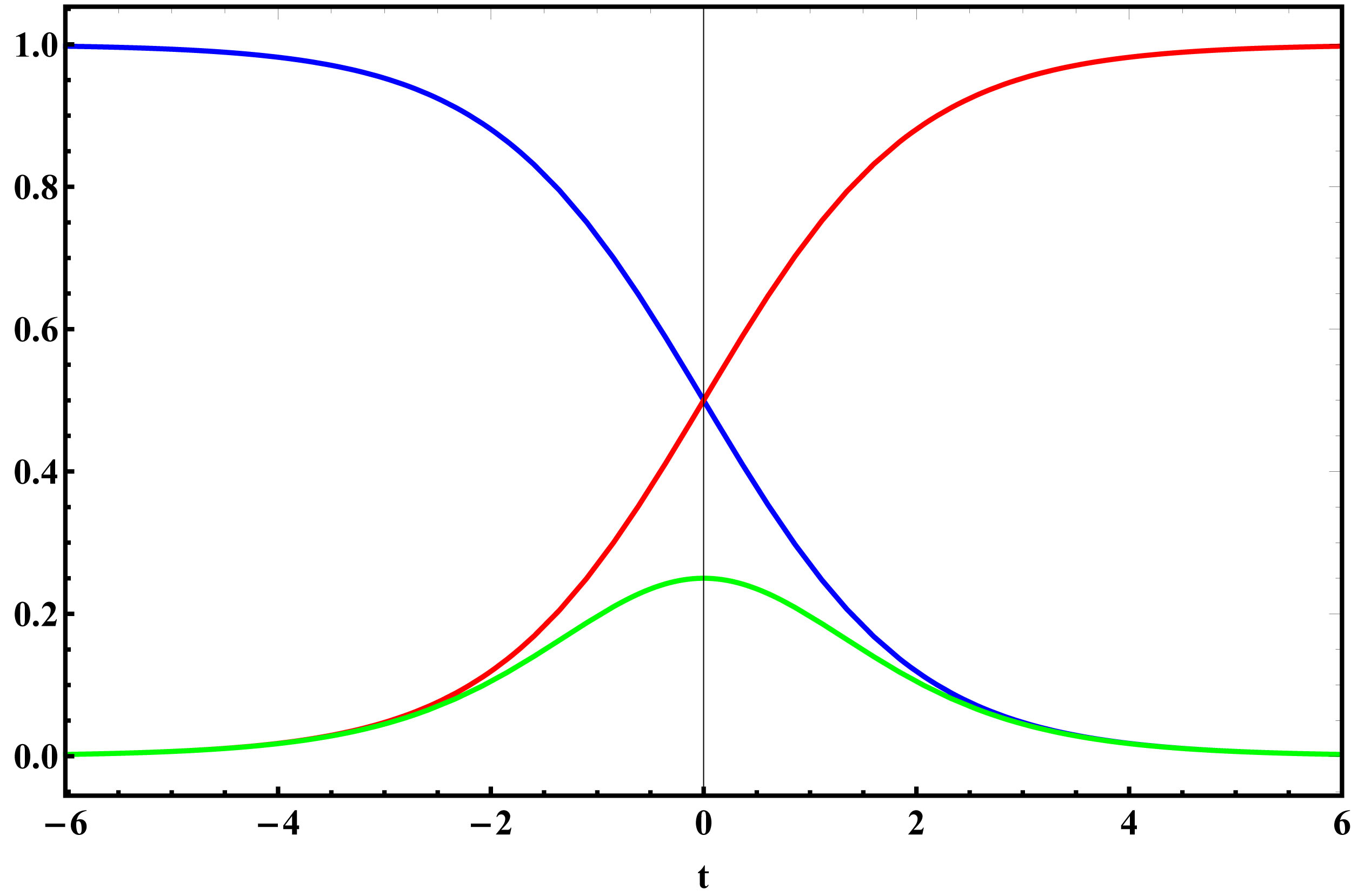}
\caption{Squared state amplitudes for atom $\alpha$: $b^2_\alpha$ (blue)
and $a^2_\alpha=b^2_\beta$ (red) for the Photon transfer
of energy $E_0=\hbar\omega_0$ from atom $\alpha$ to atom $\beta$, from Equation~\eqref{b2anda2}. 
Using the lower state energy as the zero reference, $E_0 b^2$
is the energy of the state.  The green curve shows the normalized power radiated by the atom $\alpha$ and absorbed by atom $\beta$, 
from Equation~\eqref{dEdtav5}.   The optical oscillations at $\omega_0$ are not shown, as they are normally many orders of magnitude faster
than the transition time scale $\tau$.  The time $t$ is in units of $\tau$.  In~the~next section, we will find that atoms spaced by an arbitrary distance exhibit transactions
of exactly the same form. 
\label{abandPvst}}
\end{figure}

If the starting state had been atom~$\beta$ in the excited ($b\approx 1$) state, the $\sin{\!(\phi)}=+1$ choice would have been appropriate:
\begin{equation}
{\rm Using:}\qquad\sin{\!(\phi)}=-1\qquad{\rm and}\qquad P_{\alpha \beta}\equiv \frac{\mu_0\omega_0^3d_{12}^2}{8\pi r},
\label{Pab}
\end{equation}
This rate of transferred energy is calculated for two isolated atoms suspended in space.  We have no experimental data whatsoever for such
a situation.  All optical experiments are done with some {\bf optical system} between the two atoms.  Even the simplest such arrangement
couples the two atoms {\it orders of magnitude} better than the simple $1/r$ dependence in Equation~\eqref{Pab} would indicate.  We take up the
enhancement due to an intervening optical system in Section \ref{Paths}.
Any such enhancement merely provides a constant multiplier in $P_{\alpha \beta}$.
In any case, Equation~\eqref{dEdtav3} becomes:
\begin{equation}
\begin{aligned}
\frac{\partial E_\alpha}{\partial t}=E_0\frac{\partial b^2_\alpha}{\partial t}
=-P_{\alpha \beta}\,a_\beta b_\beta a_\alpha b_\alpha
\qquad \qquad
\frac{\partial E_\beta}{\partial t}=E_0\frac{\partial b^2_\beta}{\partial t}
=P_{\alpha \beta}\, a_\alpha b_\alpha a_\beta b_\beta.
\end{aligned}
\label{dEdtav4}
\end{equation}
Remembering that our total starting energy was $E_0$ that $b^2$ is the energy of the electron in units of $E_0$ referred to the ground state,
that energy is conserved by the two atoms during the transfer, and that the wave functions~are~normalized:
\begin{equation}
\begin{aligned}
E_0\left(b^2_\alpha+b^2_\beta\right)=E_0\quad\Rightarrow \quad b^2_\alpha+b^2_\beta
=1\quad\Rightarrow \quad b_\beta&=\sqrt{1-b^2_\alpha}\cr
 a^2_\alpha+b^2_\alpha=1\quad\Rightarrow \quad a_\alpha&=\sqrt{1-b^2_\alpha}=b_\beta\cr
a^2_\beta+b^2_\beta=1\quad\Rightarrow \quad a_\beta&=\sqrt{1-b^2_\beta}= b_\alpha,
\end{aligned}
\label{NrgConsNorm}
\end{equation}
after which substitutions Equation~\eqref{dEdtav4} becomes:
\begin{equation}
\begin{aligned}
\frac{\partial\!\left( b^2_\alpha\right)}{\partial t}=b_\alpha^2 \big(1-b^2_\alpha\big)/\tau,
\quad{\rm where~the~transition~time~scale~is~} \tau\equiv\frac{E_0}{P_{\alpha \beta}}.
\end{aligned}
\label{dEdtav5}
\end{equation}
This has the solution plotted in Figure~\ref{abandPvst}:
\begin{equation}
\begin{aligned}
b^2_\alpha=a^2_\beta=\frac{1}{e^{t/\tau}+1}\qquad\qquad a^2_\alpha=b^2_\beta=\frac{1}{e^{-t/\tau}+1}.
\end{aligned}
\label{b2anda2}
\end{equation}

Note that this waveform is that of an {\it individual interaction} and has no probabilistic meaning.  It~was the subject of many intense
discussions about NCT in general, including a quite detailed one in \cite{Jaynes_1973MW}.  We refrain from such discussion here
because the dependence of the dipole moment with superposition makeup is not the only nonlinearity in the problem. 
The self-focusing nature of the matched advanced/retarded electromagnetic solutions, described in Section \ref{Global Field}, may be an
even larger nonlinearity in many cases.  Although its time dependence is much more difficult to estimate, it will almost certainly make the individual quantum transition much more abrupt. 

The direction and magnitude of the entire energy-transfer process is governed by the relative phase
$\phi$ of the electric field and the electron motion in {\it both atoms}:
When the electron motion of {\it either atom} is {\it in~phase} with the field, the field transfers energy to the electron,
and the field is said to~{\bf excite}~the~atom.  When the the electron motion has {\it opposite~phase} from the field, the electron transfers energy “to the field”, and the process is called {\bf stimulated~emission}.

Therefore, for the photon transaction to proceed the field from atom $\alpha$
must have a phase such that it ``excites'' atom $\beta$, while the field from atom $\beta$ must have a phase such
that it absorbs energy and ``de-excites'' atom $\alpha$.  In the above example, that unique combination occurs when $\sin{\!(\phi)}=-1$.

This dependence on phase makes a transaction exquisitely sensitive to the frequency match between atoms $\alpha$ and $\beta$. 
The frequency $\omega_0\approx10^{16}/$s, so for transitions in the nanosecond range, \mbox{a mismatch} of one part in $10^7$ can
cause a transaction to fail.  

\subsection{Competition between Recipient Atoms}
As discussed at the end of Section \ref{Electromagnetic Coupling}, the bi-directional vector potential coupling has allowed us
to analyze the problem of two coupled atoms, which looks like a two-electron problem, as two coupled one-electron problems,
and therefore is treatable using Schr\"odinger's wave mechanics.  The approach generalizes, so we can now examine the important three-atom case
of a single excited atom $\alpha$ that is equally coupled to two ground-state atoms $\beta1$ and $\beta2$.  Recipient atoms $\beta1$ and $\beta2$ have the same level spacing as source atom $\alpha$.   For atom $\beta1$, this corresponds to transition frequency $\omega_0$, and we assume a relative phase of $\sin{\!(\phi)}=-1$.  However, atom $\beta2$ is moving and has a slightly Doppler-shifted transition frequency $\omega_0+\Delta\omega$.  We assume that $\beta2$ has the same structure and intial phase as $\beta1$ at time $t=0$. 
Thence, Equation~\eqref{dEdtav4}, with the transition time scale $\tau\equiv E_0/P_{0\beta1}=E_0/P_{0\beta2}$, becomes:
\begingroup\makeatletter\def\f@size{8.5}\check@mathfonts
\def\maketag@@@#1{\hbox{\m@th\fontsize{10}{10}\selectfont \normalfont#1}}%
\begin{equation}
\begin{aligned}
\tau\frac{\partial b^2_{\beta1}}{\partial t}
=a_\alpha b_\alpha \big(a_{\beta1} b_{\beta1}\big)
 \qquad
\tau\frac{\partial b^2_{\beta2}}{\partial t}
=a_\alpha b_\alpha \big(a_{\beta2} b_{\beta2}\cos{\!(\Delta\omega t)}\big)
\qquad
\tau\frac{\partial b^2_\alpha}{\partial t}
=-a_\alpha b_\alpha \big(a_{\beta1} b_{\beta1}+a_{\beta2} b_{\beta2}\cos{\!(\Delta\omega t)}\big)
\end{aligned}
\label{dEdtavFrk}
\end{equation}
\endgroup
As with Equation~\eqref{NrgConsNorm}, wave functions~are~normalized and energy is conserved:
\begin{equation}
\begin{aligned}
a^2_\alpha+b^2_\alpha=1\qquad\qquad a^2_{\beta1}+b^2_{\beta1}=1\qquad\qquad a^2_{\beta2}+b^2_{\beta2}=1
 \qquad\qquad b^2_\alpha+ b^2_{\beta1}+b^2_{\beta2}=1.
\end{aligned}
\label{NrgConsFrk}
\end{equation}
After these substitutions, we obtain two simultaneous differential equations in $b^2_\beta1$ and $ b^2_\beta2$:
\begin{equation}
\begin{aligned}
\tau\frac{\partial b^2_{\beta1}}{\partial t}
&= \sqrt{b^2_{\beta1}\left(1- b^2_{\beta1}\right)\left(1- b^2_{\beta1}- b^2_{\beta2}\right)\left(b^2_{\beta1}+b^2_{\beta2}\right)}\cr
\tau\frac{\partial b^2_{\beta2}}{\partial t}
&=\sqrt{b^2_{\beta2}\left(1- b^2_{\beta2}\right)\left(1- b^2_{\beta1}- b^2_{\beta2}\right)\left(b^2_{\beta1}+b^2_{\beta2}\right)}\,\cos{\!(\Delta\omega\, t)}
\end{aligned}
\label{dEdt3varFrk}
\end{equation}
The solutions of Equation~\eqref{dEdt3varFrk} are shown in Figure~\ref{Frk} for two very small values of $\Delta\omega$:
\begin{figure}[H]
\centering
\includegraphics[height=0.33\linewidth]{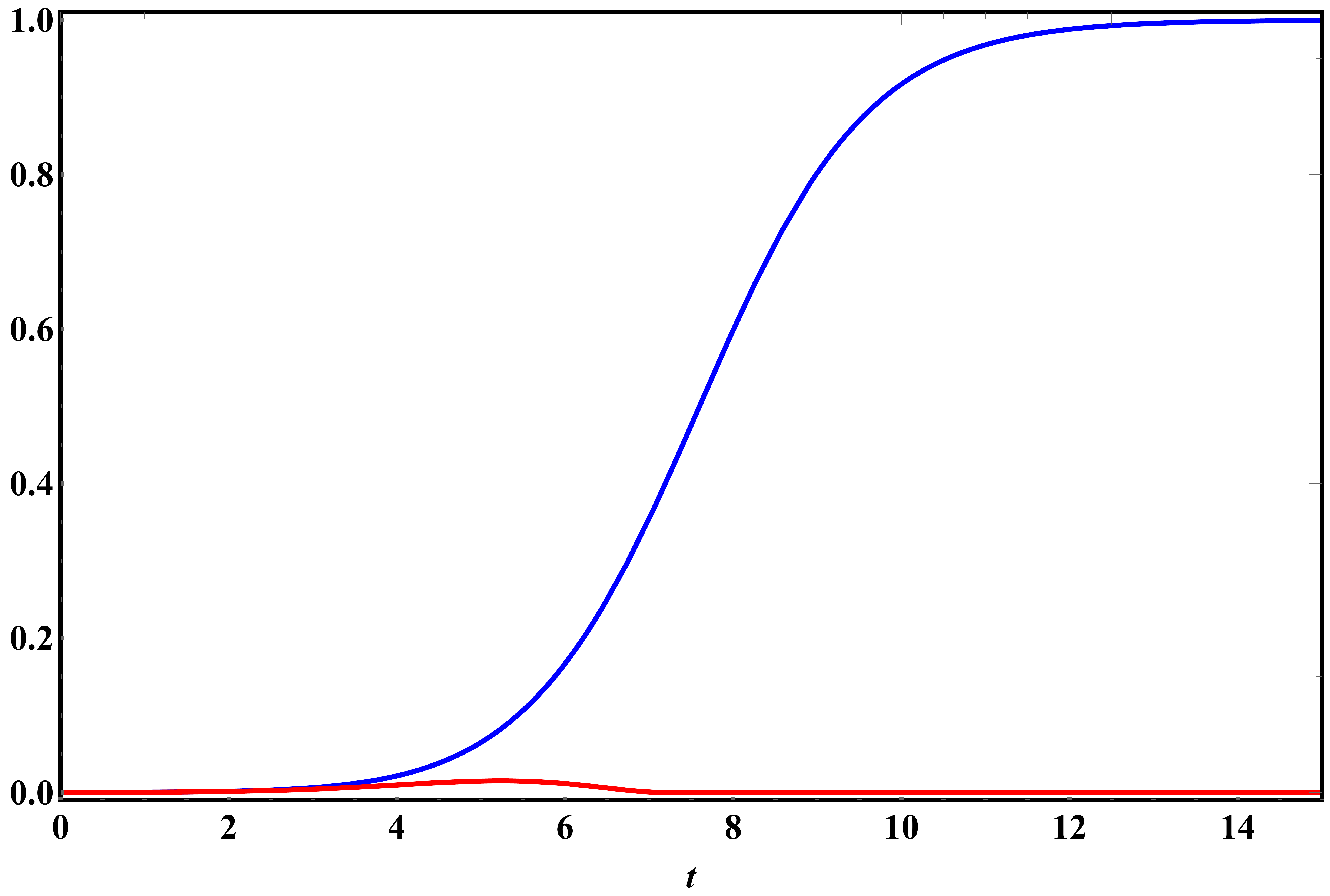}\includegraphics[height=0.34\linewidth]{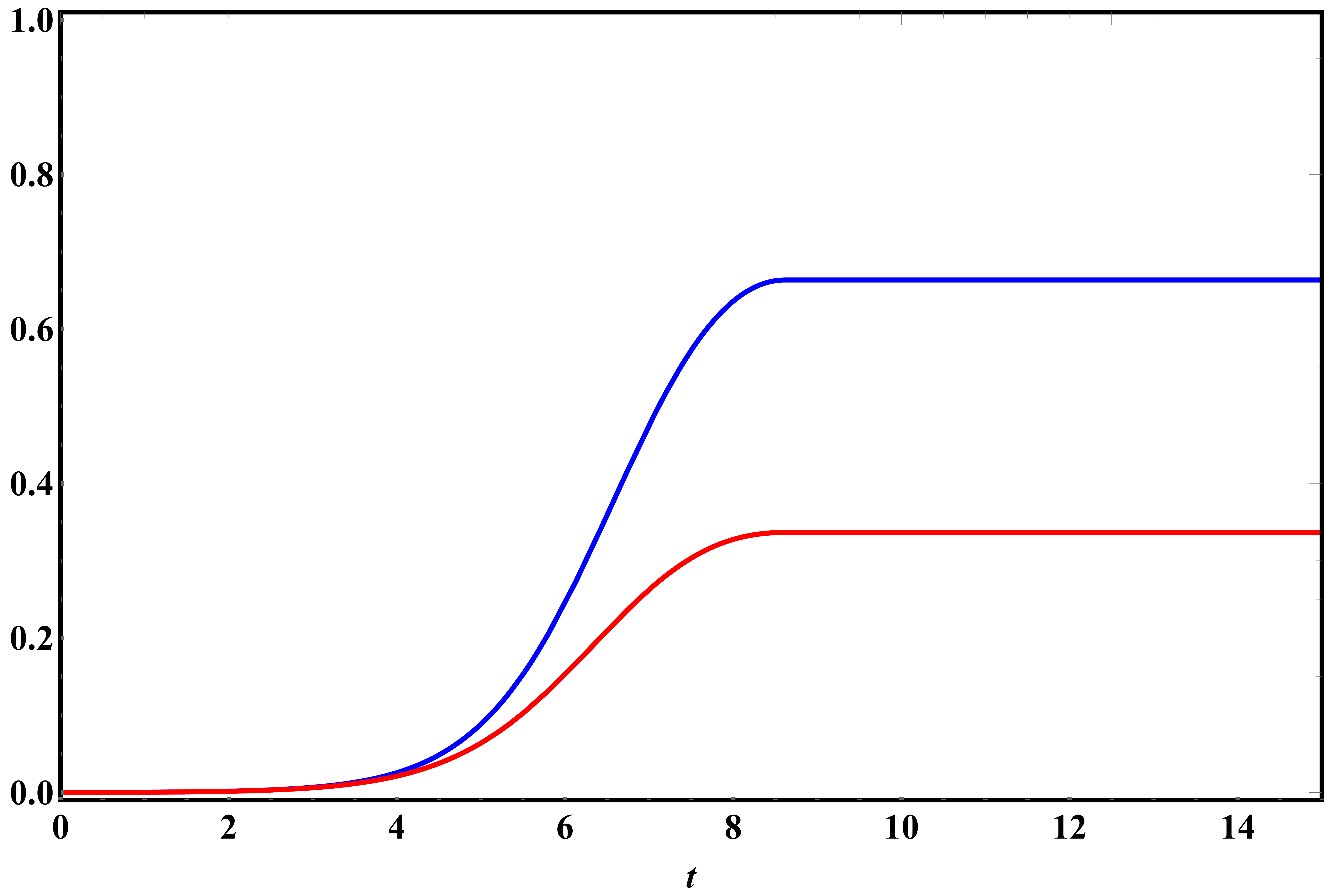}
\caption{Squared excited state amplitudes for recipient atoms $\beta1$ and $\beta2$: $b^2_{\beta1}$ (blue)
and $b^2_{\beta2}$ (red), for~the photon transfer of energy $E_0=\hbar\omega_0$ from atom $\alpha$. The time $t$ is in units of $\tau$.
The left plot, \mbox{for~$\Delta\omega=0.3/\tau$}, shows both recipient atoms being equally excited at the beginning, but the slip in phase
of the red $\beta2$ atom causes it to rapidly lose out, so the blue $\beta1$ atom hogs all the energy and proceeds to become
fully excited, much as if it were the ``only atom in town''.  Its curve is nearly identical with that for the isolated recipient atom
in Figure~\ref{abandPvst}.  The right plot, for $\Delta\omega=0.15/\tau$, shows a totally different story.  Its red $\beta2$ atom
transition frequency is just close enough to that of source atom $0$ that it ``hangs in there'' during the transition period
and ends up partially excited, leaving blue $\beta1$ atom partially excited as well.  The smaller the ``slip'' frequency  $\Delta\omega$,
the closer are the post-transition excitations of the two recipient atoms.  ``Split-photons'' of this kind are observed in
the Hanbury--Brown--Twiss Effect described in Section \ref{Historic Tests}.  As noted above, the frequency window for such events
is {\it extraordinarily narrow}, typically of order $10^{-7}\omega_0$.  Doppler shift of this magnitude requires velocity $\approx$ 30 m/s. 
Room temperature thermal velocities of gasses are typically tens to hundreds of times this value, which would eliminate such competition.  Thus, complete transactions are the most
common, with ``split'' transactions relatively rare and are likely to end as HBT-type four-atom events.
\label{Frk}}
\end{figure}

Here, again the transaction time scale $\tau$ is not to be confused with the time scale for initiating a transaction
after atom $\alpha$ is excited, which is a question of probability.  
We make no pretense here of deriving probability results for a random population of atoms, but,
from these results, we can imagine what one might look like:
The random starting point for the transaction involving one excited atom will contain small excitations of a wide range of phases. 
Equation~\eqref{dEdtav5} is a highly nonlinear equation---the amplitude of each of those excitations will initially grow exponentially
at a rate proportional to its own phase match.  
Thus, the excitation of a random recipient atom that happens to have $\sin{\!(\phi)}\approx -1$ will win in the race and become the dominant partner in the coordinated oscillation of both atoms.  {\em Thus, we have conceptualized the source of the intrinsic randomness within quantum mechanics, an aspect of statistical QM that has been considered mysterious since its inception in the 1920s.}
 
Each wave function will thus evolve its motion to follow the applied field to its maximum resonant coupling
and we can take $\sin{\!(\phi)}=-1$ in these expressions, which we have done in Equation~\eqref{dEdtav4}, Figure~\ref{abandPvst} and Equation~\eqref{dEdtavFrk}. 
[{What we have not done
is to derive the full second-order nature of phase locking in this arrangement. 
That analysis is rendered much more difficult by the potentially {\it huge}
amplification due to the self-focusing nature of the bi-directional electromagnetic coupling described in Section \ref{Paths}. 
Thus, a full derivation remains open to future generations.}]

From the TI point of view,  all three atoms start in stable states, with each having extremely small admixtures of the other state, so that they have very small dipole moments oscillating with angular frequency $\omega_0 \approx (E_2-E_1)/\hbar$.  We assume that in source atom $\alpha$ this admixture creates an initial positive-energy offer wave that interacts with the small dipole moments of absorber atoms $\beta1$ and $\beta2$
to transfer positive energy, and that in atoms $\beta1$ and $\beta2$ this admixture creates initial negative-energy confirmation waves to
the excited emitter atom $\alpha$ that interact with the small dipole moment of emitter atom $\alpha$ to transfer negative energy, as shown schematically in Figure~\ref{twoatom}.
As a result of the mixed-energy superposition of states as shown in Figure~\ref{H10H21mix}, all three atoms oscillate with very nearly the same
frequency $\omega_0$ and act as coupled dipole resonators.\

Energy transferred from source atom $\alpha$ to both recipient atoms $\beta1$ and $\beta2$ causes an increase in both minority
states of the superposition, thus increasing the dipole moment of all three states, thereby increasing the coupling and, hence,
the initial rate of energy transfer.  This behavior is self-reinforcing for any atom that can stay in phase,
giving the transition its initial exponential character.  In the usual case, only one atom is sufficiently well frequency matched to
stay in phase for the entire duration of the transition, the 
unfortunate runner-up is rudely driven out of the competition, and the winner drives the transaction to its conclusion,
as shown in the left panel  of Figure~\ref{Frk}.

In the presence of near-equal competition, one competitor either loses out to a competing transaction,
or, in case of a tie, results in a ``split-photon'',
as shown in the right panel of Figure~\ref{Frk}.  That~situation represents an {\it intermediate} state with actively oscillating dipole moments, as discussed above, in which the two confirmation waves, in the phase at the source, can precipitate the de-excitation of an additional excited atom to create a final HBT four-atom event.

The universe is full of similar atoms, all with slightly different transition frequencies due to random velocities.
There are also random perturbation by waves from other systems that can randomly drive the exponential instability in
either direction.  This random environment is the source of the intrinsic randomness in quantum processes.  Ruth Kastner \cite{Kastner_2015}
attributes intrinsic randomness to “spontaneous symmetry breaking”, which could split a ``tie'' in the absence of environmental factors.

We note here that the probability of the transition must depend on two things: the strength of the electromagnetic coupling between the two states, and the degree to which the wave functions of the initial states are superposed.  The magnitude of the latter must depend on the environment, in~which many other atoms are “chattering” and inducing state-mixing perturbations.  The more potential partner atoms
there are per unit energy, the greater the probability of a perfect match.  Thus, we see the emergence of {\bf Fermi's ``Golden Rule'' \cite{Dirac_1927}}, the assertion that a transition probability in a coupled quantum system depends on the strength of the coupling and the density of states present to which the transition can proceed.  The emergence of Fermi's Golden Rule is an unexpected gift delivered to us by the logic of the present formalism.

It is certainly not obvious a priori that the Schr\"odinger recipe for the vector potential in the momentum (Equation~\eqref{momentum}),
~together with the radiation law from a charge in motion (Equation~\eqref{AfromI2}), would conspire to enable the composition of the superposed states
of two electromagnetically coupled wave functions to reinforce in such a way that, from the asymmetrical starting state, the energy
of one excited atom could {\it explosively and completely} transfer to the unexcited atom, as shown in Figure~\ref{abandPvst} and Figure~\ref{Frk}. 

If nature had worked
a slightly different way, an interaction between those atoms might have resulted in a different phase, and no full transaction
would have been possible.  The fact that transfer of energy between two atoms has this nonlinear, self-reinforcing character makes possible
arrangements like a {\it laser}, where many atoms in various states of excitation participate in a glorious dance, \mbox{all
participating} at exactly the same frequency and locked in phase.\

{\it Why do the signs come out that way?} 
No one has the slightest idea, but the behavior is so remarkable that it has been given a name: Photons are classified as {\it bosons}, meaning that they behave that way! 

 That remarkable behavior is not due to any ``particle-like'' quantization of the electromagnetic field.  
Quantization of the photon energy is a result of the discrete nature of electron states in atoms. 

The movement of an electron in a superposed state couples to another such electron electromagnetically. 
It is essential that this electromagnetic coupling is bi-directional in space-time to conserve energy in the transaction.  
The statistical QM formulation needed some mechanism to finalize a transaction
and did not recognize the inherent nonlinear positive-feedback that nature built into a pair of coupled wave functions.
Therefore, the founders had to put in  wave-function collapse ``by~hand'', and it has mystified the field ever since.
The NCT formulations {\it did} understand the inherent nonlinear positive-feedback that nature built into a pair of coupled wave functions,
but postulated a unidirectional ``Maxwell'' treatment of the electromagnetic field that {\it did not conserve energy}, as we now~discuss.

\section{Two Atoms at a Distance}
\label{Two Atoms at a Distance}

We saw that for two atoms to exchange energy, the vector potential $A$ field at atom~$\beta$ must come from atom~$\alpha$, 
the $A$ field at atom~$\alpha$ must come from atom~$\beta$, and the oscillations must stay in coherent phase,
with a particular phase relation during the entire transition.
{\it  This phase relation must be maintained even when the two atoms are an arbitrary distance apart.}  This is the problem we now address.

To be definite, we consider the case where the two atoms are separated along the $x$-axis, atom~$\alpha$ at $x=0$ in the excited
state and atom~$\beta$ at $x=r$ in its ground state, so their separation $r$ is orthogonal to the $z$-directed current in the atoms. 
The ``light travel time'' from atom~$\alpha$ to atom~$\beta$ is thus $\Delta t =r/c$. 
What is observed is that the energy radiated by atom $\alpha$ at time $t$ is absorbed by atom $\beta$ at time $t'=t+\Delta t$:
\begin{equation}
\begin{aligned}
\left.\frac{\partial E_\beta(r,t')}{\partial t'}\right|^{t'=t+\Delta t}&=-\frac{\partial E_\alpha(0,t)}{\partial t}
\end{aligned}
\label{dEdtrtd12}
\end{equation}
This behavior is familiar from the behavior of a ``particle'', which carries its own degrees of freedom with it: 
It leaves $x=0$ at time $t$ and arrives at $x=r$ at time $t'=t+\Delta t$ after traveling at velocity $c$. 
Thus, Lewis's ``photon'' became widely accepted as just another particle, with degrees of freedom of its own. 
We shall see that this assumption violates a wide range of experimental findings.
 
 For atom~$\beta$, Equation~\eqref{dEdtav2} becomes:
\begin{equation}
\begin{aligned}
\frac{\partial E_\beta(r,t')}{\partial t'}&=\frac{1}{2\pi}\int_0^{2\pi}-q{\cal E}_{\alpha}(r,{t'})
\frac{\partial\left<z_\beta(r,t')\right>}{\partial t'} d(t')\qquad{\rm where}\qquad t'=t+\Delta t
\end{aligned}
\label{EeSch12}
\end{equation}
The retarded field from atom~$\alpha$ interacts with the motion of the electron in atom~$\beta$.  The only difference from our
zero-delay solution is that the energy transfer has its time origin shifted by $\Delta t=r/c$.

This result has required that we choose a {\it positive} sign for the $\mp r/c$ in Equation~\eqref{AfromI2}.  By doing that, we are building in an ``arrow of time'', a preferred time direction, in the otherwise even-handed formulation.  In particular, we are assuming that the retarded solution transfers positive energy.  So~far, everything is familiar and consistent with commonly held Maxwell notions:  A retarded solution carrying energy with it. 
However, we saw that the source atom required a matched vector potential to lose energy. 

 {\it The standard picture leaves no way for
atom~$\alpha$ to lose energy to atom~$\beta$. It does not conserve energy!}

When energy is transferred between two atoms spaced apart on the $x$-axis, 
the field amplitude must be ``{\it coordinate and symmetrical}'' as Lewis described. 
The field ${\cal E}_{\alpha}(x=r)$ at the second atom due to the current in the first must be exactly equal in magnitude to the the field
${\cal E}_{\beta}(x=0)$ at the first atom due to the current in the second, but separated in time by $\Delta t$: 
For atom~$\alpha$, Equation~\eqref{dEdtav2} becomes
\begin{equation}
\begin{aligned}
\frac{\partial E_\alpha(0,t)}{\partial t}&=
\int_0^{2\pi}-q{\cal E}_{\beta}(0,t)\frac{\partial\left<z_\alpha\right>}{\partial t} dt
\end{aligned}
\label{EeSch21}
\end{equation}
Thus, the field ${\cal E}_{\beta}$ from atom~$\beta$, which arises from the motion of its electron at time $t'=t+\Delta t$,
must arrive at atom~$\alpha$
at time $t$, {\it earlier} than its motion by $\Delta t$.  The only field that fulfills this condition is the {\it advanced} field
from atom~$\beta$, signified by choosing a negative sign for the $\mp r/c$ in Equation~\eqref{AfromI2}. 
That choice uniquely satisfies the requirement for conservation of energy.  It also builds complementary ``arrows of time''
into the formulation---we assume that the advanced solution transfers negative energy to the past and the retarded solution transfers positive
energy to the future.  These two assumptions create a new non-local ``handshake'' symmetry
that is not expressed in conventional Maxwell E\&M.

Once these choices for the $\mp r/c$ in Equation~\eqref{AfromI2} are made, the resulting equations for each of the energy derivatives
in Equation~\eqref{dEdtav3} are unchanged when $t'=t+\Delta t$ is substituted for $t$ in the expression for $\partial E_\beta/\partial t$. 
Thus, each transition proceeds in the local time frame of its atom---for all the world (except for amplitude) as if the other atom were local to it.
This ``locality on the light cone'' is the meaning of Lewis' comment:
\begin{quote}{\it
``In a pure geometry it would surprise us to find that a true theorem becomes false when the page
upon which the figure is drawn is turned upside down.  A dissymmetry alien to the pure geometry
of relativity has been introduced by our notion of causality.''}
\end{quote}

The dissymmetry that concerned Lewis has been eliminated.  

This conclusion is completely consistent with the 1909 formulation of Einstein \cite{Einstein_Ritz_1909},
who was critical of the common practice of simply ignoring the advanced solutions for electromagnetic~propagation:
\begin{quote}{\it ``I regard the equations
containing retarded functions, in contrast to Mr.\ Ritz, as merely
auxiliary mathematical forms. The
reason I see myself compelled to take this view is first of all that those
forms do not subsume the energy principle, while
I believe that we should
adhere to the strict validity of the energy principle until we have
found important reasons for renouncing this guiding star.''}
\end{quote}

After defining the retarded solution as $f_1$, and the advanced solution as $f_2$, he elaborates:
\begin{quote}{\it
``Setting $ f(x,y,z,t)=f_1$
amounts to calculating the electromagnetic effect at the point $x,y,z$
from those motions and configurations of the electric quantities that took place
{\it prior to} the time point $t.$\''\
"Setting $f(x, y,z,t)=f_2,$
one determines the electromagnetic
effects from the motions and configurations that take place {\it after}
the time point $t.$''

``In the first case the electric field is calculated from the totality of the processes
producing it, \\and in the second case from the totality of the processes absorbing it...\\
Both kinds of representation can always be used, regardless of how
distant the absorbing bodies are  imagined to be.''}
\end{quote}

The choice of advanced or retarded solution cannot be made a priori:
It depends upon the {\it boundary conditions} of the particular problem at hand. 
The quantum exchange of energy between two atoms just happens to require
one advanced solution carrying negative energy and one retarded solution carrying positive energy to satisfy its boundary conditions
at the two atoms, which then guarantees the conservation of energy.

Thus, the even-handed time symmetry of Wheeler--Feynman electrodynamics \cite{Wheeler_1945,Wheeler_1949} and of the Transactional Interpretation of quantum mechanics \cite{Cramer_2016}, as most simply personified in the two-atom photon transaction considered here, arises from the symmetry of the electromagnetic propagation equations, with boundary conditions imposed by
the solution of the Schr\"odinger equation for the electron in each of the two atoms, as foreseen by Schr\"odinger. 
We see that the missing ingredients in previous failed attempts, by Schr\"odinger and others,
to derive wave function collapse from the wave mechanics formalism were that
advanced waves were not explicitly used as a part of the process.

To keep in touch with experimental reality, we return to our two H atoms spaced a distance $r$ apart.  We can estimate the
"transition time'' $\tau$ from Equations~\eqref{dEdtav4} and~\eqref{Pab}:
\begin{equation}
\begin{aligned}
\frac{\partial E_\beta}{\partial t}=P_{\alpha \beta}\, a_\alpha b_\alpha a_\beta b_\beta
= \frac{\mu_0\omega_0^3d_{12}^2}{8\pi r}\, a_\alpha b_\alpha a_\beta b_\beta.
\end{aligned}
\label{dEdtav4a}
\end{equation}
From the green curve in Figure~\ref{H10H21mix}, we can estimate the dipole strength, which is $q$ times the ``length'' between
the positive and negative ``charge lumps'', say $d_{12}\approx 3q a_0$.  
At the steepest part of the transition, all the $a$ and $b$ terms will be $1/\sqrt{2}$, so
\begin{equation}
\begin{aligned}
\left.\frac{\partial E_\beta}{\partial t}\right|_{\rm max}
\approx \frac{\mu_0\omega_0^3 (3qa_0)^2}{32\pi r}.
\end{aligned}
\label{dEdtav4tau}
\end{equation}

From any treatment of the hydrogen spectrum, we obtain, for the 210$\rightarrow$100 transition:
\begin{equation}
E_{0}=\hbar\omega_0= \frac{9 q^2}{128 \pi \epsilon_0 a_0}
\label{ERy21}
\end{equation}
so the transition time will be:
\begin{equation}
\begin{aligned}
\tau_{12}\approx\frac{E_0}{\left.\frac{\partial E_\beta}{\partial t}\right|_{\rm max}}
\approx \frac{r}{4\,a_0^3 \omega_0^3 \epsilon_0\mu_0}=\frac{r\, c^2}{4\,a_0^3 \omega_0^3}\approx r \times 0.04~{\rm\frac{sec}{m}},
\end{aligned}
\label{tau21}
\end{equation}

Thus,  if the assumption of the $1/r$ dependence of the vector potential ($r$ dependence of the transition time)
were the whole picture, it would take $1/25$ of a second for a transaction to complete if the atoms were suspended one meter apart. 
Such a long transition time would allow the excited atom's energy to be frittered away by the many possible competitive paths,
thus making any modern optical experiment virtually impossible, so {\it no experiments are done that way!}
In real experiments, atoms are coupled by some {\it optical system}, composed of lenses, mirrors, and the like.  That
optical system will have some {\bf solid angle} containing paths from one atom to the other.  The effect of the optical system
is to replace the $1/r$ dependence with Solid Angle/$\lambda$, as described in Section \ref{Paths}.  Thus, the 0.04 s transition
~given by Equation~\eqref{tau21} for two isolated atoms 1 m apart becomes $2\times 10^{-9}$ s when a 1 steradian optical system is used.

Once again, we caution that the time estimated here is {\it the time course of the single transaction after a handshake is formed}, which
{\it must not} be confused with the probabilistic time for a transaction to be initiated after excitation of the source atom.

\section{Paths of Interaction}
\label{Paths}
We have developed a simple conceptual understanding of how a single quantum $\hbar\omega_0$ of energy is transferred
from one isolated atom to another by way of a ``photon'' transaction.  Real experiments with such transactions measure
the statistics of many such events as functions of intensity, polarization, time~delay, and other variables. 
Much has been discovered in the process, some results quite surprising, 
as described for the Freeman--Clauser experiment in Section \ref{Historic Tests}. 
Thus, the time has come for us to discuss, at a conceptual level, where the probabilities come from. 
In the wonderful little book {\it QED}~\cite{Feynman_1985}, our~Caltech colleague the late Richard Feynman
gives a synopsis of the method by which light propagating along multiple paths initiates a transaction, which he calls an event:

\begin{quote}{\it
``Grand Principle:  The probability of an event is equal to the square of the length of an arrow called the 'probability amplitude.'...'' 

``General Rule for drawing arrows if an event can happen in alternative ways: 
Draw an arrow for each way, and then combine the arrows ('add' them) by hooking the head of one to the tail of the next.''

``A 'final arrow' is then drawn from the tail of the first arrow to the head of the last one.''

``The final arrow is the one whose square gives the probability of the entire event."}
\end{quote}

Feynman's ``arrow'' is familiar to every electrical engineer as a {\bf phasor}, introduced in 1894 by Steinmetz
 \cite{Steinmetz_1894,Steinmetz_1900}
as an easy way to visualize and quantify phase relations in alternating-current systems.
In physics, the technique is known as the {\bf sum over histories} and led to {\bf Feynman path integrals}.
His ``probability amplitude'' is the amplitude of our vector potential, whose square is the {\bf probability} of a photon.

\begin{figure}[!ht]
\begin{center}
\includegraphics[height=5cm]{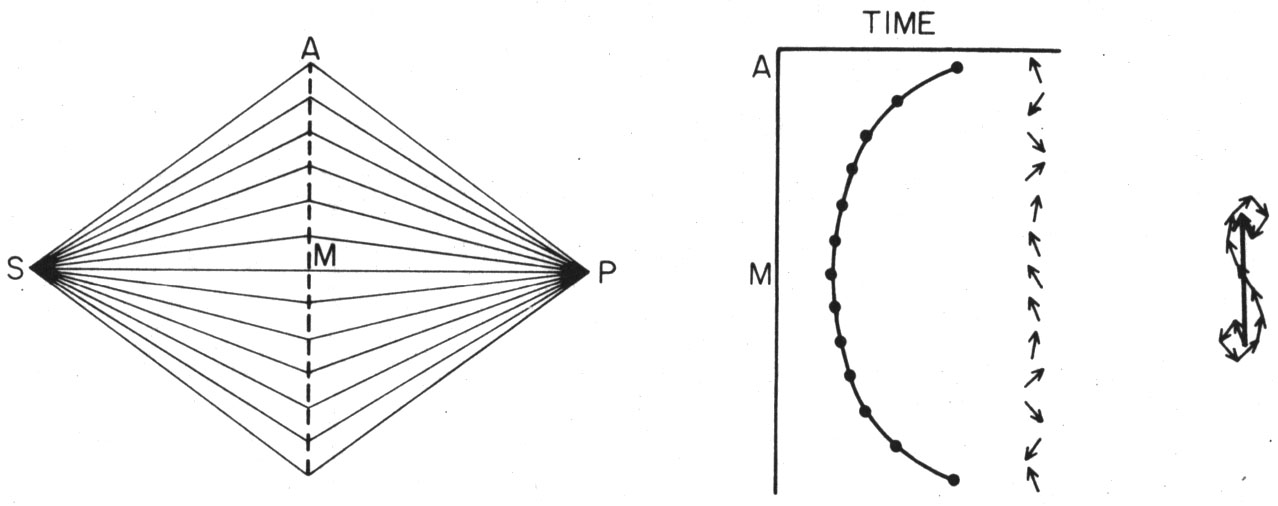} 
\caption{All the paths from coherent light source S to detector P are involved in the transfer of energy.
The solid curve on the ``TIME'' plot shows the propagation time, and hence the accumulated phase, of the corresponding path. 
Each small arrow on the ``TIME'' plot is a {\bf phasor}
that shows the magnitude (length) and phase (angle) of the contribution of
that path to the resultant total vector potential at P. The ``sea~horse'' on the far right shows how these contributions
are added to form the total amplitude and phase of the resultant potential.
(From Fig. 35 in Feynman's {\bf QED}).
\label{LensPhase_1}}
\end{center}
\end{figure}

Feynman then illustrates his Grand Principle with simple examples how a source of light S at one point in space and time
influences a receptor of that light P at another point in space and time, as~shown in Figure~\ref{LensPhase_1}. 
It is somewhat unnerving to many people to learn from these examples that the resultant intensity is dependent
on {\it every possible path} from S to P.  We strongly recommend that little book to everyone. 
That discussion, as well as what follows, details the behavior of highly coherent electromagnetic
radiation with a well-defined, highly stable frequency $\omega$ and wavelength $\lambda$. 

Of course, we have all been taught that light travels in straight lines which spread out as they radiate
from the source, and so the resultant intensity decreases as $1/r^2$, where $r$ is the distance from
the source S.  However,  if the light intensity at P depends on {\it all of the paths}, how can this $1/r^2$
dependence come about?  Well, let's follow Feynman's QED logic:  
[{If the reader does not have
a copy of QED handy, there is a condensed version in Chapter I-26 of the
{\it Feynman Lectures on Physics} at \url{https://www.feynmanlectures.caltech.edu/I_26.html}}]. 

We can see from the ``seahorse'' phasor diagram at the right of Figure~\ref{LensPhase_1} that the vast majority of the length of the
resultant arrow is contributed by paths very close to the straight line S-M-P.  Thus, let's make a rough
estimate of how many paths there are near enough to ``count.''  We can see from the diagram that,
once the little arrows are plus or minus $90^\circ$ from the phase of the straight line, additional paths
just cause the resultant to wind around in a tighter and tighter circle, making no net progress.  Thus,
the uppermost and lowermost paths that ``count'' are about a quarter wavelength longer than the straight line. 
Let's use $r$ for the straight-line distance S-P, $\lambda$ for the wavelength, and $y$ for the vertical
distance where the path intersects the midline above M.  Then, Pythagoras tells us that the length $l/2$
of either segment of the path is  
\begin{equation}
\begin{aligned}
\frac{l}{2}=\sqrt{\left(\frac{r}{2}\right)^2 + y^2}
\end{aligned}
\label{straightpath1}
\end{equation}
Therefore, the entire path length $l$ is 
\begin{equation}
\begin{aligned}
l=\sqrt{r^2 + 4y^2}=r\sqrt{1 + 4\frac{y^2}{r^2}}
\end{aligned}
\label{straightpath2}
\end{equation}
We are particularly interested in atoms at a large distance from each other, and will guess that this means that $y$ is very small
compared to $r$, so all the paths involved are very close to the straight line.  We can check that assumption later. 
Since $y^2/r^2\ll 1$, we can expand the square root:
\begin{equation}
\begin{aligned}
l\approx r\left(1+2\frac{y^2}{r^2}\right)= r+2\left(\frac{y^2}{r}\right)
\end{aligned}
\label{straightpath3}
\end{equation}
Thus, the outermost path that contributes is $2y^2/r$ longer than the straight line path. 
We already decided that maximum extra length of a contributing path would be about a quarter wavelength: 
\begin{equation}
\begin{aligned}
2\left(\frac{y^2}{r}\right)\approx \frac{\lambda}{4}\quad\Rightarrow\quad
\left(\frac{y}{r}\right)^2\approx \frac{\lambda}{8r}
\end{aligned}
\label{straightpath4}
\end{equation}

We can now check our assumption that $y^2/r^2\ll 1$.  If $r=1$~m and our $210\rightarrow100$ transition has $\lambda\approx10^{-7}$m,
then $y^2/r^2\approx 10^{-8}$, so our assumption is already very good, and gets better
rapidly as $r$ gets larger.  

How do we estimate the number of paths from S to P?  Well, no matter how we choose the path
spacing radiating out equally in all directions from S, the number of paths that ``count'' will be proportional
to the solid angle subtended by the outermost such paths.  Paths outside that ``bundle'' will have phases
that cancel out as Feynman describes.  The angle of the uppermost path is $y/r$.  Paths~also radiate out perpendicular to the page
to the same extent, so the total number that ``count'' goes as the solid angle $=\pi (y/r)^2$. 
Feynman tells us that the resultant amplitude $A$ is proportional to the total number of paths that ``count,''
so we conclude from Equation~\eqref{straightpath4}:
\begin{equation}
\begin{aligned}
A\propto {\rm Solid\ Angle}=\pi\left(\frac{y}{r}\right)^2\approx \frac{\pi\lambda}{8\,r}
\end{aligned}
\label{inverserlaw}
\end{equation}
{\bf Thus, this is the fundamental origin of the $1/r$ law for amplitudes.}

The intensity is proportional to the square of the amplitude, and therefore goes like $1/r^2$,
as we all learned in school.  Thus,  instead of lines of energy radiating out into space in all directions,
Feynman's view of the world encourages us to visualize the source of electromagnetic waves as
``connected'' to each potential receiver by all the paths that arrive at that receiver in phase.
Just to convince us that all this ``path'' stuff is real, Feynman gives numerous fascinating examples
where the $1/r^2$ law doesn't work at all.  Our favorite is shown in Figure~\ref{LensPhase_2}:
\begin{figure}[H]
\begin{center}
\includegraphics[height=4.5cm]{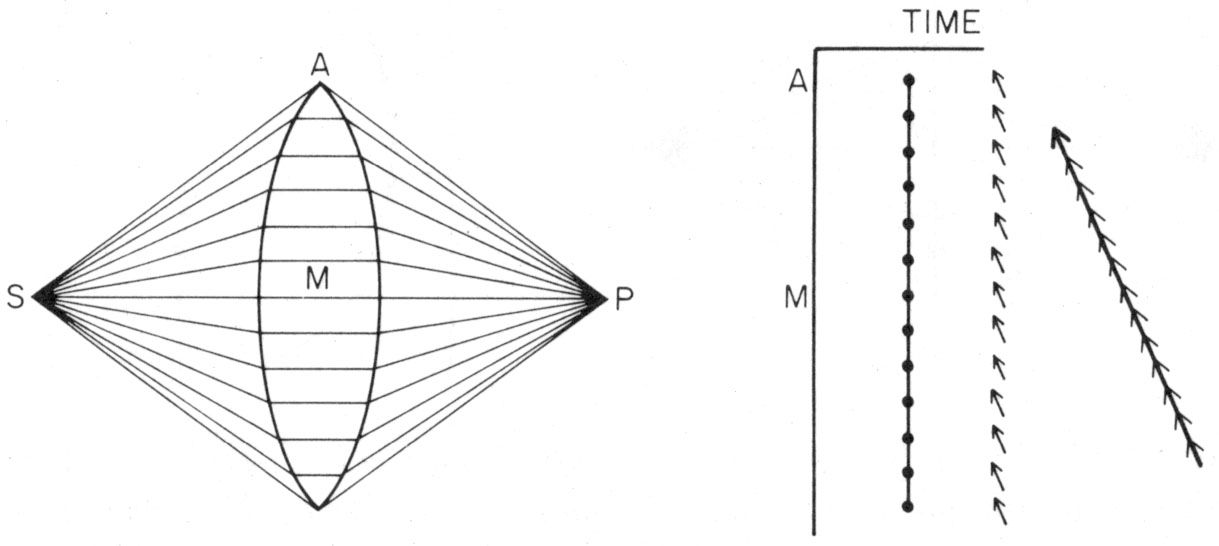} 
\caption{Situation identical to Figure~\ref{LensPhase_1} but with a piece of glass added. 
The shape of the glass is such that all paths from the source S reach P in phase.
The result is an {\it enormous} increase in the amplitude reaching P.
(From Fig.~36 in Feynman's {\bf QED}).
\label{LensPhase_2}}
\end{center}
\end{figure}

The piece of glass does not alter the amplitude of any individual path very much---it might lose
a few percent due to reflection at the surfaces.  However, it {\it slows down} the speed of propagation
of the light.  In addition, the thickness of the glass
has been tailored to slow the shorter paths more than the longer paths, so {\it all paths take
exactly the same time.}  The net result is that {\it the oscillating potential propagating along
every path reaches P in phase with all the others!}  Now, we are adding up all the little phasor arrows
{\it and they all point in the same direction!}  The amplitude is enhanced by this little chunk of glass  by the factor: 
\begin{equation}
\begin{aligned}
{\rm Enhancement Factor}=\frac{\rm Lens\ Solid\ Angle}{\rm Solid\ Angle\ from\ Equation~\eqref{inverserlaw}}
\approx \frac{8\,r}{\pi\lambda}{\rm Lens\ Solid\ Angle}
\end{aligned}
\label{EnhancementFactor}
\end{equation}

For the arrangement shown in Figure~\ref{LensPhase_2}, at the
size it is printed on a normal-sized page, $r\approx 5$~cm, so~from From Equation~\eqref{inverserlaw}, the solid angle 
of the ``bundle'' of paths without the glass was of the order of $10^{-7}$.  The solid angle enclosing the paths through
the glass lens is $\approx$1 steradian (sr), so the little piece of glass has increased the potential at $P$ by {\it seven orders of magnitude}, and the intensity of the light by {\it fourteen orders of magnitude!}  In this and the more general case, we get an important principle:
\begin{equation}
{\bf Insertion\ of\ an\ optical\ system\ replaces\ the\ \frac{1}{r}\ factor\ by}\ \frac{8}{\pi\lambda}{\bf \times Lens\ Solid\ Angle}
\end{equation}
Returning to our two H atoms spaced 1 meter apart, we found in Equation~\eqref{tau21} that, using the standard $1/r$ potential, the
"transition time'' for the quantum transaction was $\approx$0.04 s.  For the $\omega_0$ wavelength the Rate Enhancement Factor
with a 1st optical system is $\approx$$2\times 10^{7}$, thereby shortening the transition time $\tau$ by a factor of $\approx$$5\times 10^{-8}$,
making $\tau \approx\ 2$ ns.

We have learned an important lesson from Feynman's characterization of propagation phenomena:
Changing the configuration of components of the arrangement in what appear to be innocent ways
can make {\it drastic} differences to the resultant potential at certain locations.  The reader will find
many other eye-opening examples in QED and FLP I-26.  We will find in Section \ref{Global Field} that
two atoms in a ``quantum handshake'' form a pattern of paths that greatly increases the potential
by which the atoms are coupled, and hence can shorten the transition beyond what is possible with just the optical system.

All the results in statistical QM are probabilities because Heisenberg denied that there was any physics in the transactions. 
That denial has left the field in a conceptual mess.  There is no doubt that statistical QM makes it easy to calculate probabilities
of a wide variety of experimental outcomes, and~that these predicted outcomes overwhelmingly agree with reality.  However that discipline
is, by design, powerless to provide reasoning for {\it how} those outcomes come about.  The object of this paper is to {\it understand
the individual transaction, not to calculate probabilities.}   Thus, the times quoted above are the times required for the individual event,
once initiated, {\it not} the time constant of some statistical distribution.  
{\bf We deal with a realm of which statistical QM denies the existence.}

\section{Global Field Configuration}
\label{Global Field}

We are now in a position to visualize the field configuration for the quantum exchange of energy between two atoms, as analyzed
in Section \ref{Two Atoms at a Distance}, using the locations and coordinated defined there.
From~Equations~\eqref{AandEfromI} and~\eqref{2dmoments}, and using $\sin{\!(\phi)}=-1$,
the total field is composed of the sum of the retarded solution ${A}_\alpha$, at distance $r_\alpha$ from atom~$\alpha$
and the advanced solution ${A}_\beta$, at distance $r_\alpha$ from atom~$\beta$: 

\begin{equation}
\begin{aligned}
A_{\alpha} &\propto\frac{1}{r_\alpha}\frac{\partial\left<z_\alpha\right>}{\partial t}
= - \frac{1}{r_\alpha}\sin{\big(\omega_0( t-r_\alpha/c)\big)}\cr
A_{\beta} &\propto\frac{1}{r_\beta}\frac{\partial\left<z_\beta\right>}{\partial t}
\propto \frac{1}{r_\beta}\cos{\!\big(\omega_0( t+\Delta t+r_\beta/c)\big)}
\end{aligned}
\label{AandEfromI2}
\end{equation}


Including both $x$ and $y$ coordinates in the distances ${r_\alpha}$ and ${r_\beta}$ from the two atoms,
the vector potentials from the two atoms anywhere in the $x-y$ plane are
\begin{equation}
\begin{aligned}
A_\alpha(x,y,t)&\propto -\frac{1/\tau}{\sqrt{x^2+y^2}}\, 
\sin{\!\left(\omega_0 \!\left(t- \frac{\sqrt{x^2+y^2)}}{c}\right)\right)}\cr
A_\beta(x,y,t)&\propto \frac{1/\tau}{\sqrt{(x-\Delta x)^2+y^2}}\, 
\cos{\!\left(\omega_0 \!\left((t+ \Delta t)+\frac{\sqrt{(x-\Delta x)^2+y^2)}}{c}\right)\right)}\cr
\end{aligned}
\label{A4int12b2}
\end{equation}

\begin{figure}[!ht]
\begin{center}
\includegraphics[height=5cm]{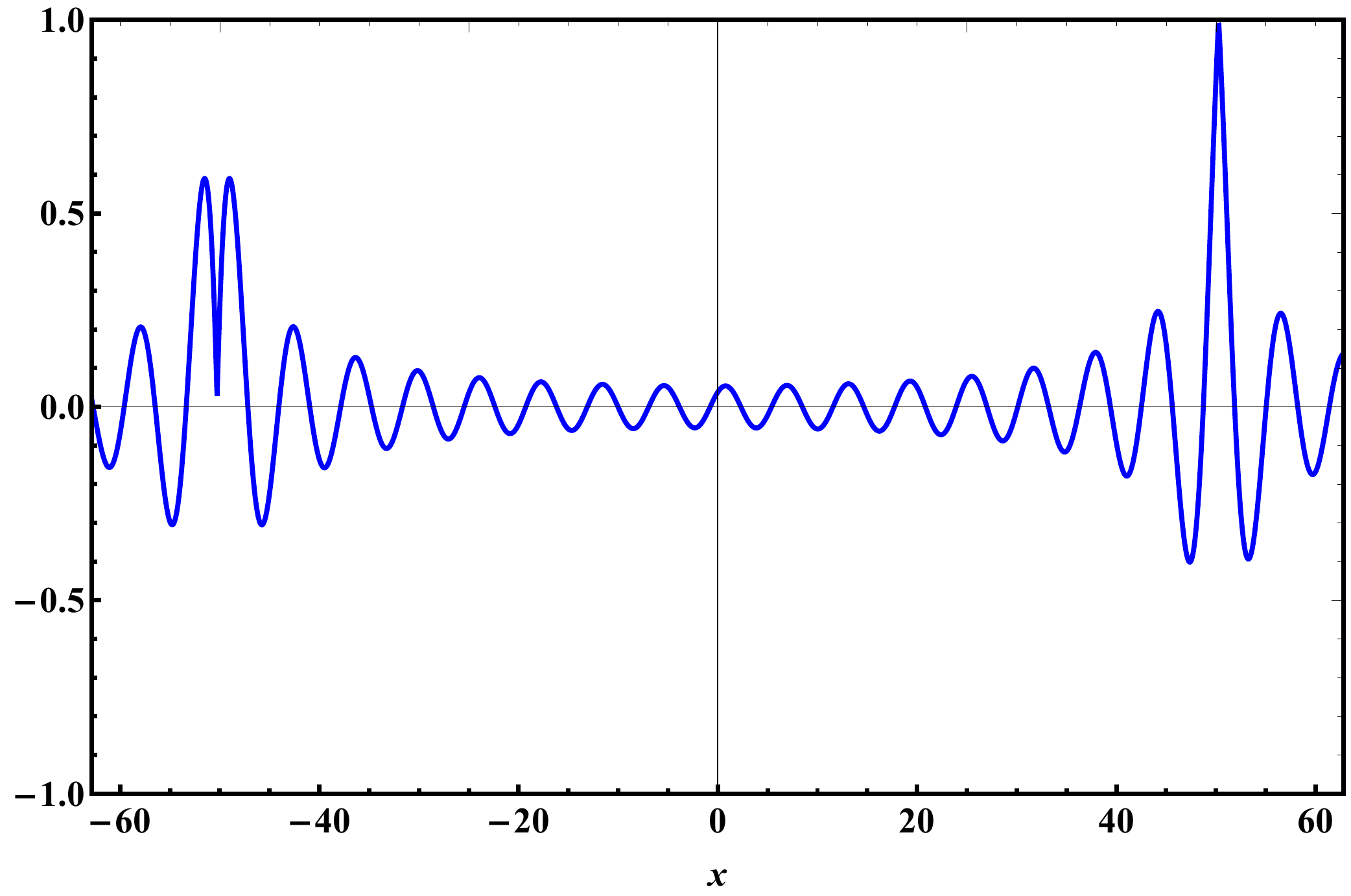} 
\caption{Normalized vector potential along the $x$-axis (in wavelength/2$\pi$) between two atoms in the ``quantum handshake'' of Equation~\eqref{A4int12b2}. 
The wave propagates smoothly from atom~$\alpha$ (left) to atom~$\beta$ (right). 
\href{run:Mov3aCM.mov}{Animation here} \cite{AnimLink}.
\label{AdvRtd1D}}
\end{center}
\end{figure}

An example of the total vector potential $A_{\rm tot}=A_\alpha + A_\beta$ along the $x$-axis is shown in Figure~\ref{AdvRtd1D}.\\ 

\begin{figure}[H]
\begin{center}
\includegraphics[height=7cm]{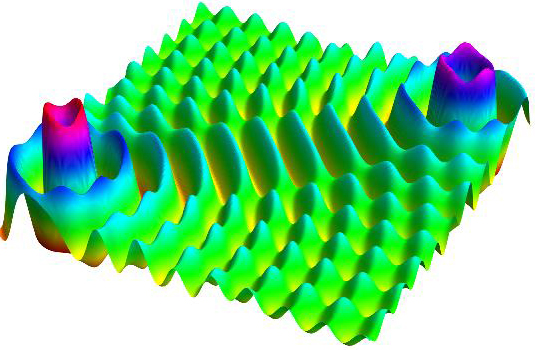} 
\caption{Two atoms in the ``quantum handshake'' of Equation~\eqref{A4int12b2}.     
\href{run:Mov4aCM.mov}{Animation here} \cite{AnimLink}.
\label{Still3D}}
\end{center}
\end{figure}

A ``snapshot'' of the potential of Equation~\eqref{A4int12b2}  at one particular time 
for the full $x-y$ plane is shown in Figure~\ref{Still3D}. 

The still image in this figure looks like a typical interference pattern from two sources---a ``standing wave.''
There are high-amplitude regions of constructive interference which appear light blue and yellow on this plot. 
These are separated from each other by low-amplitude regions of destructive interference, which appear green. 
In a standing wave, these maxima would oscillate at the transition frequency, with no net motion.
The animation, however, shows a totally different story:  Instead of oscillating in place
as they would in a standing wave, {\it the maxima of the pattern are moving steadily from the source atom (left) to the receiving atom (right).} 
This movement is true, not only of the maxima between the two atoms, but of maxima well above and below the line
between the two atoms.  These maxima can be thought of as Feynman's paths, each carrying energy along its trajectory
from atom~$\alpha$ to atom~$\beta$. \\ For those readers that do not have access to the animations, the same story
is illustrated by a stream-plot of the Poynting vector in the $x$-$y$ plane, shown in Figure~\ref{Streamplot}: 
\begin{figure}[H]
\begin{center}
\includegraphics[height=5cm]{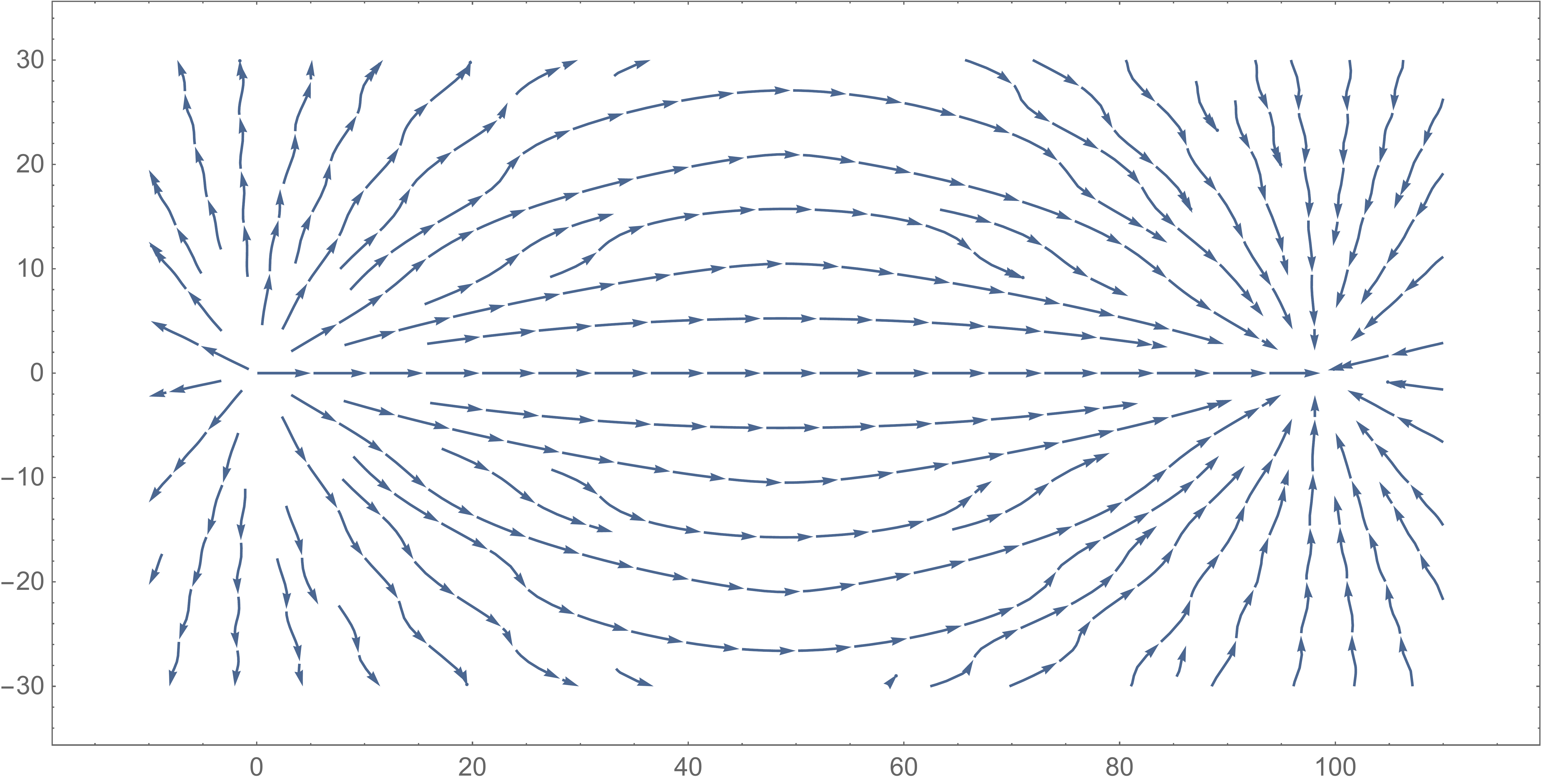} 
\caption{Poynting vector stream lines of the ``quantum handshake'' of Equation~\eqref{A4int12b2}.     
\label{Streamplot}}
\end{center}
\end{figure}

\begin{figure}[H]
\begin{center}
\includegraphics[height=7cm]{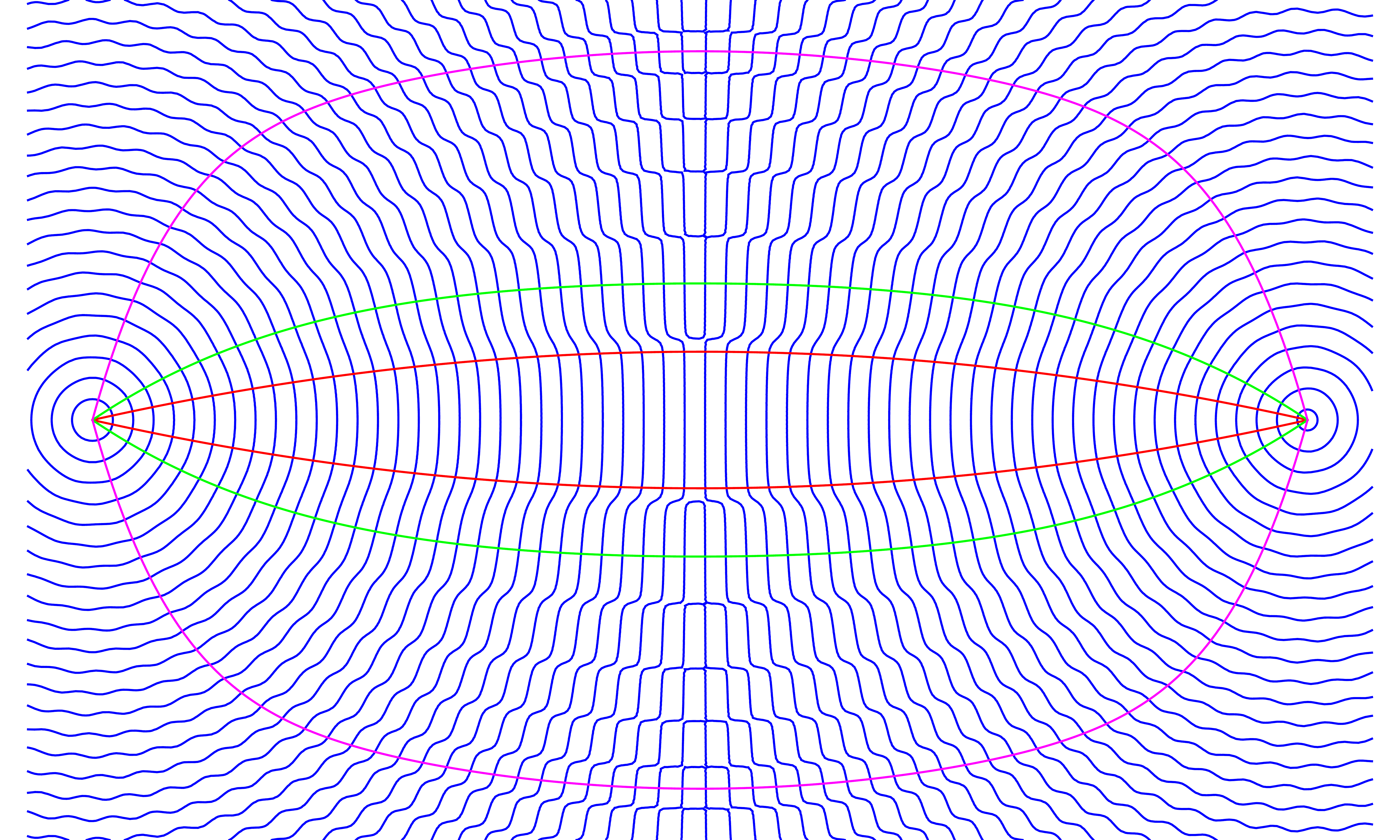} 
\caption{The zero crossings of the handshake vector potential at $t=0$.  Paths near the axis (between the two red lines)
are responsible for the conventional $1/r$ dependence of the potential.
Paths shown through high-amplitude regions have an even number of zero crossings,
and thus the potentials traversing these paths all arrive in phase, thus adding to the central potential.
\label{ZeroX}}
\end{center}
\end{figure}

We can get a more precise idea of the phase relations by
looking at the zero crossings of the potential at one particular time, as shown in Figure~\ref{ZeroX}.  
Paths from atom~1 to atom~2 can be traced through either the high-amplitude regions
or the low-amplitude regions.  The paths shown in Figure~\ref{ZeroX} are traced through high-amplitude regions. 

\vspace{-12pt}

The central set of paths, delimited by the red lines, are responsible for the conventional $1/r$ dependence of the potential,
as described with respect to Equation~\eqref{inverserlaw}.  Working outward from there,
each high-amplitude path region is separated from the next
by a slim low-amplitude region.  It is a remarkable property of this interference pattern that each low-amplitude path has $\pi$ more
phase accumulation along it than the prior high-amplitude path and $\pi$ less than the next high-amplitude~path. 
The low-amplitude paths are the ones that contribute to the ``de-phasing'' in this arrangement, but~they are very slim and of low amplitude,
so they don't de-phase the total signal appreciably.
In addition, the phases of the paths through the high-amplitude regions are separated by $2\pi n$, where $n$ is an integer.
{\bf All waves propagating from atom~$\alpha$ to atom~$\beta$ along high-amplitude paths arrive in phase!}

In Feynman's example shown in Figure~\ref{LensPhase_1}, there are an equal number of paths of any phase,
so every one has an opposite to cancel it out.  In Figure~\ref{LensPhase_2}, the lens makes all paths have {\it equal time delay},
which then enables them to all arrive with the same phase.

The phase coherence of the advanced-retarded handshake creates a pattern of potentials that has a unique property:
It is not like either of Feynman's examples in Figure~\ref{LensPhase_1} or Figure~\ref{LensPhase_2}. 
Its high-amplitude paths do arrive {\it in phase}, but by
a completely different mechanism.  It all starts with the bundle of paths between the red lines, which has the $1/r$ amplitude,
just as if there were no quantum mechanism.  Then, as the handshake begins to form, additional paths are drawn
into the process. The process is self-reinforcing on two levels---the increase in dipole moment and the increase in number of paths
that arrive in phase.  Paths that formerly would have cancelled the in-phase ones
are ``squeezed'' into extremely narrow regions, all of low amplitude, as can be seen in Figure~\ref{ZeroX}.  Thus, a large
fraction of the solid angle around the atoms is available for the in-phase high-amplitude paths.

For optical systems with large solid angle, the self-focusing enhancement may still be noticeable in the shape of the transition waveform,
but for one-sided systems, like an astronomical telescope, we~expect it to be be dominant.

Following Feynman's program has led us to conclude that:
{The vector potential from all paths sum to make a highly-amplified connection between distant atoms.}
The advanced-retarded potentials form nature's very own phase-locked loop, which forms nature's own
{\it Giant Lens in the Sky!} 

The consequences of this fact are staggering:  Once an initial handshake interference pattern is formed between two atoms
that have their wave functions synchronized, the strength grows explosively:  Not only because the dipole moment of each atom grows exponentially, but, in addition, a substantial fraction of the possible interaction paths between the two atoms
propagate through high-amplitude regions, {\it independent of the distance between them!} 
Although we have not worked out the difficult second-order dynamics of phase-locking between coupled atoms,
we believe that here is the solution to the long-standing mystery of the ``collapse of the wave function'' of the ``photon''. 
  
The interaction depends critically
on the advanced-retarded potential handshake to keep all paths in phase.  Ordinary propagation over very long paths becomes ``de-phased"
due to the slightest variations of the propagation properties of the medium.  By contrast, the advanced and retarted fields are precise negative images of each other on exactly the same light cone, so the phase of high-amplitude handshake paths
are always related by an even number of $\pi$ to the phase of other such paths.  \mbox{Paths having }odd numbers of $\pi$ phase
are always of low amplitude, and do not cancel the even-$\pi$ phase paths as they would in a one-way propagating wave like
Feynman used in his illustrations.  [{A detailed analysis of these properties has not been done.  It is a wonderful project for the future.}]

The interaction proceeds in the local time frame of each atom because they are linked with the advanced-retarded potential. 
The waves carrying positive energy from emitter to absorber are retarded waves with positive transit time;  they reach the absorber
after a single transit time $\Delta t=r/c$.  Once~they have established a phase-coherent ``handshake'' connection,
Lewis' ``{\it coordinate and symmetrical}'' advanced waves with negative transit time are launched toward the emitter, arriving
at the precise time and in the precise phase to withdraw energy from the emitter.  During the transaction, as long as the
"handshake'' connection is active, any change in the state of one atom will be directly reflected in the state of the other.
Aside from the time-of-flight propagation time to establish the ``handshake'', there is no additional ``round trip'' time delay in the
quantum-jump process, which proceeds as if the two atoms were local to each other.  

Thus, the Transactional Interpretation allows us to conceptualize Niels Bohr's ``instantaneous'' quantum jump \cite{Heisenberg_1985}
concept that Schr\"{o}dinger, who expected time-extended classical transitions, found impossible to accept \cite{Schrodinger_1952}.\\
\vspace{-12pt}

\section{Relevance to the Transactional Interpretation}
\label{Relevance to TI}

The calculation that we have presented here, with its even-handed treatment of advanced and retarded four-potentials, was inspired by WFE and the Transactional Interpretation, but it also provides interesting insights that clarify and modify the mechanism by which a transaction forms.  Wheeler--Feynman electrodynamics suggests that a retarded wave, arriving at a potential absorber, stimulates that entity to generate a canceling retarded wave accompanied by an advanced wave.  The TI suggests that this advanced wave arrives back at the emitter with amplitude $\psi\psi*$, a relation suggesting the Born rule.

However, from our calculation, we see that a slightly different process is described. \mbox{Both emitter} and absorber initially have small admixtures of the opposite eigenstate, giving them dipole moments that oscillate with the {\it same} difference-frequency $\omega_0$. The two oscillations each have an environment-induced random phase.  If the phases have the correct relation [$\sin(\phi)\approx-1$], the dipole moments of both atoms initially increase exponentially, the system becomes a phase-locked loop, and~it avalanches to a final state of multi-path energy transfer that satisfies all boundary conditions.  In~that scenario, the initial confirmation wave is likely to have an amplitude much weaker than WFE would suggest, and the quantity $\psi\psi*$ becomes that used by Schr\"odinger to provide the electron density~function.

How can a linear system generate such nonlinear behavior?  While the Schr\"odinger equation is indeed linear,  Equation~\eqref{dEdtav5} governing the evolution of transaction formation and wave function collapse is highly nonlinear.  Thus, the TI's assertion that the offer wave ``stimulates the generation of the confirmation wave'' must be modified.  Rather, advanced and retarded potentials, boundary conditions at both ends, and a fortuitous matching of phase trigger the nonlinear avalanche in both atoms and brings about the transaction.
 
We also note that the advanced and retarded waves do not carry ``information'' in the usual sense in either time direction, but only deliver a pair of oscillating four-potentials to the sites of a pair of oscillating charges, leading dynamically to an initially exponential rise in coupling, a focusing of alternative paths, the formation of a transaction, a transfer of energy, and the enforcement of conservation laws.  For the Transactional Interpretation, this phase selection process clarifies the randomizing mechanism by which, in the first stage of transaction formation, the emitter makes a random choice between competing offer waves arriving from many potential absorbers.  The offer wave with the best phase is likely to win, even if it comes from far away.  [It is sometimes asserted that this handshake situation is only possible in a frozen deterministic four-dimensional ``block universe'' because of the two-way connections between present and future.  We reject this assertion, which is dissected in some detail in {Section 9.2 }of Ref. \cite{Cramer_2016}.  While it is true that the assumption of a block universe would dispel or bypass many of the quantum paradoxes, it would only do so at the terrible price of imposing complete determinism on the universe.].

We saw in the derivation of the coupling of two separated atoms that it was necessary to use the advanced 4-potential {\it in order to satisfy the law of conservation of energy}.  This, not ``information transfer'', is the role of the quantum handshake in the Transactional Interpretation.  The quantum handshakes act to enforce conservation laws and do not form unless all conserved quantities are properly transferred and conserved.  This is what is going on in {\it quantum entanglement}: the separated parts of a quantum system are linked by conservation laws that are enforced by V-shaped three-vertex advanced-retarded quantum handshakes \cite{Cramer_2016} and cannot emerge as a completed transaction unless those conservation laws are satisfied.  In this context, we note that the Transactional Interpretation, using such linked advanced-retarded handshakes, is able to explain in detail the behavior of over 26 otherwise paradoxical and mysterious quantum optics experiments and {\it gedankenexperiments} from the literature.  {See Chapter 6 of Ref. \cite{Cramer_2016}}.   If we cannot dismiss the plethora of competing QM interpretations based on their failure in experimental tests, {\it we should eliminate them when they fail to explain paradoxical quantum optics experiments} (as almost all of them do.)

\section{Historic Tests}
\unskip
\label{Historic Tests}
\subsection{The Hanbury--Brown--Twiss Effect and Waves vs. Particles}
It is often said that particles in quantum mechanics ``travel as waves but  arrive as particles''.  The~Hanbury--Brown--Twiss effect \cite{Hanbury_1953} (HBT) is an example of this principle.  It demonstrates that, in the second-order interference of incoherent wave sources, photons are divisible and are not  electromagnetic ``billiard balls'' that maintain individual identities.  The HBT technique was first applied to the measurement of the diameters of nearby stars, e.g., Betelguese, using intensity interferometry with radio waves.  The original experiments involved large parabolic radio dishes mounted on rail cars.

\begin{figure}[H]
\begin{center}
\includegraphics[height=7cm]{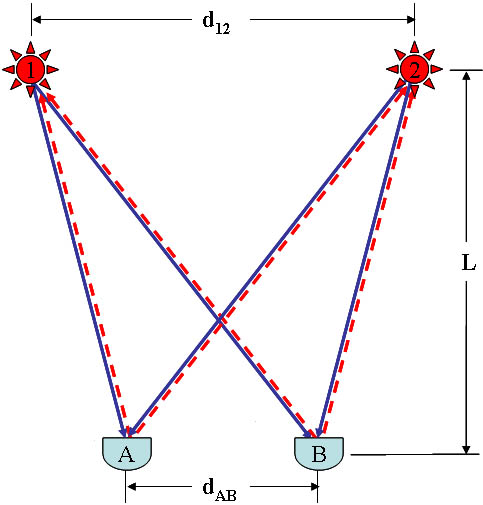} 
\caption{Schematic diagram of the Hanbury--Brown--Twiss effect, with excited atoms $1$ and $2$ in distant separated sources simultaneously
exciting ground state atoms $A$ and $B$ in two separated detectors.
\label{HBT}}
\end{center}
\end{figure}

A simplified version of an HBT interference measurement is illustrated in Figure~\ref{HBT}.  Sources $1$ and $2$ are separated by a distance $d_{12}$.  Both sources emit waves of the same wavelength $\lambda = 2\pi c/\omega$, but~are not causally connected.
Radiation from the two sources is received by detectors $A$ and $B$, which are separated by a distance $d_{AB}$. The line between the sources is parallel to the line between the detectors, and the two lines are separated by a distance $L$.  Hanbury-Brown and Twiss showed that a significant rate of coincident detections---up to a factor of 2 larger than could be ascribed to chance---was observed in this arrangement. 
For this simple configuration, the probability of a coincident detection in the two detectors, for small $d_{AB}$ and very large $L$, is proportional to $1+\cos[2\pi ~ d_{AB}~ d_{12}/(\lambda L)]$, which is maximum when $d_{AB}=0$ and falls off as the detector separation increases, the rate of falloff indicating the value of $d_{12}$.  

\vspace{-12pt}

The very fact that coincidences are observed in this experiment reveals a deep truth about electromagnetic coupling: 
Photons cannot be consistently
described as little blobs of mass-energy that travel uniquely from a single source point to a single detector point. In the HBT effect, a whole
photon's worth of energy $\hbar\omega$ is assembled at each detector out of fractional energy contributions from each of the two sources.

In the TI description of the HBT event described above, a retarded offer wave is emitted by the source $1$ and travels to both detectors $A$ and $B$. Similarly, a retarded offer wave is emitted by the source $2$ and travels to both detectors. Detector $A$ receives a linear superposition of the two offer waves and seeks to absorb the ``offered'' energy by producing an advanced confirmation wave.  If the phases match, as they will in a coincident event, the energy transfer begins with an exponential increase in the dipole moment of each source atom.   Atoms in detectors $A$ and $B $ respond similarly, their oscillating dipole moments producing advanced confirmation waves that travel back to the two sources, each~of which responds with an increasing dipole moment that enhances the offer waves.  A four-atom transaction of the form shown in Figure~\ref{HBT} is formed that removes one photon's worth of energy $\hbar\omega$ from each of the two sources $1$ and $2$ and delivers one photon's worth of energy $\hbar\omega$ to each of the two detectors \mbox{$A$ and $B$. }


Neither of the detected ``photons'' {\it can be said to have originated uniquely in one of the two sources}.  The~energy arriving
at each detector originated partly in one source and partly in the other. It might be said that each source produced two ``fractional photons"
and that these fractions from two sources combined at the detector to make a full size ``photon''. The ``particles'' transferred have no separate identity that is independent from the satisfaction of the quantum mechanical boundary conditions.  The boundary conditions here are those imposed by the HBT geometry, conservation laws, and the detection criteria.

Finally, we note that in many experiments published in the physics literature the HBT effect has been observed and demonstrated not only for photons, but also in the detection of charged $\pi$ mesons emitted in the ultra-relativistic collision of heavy nuclei.  It is observed that the detection probability doubles when the detected particles are close in momentum and position.  Furthermore, in nuclear physics experiments with pairs of half-integer-spin neutrons or protons, a Pauli-Exclusion version of the HBT effect has been observed, in which the detection probability drops to zero when the detected particles are close in momentum and position \cite{Padula_2005}.  All of these particle-like entities ``travel as waves but  arrive as particles''.

\subsection{Splitting Photons}

At the 5th Solvay Conference in 1927, Albert Einstein posed a riddle, sometimes called ``Einstein's Bubble Paradox'', to the assembled founders of quantum mechanics \cite{Jammer_1974,Cramer_2016}.  Einstein's original language was rather convoluted and technical (and in German), but his question can be simply stated as follows:
\begin{quote}
{\em A source emits a single photon isotropically, so that there is no preferred emission direction. According to the quantum formalism, this should produce a spherical wave function $\Psi$  that expands like an inflating bubble centered on the source. At some later
time, the photon is detected.  Since the photon does not propagate further, its wave function bubble should ``pop'', disappearing instantaneously from all locations except the position of the detector.  In this situation, how do the parts of the wave function away from the detector ``know'' that they should disappear, and how is it enforced that only a single photon is always detected when only one photon is emitted?}
\end{quote}

The implication of his question is that, if a photon is an indivisible particle, it should not be possible to divide one. 
 However, we have already seen in the Hanbury--Brown--Twiss effect that photons are {\it not} little indivisible billiard balls and that they {\it can} divide their energy between two receiving atoms.  How, then, is it possible that for one photon emitted there is always only one received?

\begin{figure}[H]
\centering
\includegraphics[width=0.5\linewidth]{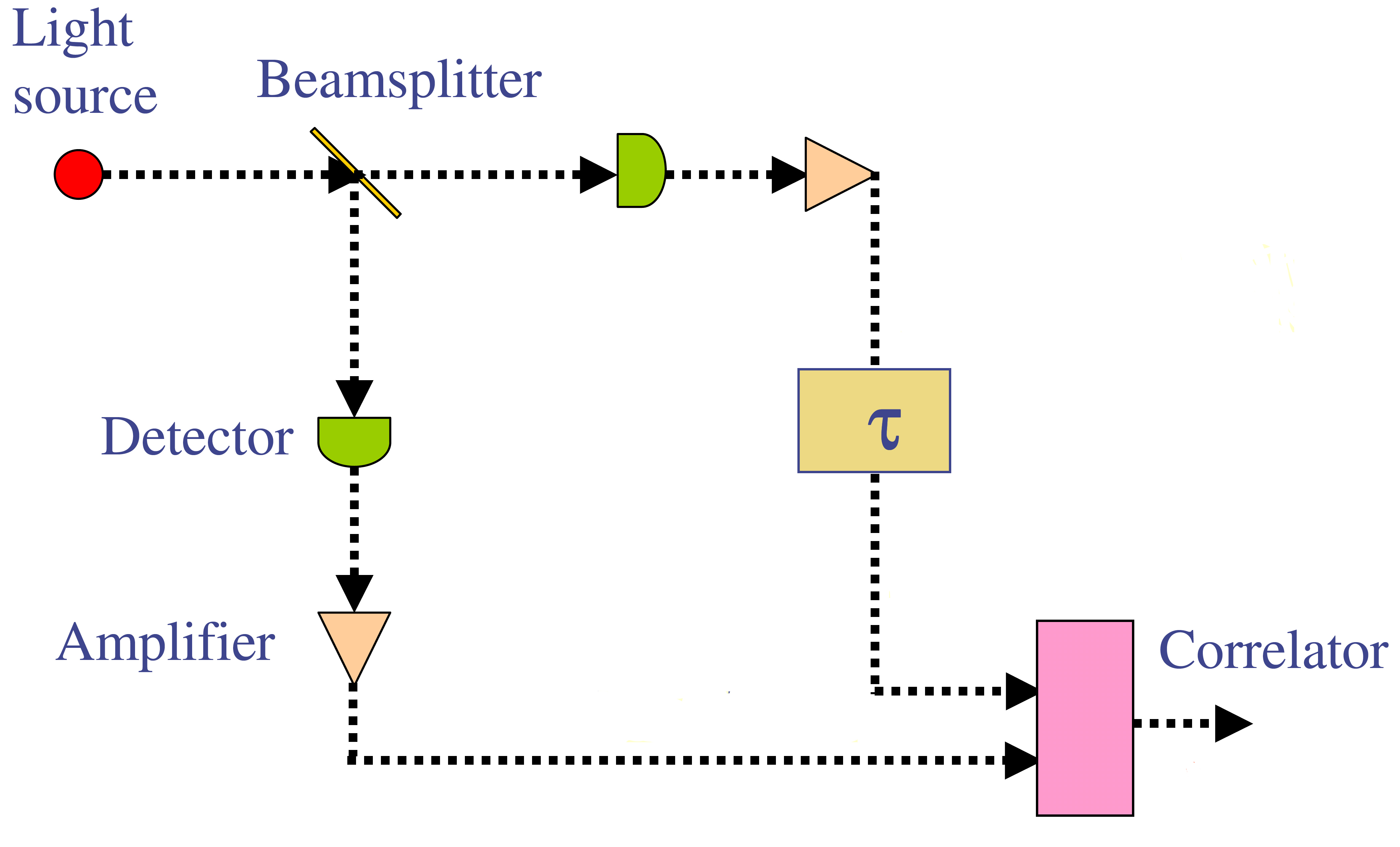}\hbox to 0.5in{}\includegraphics[width=0.35\linewidth]{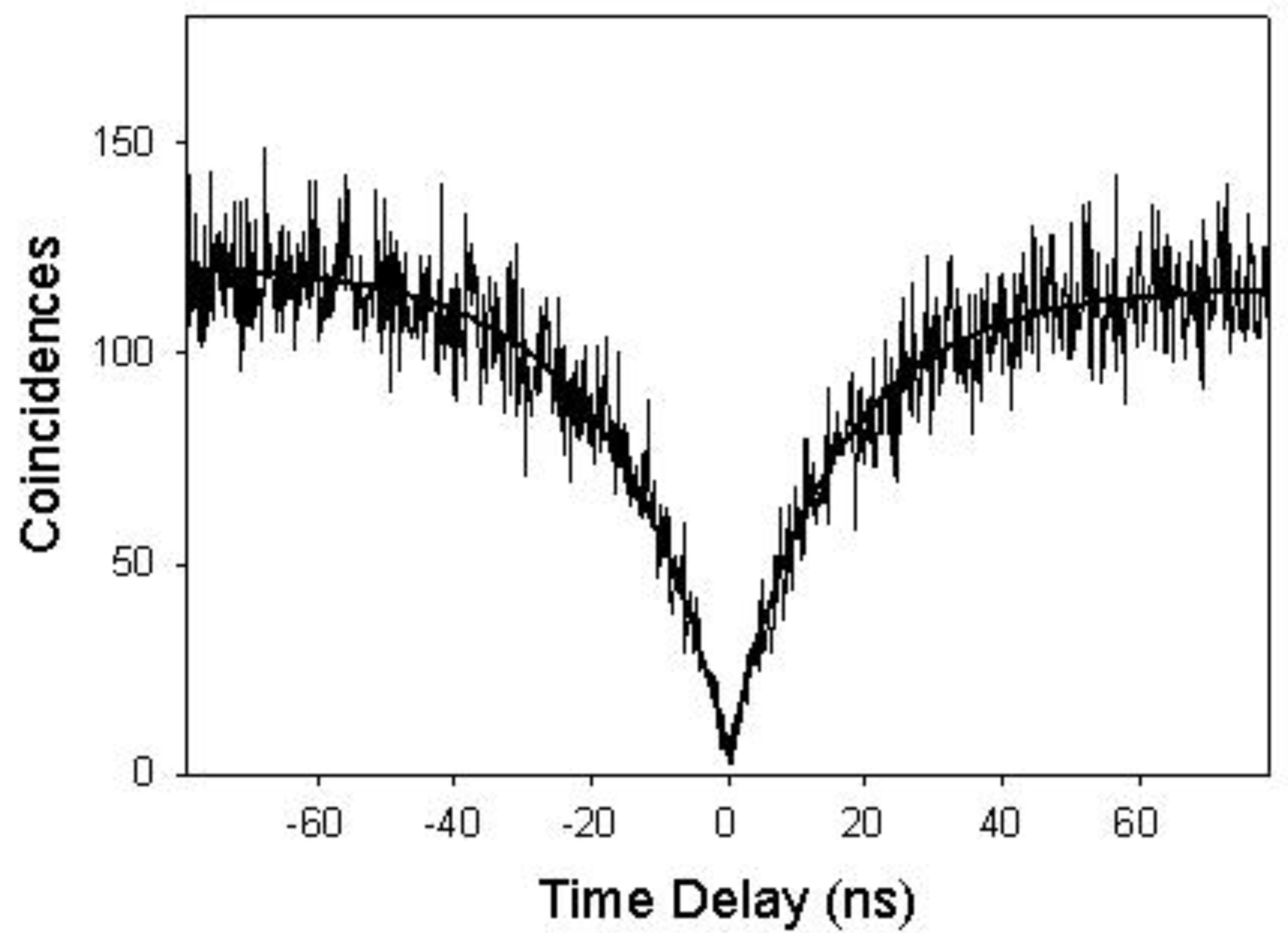}
\caption{\textbf{Left}:  Schematic of the photon-splitting experiment.  \textbf{Right}:  Plot of the number of coincidences vs.
time delay between the arriving pulses.  The source on average generates a photon per $\approx$12 nsec. The finite number counted at zero delay is consistent with the accidental presence of two excited atoms.
\label{PhotSplit}}
\end{figure}

A search for this hypothetical divided-photon  behavior is implemented in the setup shown in the left panel in
Figure~\ref{PhotSplit}, which was enabled once it became possible to build sources of single photons.
The idea is that, if a photon is just a short pulse of light, half of it should go through each of the dotted paths, and both halves
should be counted at the same time, registering as a coincidence.  Of course, the original pulse must have twice the energy required
to trigger a detector, so either half by itself would have just enough.  However, if the photon was indivisible, as mandated by certain versions of QM,
it would make a random decision on which path to take, and no coincidences would be observed. In practice, the number of coincidences is
counted for a certain counting period, with the time delay $\tau$ between the two detector output pulses as a parameter.  Since
the photons are generated randomly, the time between successive photons can accidentally range from zero to large, and a plot
of the number of correlator outputs vs. time delay $\tau$ gives information about the statistics of the source.

Modern versions of the experiment \cite{Lounis_2000} give plots like that shown in the right panel in Figure~\ref{PhotSplit}.  [{Early versions of
this kind of experiment suffered from certain defects that made them inconclusive.  \mbox{A definitive }version using ``heralded photons'' was finally accomplished by Clauser in 1973 \cite{Clauser_PRD74}.}]
Let~us look at the experiment from a TI perspective:  The source excites one atom, which, due to random coupling,
develops a superposition with a tiny presence of ground state and nearly unity excited state.  As described for the
two-atom photon, the tiny presence in the superposition enables the dipole moment to oscillate, thereby generating
a radiating vector potential that propagates along both dotted paths to both detectors.  To find a perfectly matched
partner atom is rare, but, when one matches up, a quantum handshake grows up connecting them.

There are then three possible outcomes:
\begin{enumerate}[leftmargin=.2in]
 \item The handshake goes to completion and the partner atom is in one of the detectors, in which case only the chosen detector registers an output.
 \item The handshake goes to completion, but the partner atom is not in one of the detectors.  In this case, no output is registered from either detector.
 \item The initial stages of a handshake begins in {\em two} partner atoms, one in each detector.  When the source atom has de-excited, both of the detector atoms are left in mixed states with roughly equal components of ground-state and excited state wave functions, as was illustrated in Figure~\ref{Frk}.  This~is not a stable configuration because both of the detector atoms have oscillating dipole moments sending strong ``unrequited'' advanced confirmation waves.  These waves are in phase at and focused on the source, and they are likely to find another well-phased excited-state atom there or nearby that will complete the four-way transaction.  Thus, a four-atom HBT event should be created, in which there are two emissions and two detections.  Such an unlikely event would register as an ``accidental'' case of two simultaneous emissions in the same time window.  There will never be an event with a single emission and two detections.
\end{enumerate}

Thus, we see that the outcomes of the experiment predicted by TI and QM are essentially identical.
Certainly, no solid conclusion can be drawn from this experiment as to whether quantization occurs in the field or in the transaction. 

\subsection{Freedman--Clauser Experiment}
\label{F-C Expt}
\begin{figure}[H]
\centering
\includegraphics[width=0.65\linewidth]{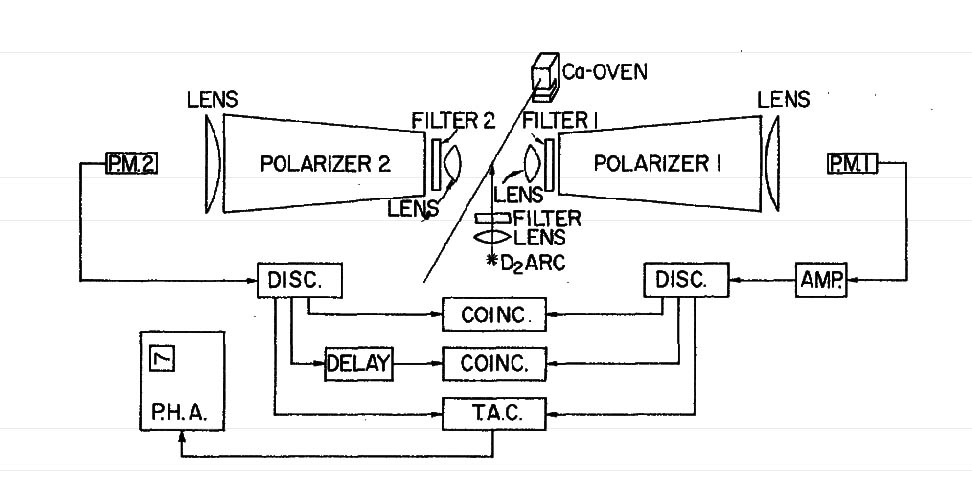}\hbox to 0.05\linewidth{}\includegraphics[width=0.3\linewidth]{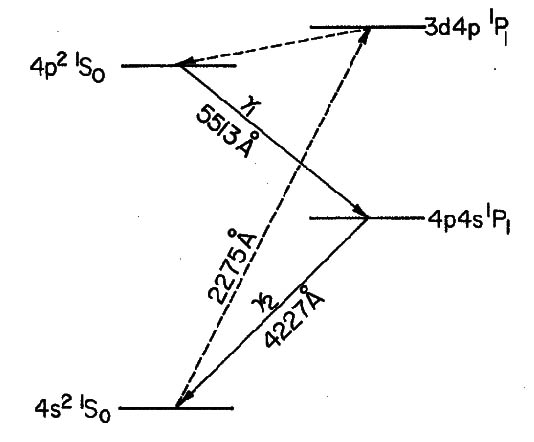}
\caption{Left:  Schematic of the polarization-correlation experiment.  Right:  Energy levels of the Ca atoms used in this experiment.
From Freedman and Clauser 1972 \cite{FreedmanClauser_1972}.
\label{FreedClauser}}
\end{figure}

In our introductory discussion of Schr\"odinger's visualization of his newly-invented Wave Mechanics in Section \ref{Quantum States},
we described how Clauser and his colleagues, through experiments that were heroic at the time, were able to show that no
``local, realistic theory'' was compatible with their results \cite{FreedmanClauser_1972,Fry_1976,Clauser_1976,Aspect_1981,Aspect_1982,Aspect_1982a}. 
We now describe the earliest conclusive version of these EPR experiments and show how our TI approach gives a
simple and natural explanation of the otherwise mysterious outcome.  A~sketch of the arrangement is shown in Figure~\ref{FreedClauser}.

\noindent
The atomic configuration used was the three-level system of the Ca atom, shown at the right of the figure. 
The atomic wave functions were an upper $4p^2\, S_0$ state $\Psi_3$ of frequency $\omega_3$,
a middle $4p4s\, P_1$ state $\Psi_2$ of frequency $\omega_2$,
and a $4s^2\, S_0$ ground state $\Psi_1$ of frequency $\omega_1$,
so that the cascade starts and ends in an $S_0$ state of zero angular momentum.\\

The state wave functions are of the same form as those given in Equation~\eqref{SchEqSols}:
\begin{equation}
\begin{aligned}
\Psi_{3} &= R_3(r)\,e^{-i\omega_3 t}\quad\quad
\Psi_{2} = f_2(r)\cos{\!(\theta)}\,e^{-i\omega_2 t}=R_2(r,\theta)\,e^{-i\omega_2 t}\quad\quad
\Psi_{1} &= R_1(r)\,e^{-i\omega_1 t}
\end{aligned}
\label{SchEq3Sols}
\end{equation}

The transition state is a superposition of these three states:
The analysis is a simple extension of that for the two-level system, starting with Equation~\eqref{mixedphicomplex}:
 \begin{equation}
\Psi=ae^{i\phi_a}R_1e^{-i\omega_1 t}+be^{i\phi_b}R_2e^{-i\omega_2 t}+ce^{i\phi_c}R_3e^{-i\omega_3 t},
\label{mixed3complex}
\end{equation}
The charge density of the mixed state is then:
\begingroup\makeatletter\def\f@size{9}\check@mathfonts
\def\maketag@@@#1{\hbox{\m@th\fontsize{10}{10}\selectfont \normalfont#1}}%
\begin{equation}
\begin{aligned}
\rho&=q\Psi^*\,\Psi \cr
\frac{\rho}{q}&=\left(ae^{-i\phi_a}R_1e^{i\omega_1 t}+be^{-i\phi_b}R_2e^{i\omega_2 t}+ce^{-i\phi_c}R_3e^{i\omega_3 t}\right)
\left(ae^{i\phi_a}R_1e^{-i\omega_1 t}+be^{i\phi_b} R_2e^{-i\omega_2 t}+ce^{i\phi_c} R_3e^{-i\omega_3 t}\right)\cr
&=a^2 R_1^2+b^2 R_2^2+c^2 R_3^2\cr
&+2abR_1R_2\cos{\!\big((\omega_2-\omega_1) t+(\phi_b-\phi_a)\big)}\cr
&+2acR_1R_3\cos{\!\big((\omega_3-\omega_1) t+(\phi_a-\phi_c)\big)}\cr
&+2bcR_2R_3\cos{\!\big((\omega_3-\omega_2) t+(\phi_b-\phi_c)\big)}
\end{aligned}
\label{rhomix3wf}
\end{equation}
\endgroup

The dipole srength $d_{ij}$ for the three terms is given by straightforward extension of Equation~\eqref{d12eq}
\begin{equation}
\begin{aligned}
d_{12} = 2q\int R_1 R_2 z \qquad
d_{23} = 2q\int R_3 R_2 z\qquad
d_{13} = 2q\int R_1 R_3 z=0
\end{aligned}
\label{d123eqs}
\end{equation}
By symmetry around the ($+z$ axis) of the spherical coordinate system, both $d_{12}$ and $d_{23}$ are in the $\vec z$ direction
determined by the direction of the $\Psi_{2}$ wave function.  In general, that direction will shift around in space
depending on the coupling of the atom to others.  However, in any given situation, there is only one $z$-axis
that defines the ``North pole'' direction, and both dipole moments are oscillating in that direction. 
Thus, it is the $\Psi_{2}$ state, shared by
both transactions of the 3-state ``cascade'' that aligns the linear polarizations of the two interlocking transactions.

\subsubsection{Dynamics of the Transaction}

The cascade atom is emitted from the oven shown at the top of the figure.  It is is initially ``pumped'' by the $D_2$ arc when it is
centered between the two lenses.  The excitation is along the 2275\AA $\,$ dashed path on the energy diagram, from which
it relaxes into the $4p^2\,S_0$ excited state where~$a\approx 0,\, b\approx 0,\,c\approx 1$,
leaving a small residual admixture of the middle and ground states.
By Equation~\eqref{rhomix3wf} and the obvious generalization of Equation~\eqref{abmixdipole}, the superposition begins to oscillate with dipole moment:
\begin{equation}\begin{aligned}
{\rm dipole\ moment}=q\left<z\right>=d_{12}ab\,\cos{\!(\omega_{12} t +\phi_{ab})}+d_{23}bc\,\cos{\!(\omega_{23} t +\phi_{bc})}
\end{aligned}
\label{abmix2dipoles}
\end{equation}
where we have included only non-zero time-varying terms terms and have used:
\begin{equation}
\begin{aligned}
\omega_{12}=(\omega_2-\omega_1)\qquad \omega_{23} =(\omega_3-\omega_2)\qquad
\phi_{ab}=(\phi_b-\phi_a)\qquad \phi_{bc}=(\phi_c-\phi_b)
\end{aligned}
\label{omphdefs}
\end{equation}
Because $\omega_{12}\neq \omega_{23}$, the two terms in the dipole moment do not interact over periods of many cycles,
and each term can couple to a separate atom to form its own quantum handshake.  The analysis has already been done
starting with Equation~\eqref{2dmoments} and ending with Equation~\eqref{dEdtav4}, and applies directly to the upper ($\omega_{23}$) transaction
with atom we will call $\alpha$, which has level spacing exactly equal to $E_{23}$.

Since $c$ and $a_\alpha$ both start almost equal to 1 and have the same derivative, we can set $c=a_\alpha$. \\
\begin{equation}
\begin{aligned}
\frac{\partial E_\alpha}{\partial t}=-E_{23}\frac{\partial c^2}{\partial t}=-E_{23}\frac{\partial a_\alpha^2}{\partial t}
=-P_{\alpha}\,c\, b\, a_\alpha b_\alpha\sin{\!(\phi_{\alpha})}=-P_{\alpha}\,c^2\, b\,  \sqrt{1-c^2}\,\sin{\!(\phi_{\alpha})}
\end{aligned}
\label{dEdthi}
\end{equation}
Similarly, we describe the fraction $a^2$ of the superposition in the ground state due to the lower 5227\AA $\,(\omega_{12})$ transaction
with atom we will call $\beta$, which has level spacing exactly equal to $E_{12}$.\\
Since $a$ and $b_\beta$ both start almost equal to 0 and have the same derivative, we can set $a=b_\beta$.\\ 
\begin{equation}
\begin{aligned}
\frac{\partial E_\beta}{\partial t}=E_{12}\frac{\partial a^2}{\partial t}=E_{12}\frac{\partial b_\beta^2}{\partial t}
=P_{\beta}\,a\, b\, a_\beta b_\beta\sin{\!(\phi_{\beta})}=P_{\beta}\,a^2\, b\,  \sqrt{1-a^2}\,\sin{\!(\phi_{\beta})}
\end{aligned}
\label{dEdtlo}
\end{equation}
These relations may then be expressed more compactly as:
\begin{equation}
\begin{aligned}
\frac{\partial c^2}{\partial t}=-\frac{1}{\tau_\alpha}\,c^2\, b\,  \sqrt{1-c^2}\quad\quad
\frac{\partial a^2}{\partial t}=\frac{1}{\tau_\beta}\,a^2\, b\,  \sqrt{1-a^2}\quad\quad
a^2+b^2+c^2=1
\end{aligned}
\label{docdt}
\end{equation}
where $1/{\tau_\beta}=P_{\beta}\,\sin{\!(\phi_{\beta})}/E_{12}$ and $1/{\tau_\alpha}=P_{\alpha}\,\sin{\!(\phi_{\alpha})}/E_{23}$ express the respective strength of coupling to each atom,
and the last relation constrains the superposition to contain exactly one electron. 
Equation~\eqref{docdt} is identical in form to Equation~\eqref{dEdtav5} with the exception of the shared state amplitude $b$ occurring in both derivatives.

\begin{figure}[H]
\centering
\includegraphics[width=0.45\linewidth]{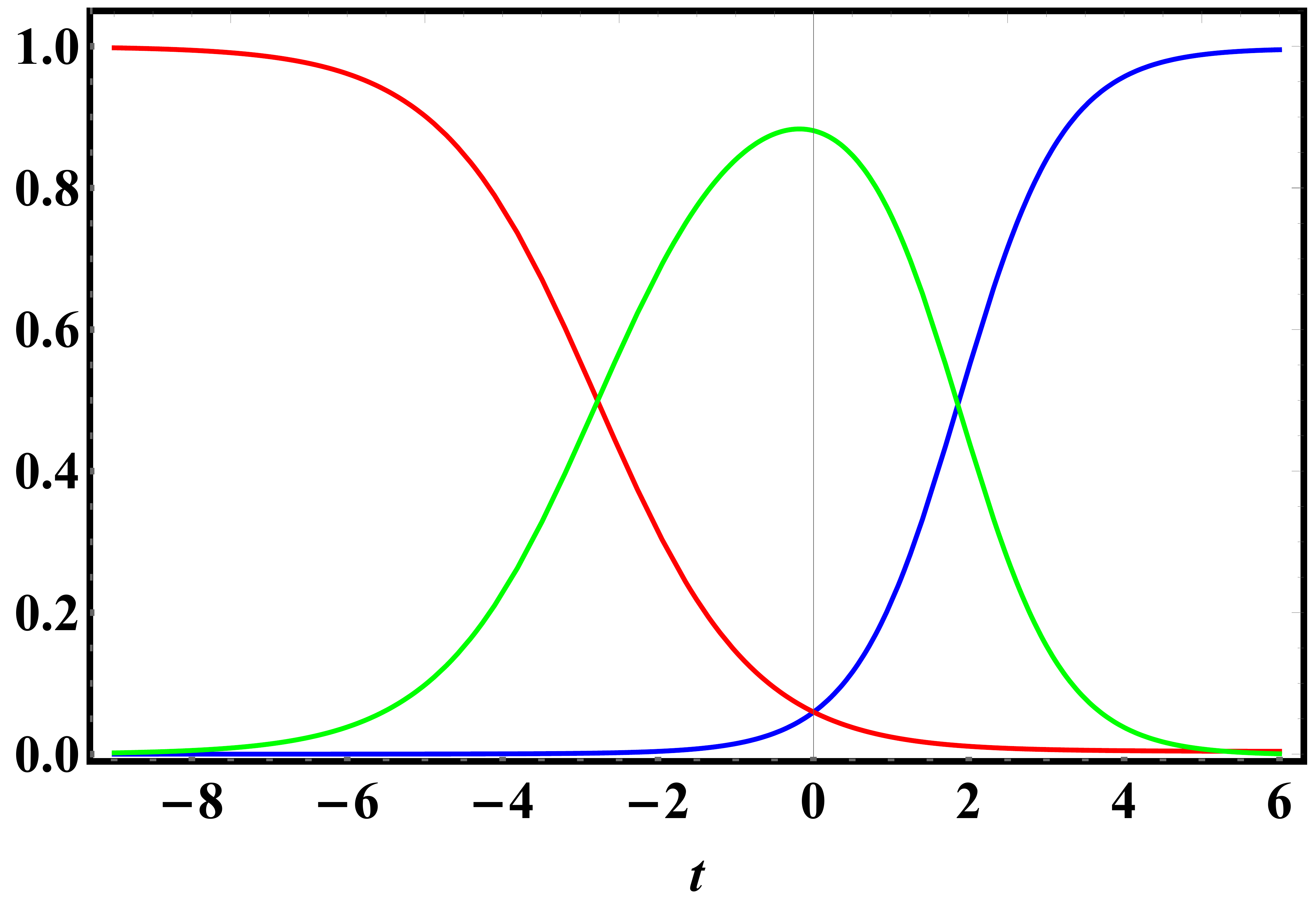}\includegraphics[width=0.5\linewidth]{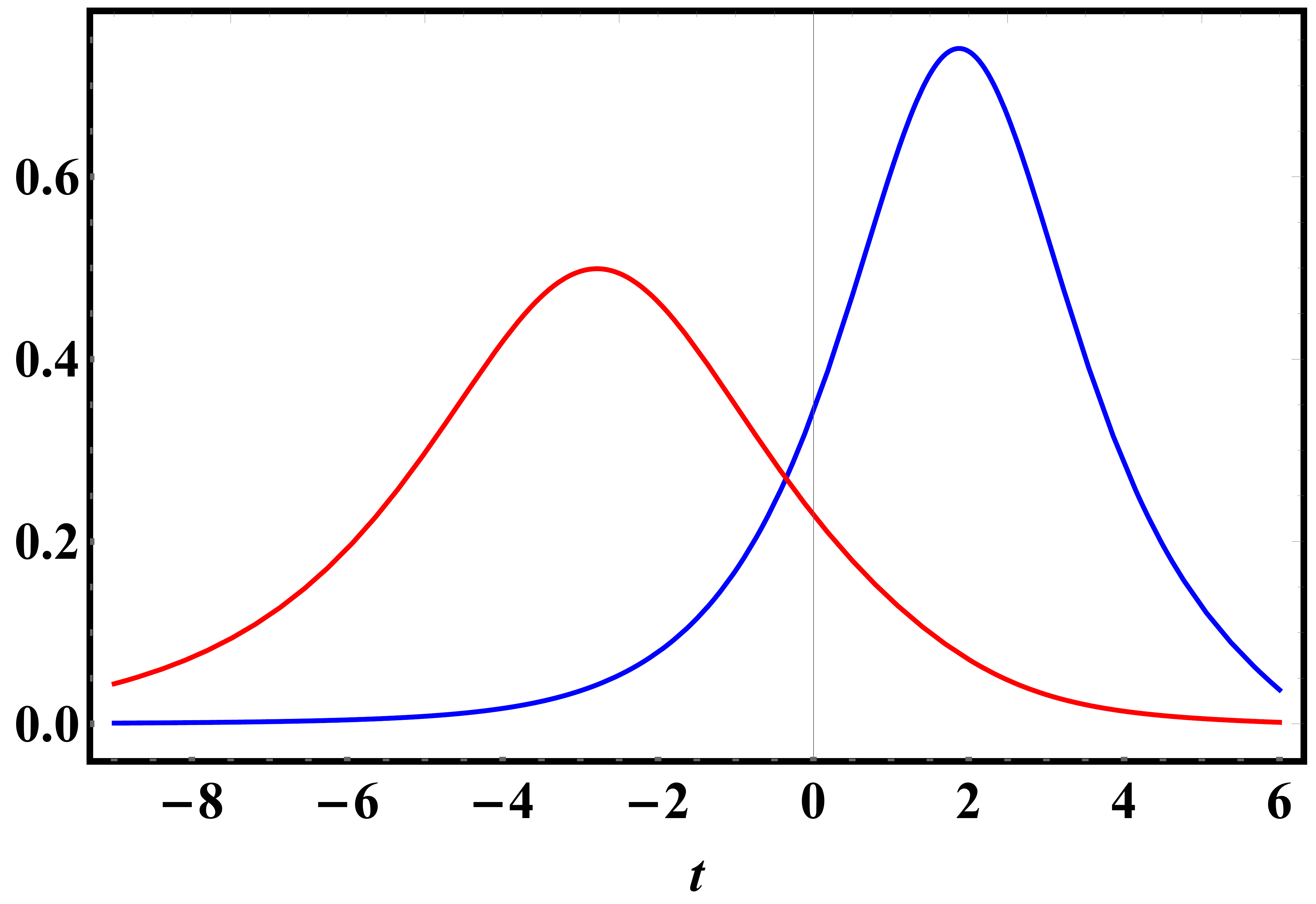}
\caption{\textbf{Left}:  Superposition contributions of the upper state $c^2$ (red), shared middle level $b^2$ (green) and ground state $a^2$ (blue). \textbf{Right}:  Amplitude of dipole oscillations due to upper transition at $\omega_{23}$ (red) and lower transition at $\omega_{12}$ (blue),
from .  The horizontal axis in both plots is time in units of $\tau_\alpha=1.5\,\tau_\beta$.  
\label{ForkTransition}}
\end{figure}

The behavior of this arrangement, shown in Figure~\ref{ForkTransition}, is very instructive.  To review in brief:
The cascade atom is prepared by providing the energy to promote the electron to the upper state: $c\approx 1,\, a\approx 0,\, b\approx 0$. 
The preparation is never perfect, so there is always a small residual of $b$ and $a$ components in the initial superposition.  The small
admixture of $b$ and $c$, by Equation~\eqref{abmix2dipoles}, creates an oscillating dipole moment at frequency $\omega_{23}$, the amplitude of
which is shown as the red curve in the right panel of  Figure~\ref{ForkTransition}.  
That oscillating dipole moment initiates a vector potential ``offer wave'' which propagates
outward according to Equation~\eqref{AandEfromI}.  When that vector potential couples to the wave function of another atom of the same frequency,
its wave function oscillates with the vector potential in the correct phase to withdraw energy according to Equation~\eqref{EeSch21}. 
The vector potential from that oscillation propagates {\it backwards in time}, so it is ``felt'' by the cascading atom as if it had been there all~along.  

\begin{figure}[H]
\centering
\includegraphics[width=0.5\linewidth]{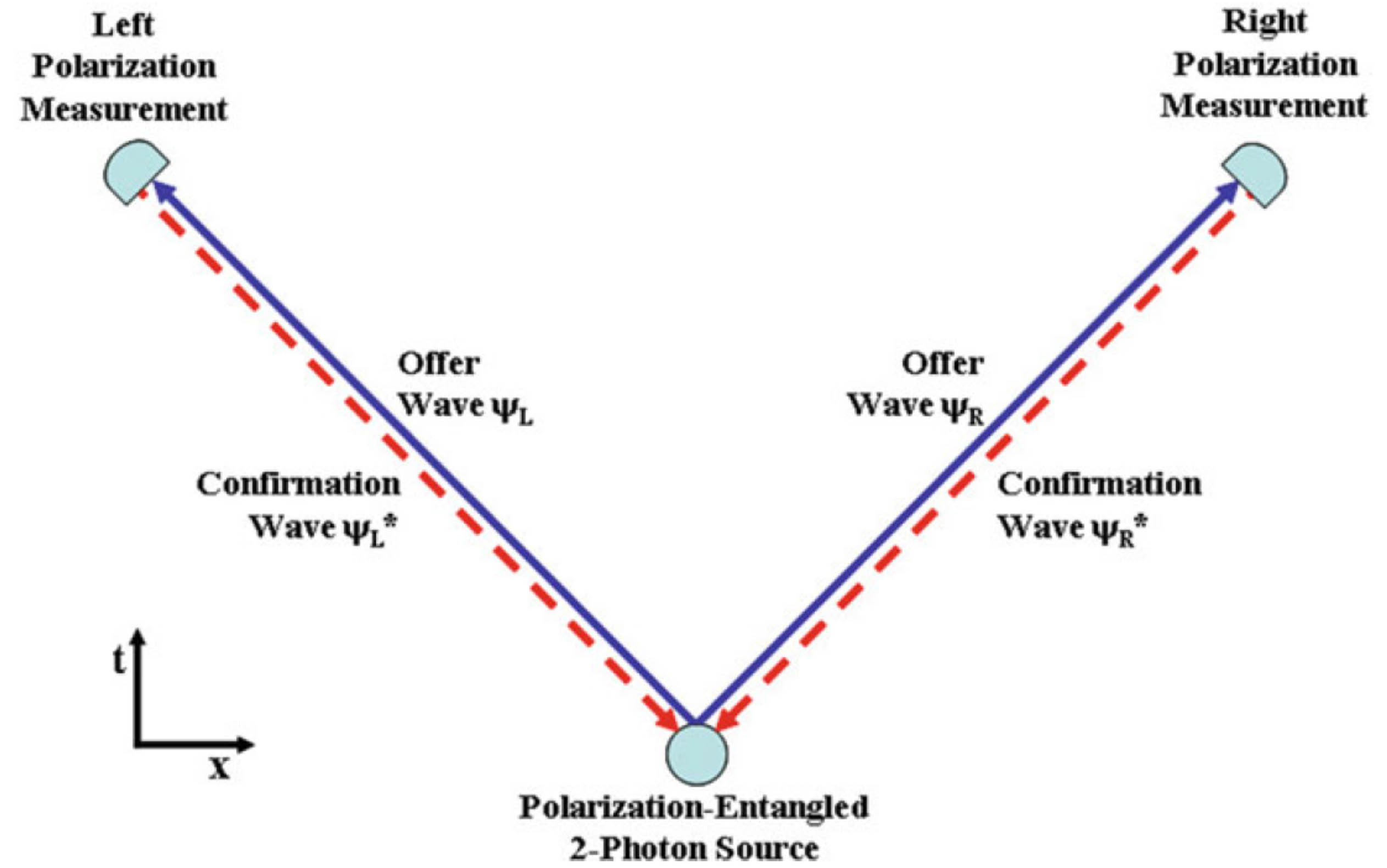}
\caption{Three-vertex transaction formed by a detection event in the Freedman--Clauser experiment.  Linked transactions between the source and the two polarimeters cannot form unless the source boundary condition of matching polarization states is met.\label{F-C_TI}}
\end{figure}

The Transactional Interpretation is applied to the Freedman--Clauser experiment as shown in Figure~\ref{F-C_TI}.  Two-way handshakes between the source and the two polarimeters are joined at the source and must satisfy the boundary condition, based on conservation of angular momentum in a system that begins and ends in a state of zero angular momentum that the polarizations must match for the two offer waves emitted back-to-back.  This V-shaped transaction holds the key to understanding the mechanism behind quantum nonlocality in EPR expertiments.

It was pointed out by one of our reviewers that the kind of description just given might appeal to readers who are not accustomed to
standard relativistic space-time diagrams, on which events on the ``light cone'' are local in the sense that $r^2-c^2 t^2=0$, and that
we had, in some measure, ignored G.N. Lewis' chiding quoted earlier ``A dissymmetry alien to the pure geometry
of relativity has been introduced by our notion of causality.''  Thus, let's try it again:

\vspace{6pt}

\noindent  {\bf A quantum handshake is an antisymmetric bidirectional
electromagnetic connection between two atoms on a light cone, whose direction of time is the direction of positive energy transfer.}
\vspace{6pt}

Either way we look at it, the cascade atom and atom $\alpha$ are locked in phase and amplitude at frequency $\omega_{23}$,
and the locked amplitude of oscillation of the wave
functions of the two atoms is growing with time.  In the process, the $z$-axes of both atoms becomes better and better aligned. 
Meanwhile, the small superposition amplitudes $b$ and $a$ are growing, thus developing a growing oscillation at $\omega_{12}$,
shown by the blue curve in the right panel of Figure~\ref{ForkTransition}.  Since both the upper and lower levels are $S$ states, they have
no effect on the orientation of the oscillation, which is determined by the shared middle level, which is a $P$ state that has a definite
direction in space.  That direction determines the direction of oscillation of any superposition involving that $P$ state. 
The vector potential from the nascent $\omega_{12}$ oscillation recruits a
willing atom $\beta$ whose level spacing is precisely match to $\omega_{12}$, and forms an embryonic quantum handshake of its own 
whose $z$-axis is already determined by the fully developed $\omega_{23}$ oscillation.  From there, both transactions were completed,
with $z$ axes aligned, in very much the same way we have described for a single photon. 

\subsubsection{The NCT-Killer Result}

\begin{figure}[H]
\centering
\includegraphics[width=0.4\linewidth]{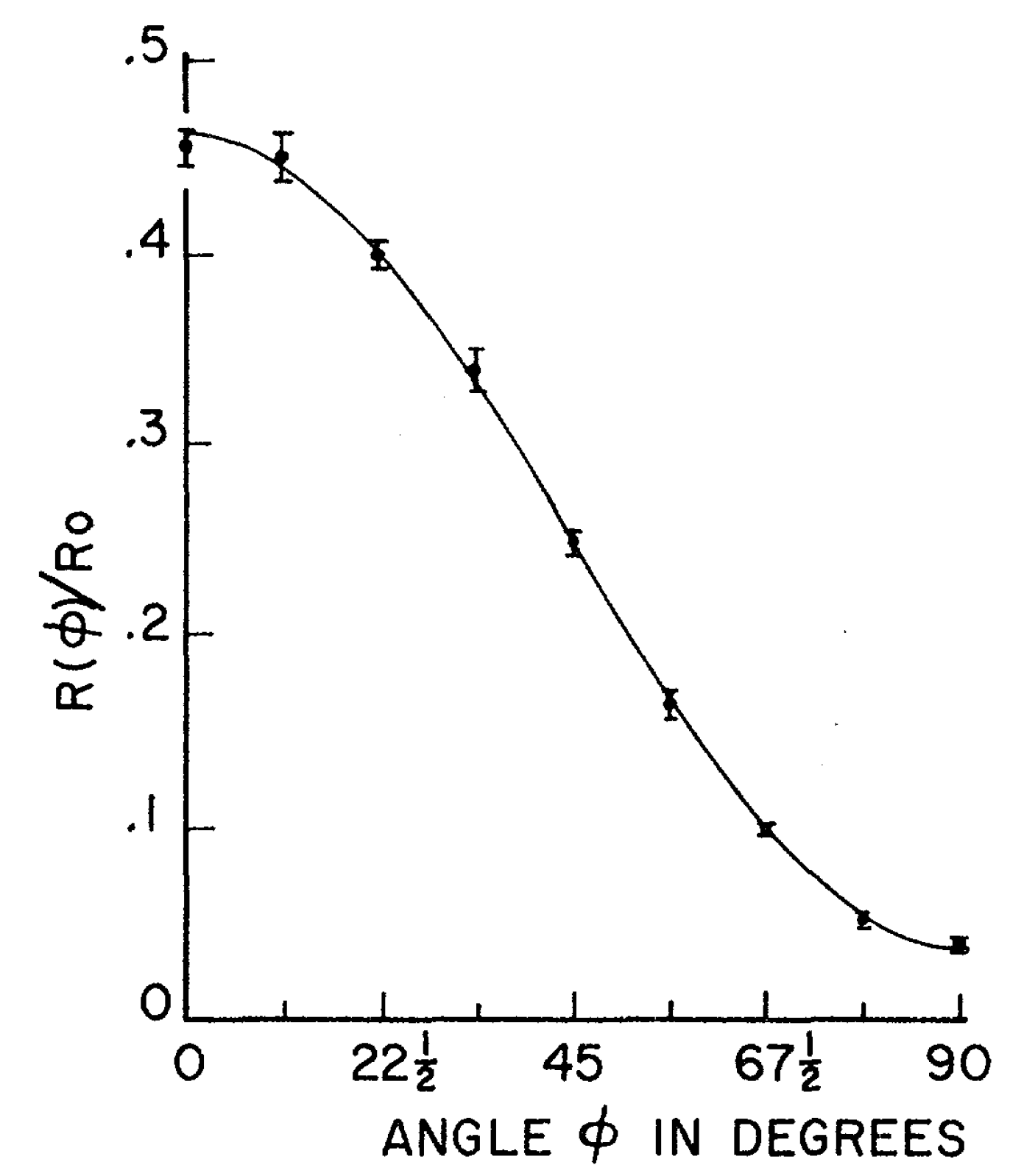}
\caption{Coincidence rate vs. angle $\phi$ between the
polarizers, divided by the rate with both polarizers removed. 
The solid line is the prediction by quantum mechanics, calculated using
the measured efficiencies of the polarizers and solid
angles of the experiment.\label{FreedClausResult}}
\end{figure}

The unique aspect of the Freedman--Clauser experiment was separating the back-to-back paths of the two wave propagations with filters
that selected the $\omega_{23}$ path to the left and the $\omega_{12}$ path to the right.  Each path was equipped with its
own glass-plate-stack linear polarizer, whose angle could be adjusted.  The number of coincidences of signals from the left and right photomultipliers
P.M.1 and P.M.2 were plotted as a function of the angle $\phi$ between the polarization axes of the polarizers, and is shown in Figure~\ref{FreedClausResult}.

The key aspect of Figure~\ref{FreedClausResult} is that at $\phi = 90^\circ$ the coincidence rate drops to essentially zero.  The~graph does show a small non-zero coincidence rate at $90^\circ$, but this is because of the imperfect linear polarization discriminations of the glass-plate-stack polarizing filters and the finite solid angles of the detection paths.  Jaynes' NCT approach failed to reproduce \cite{Clauser_1973MW} this result. That failure, in the view of most of the field, ``falsified'' the NCT approach to quantum phenomena and caused it to be subsequently ignored.

We can understand this result very simply from our TI perspective, reasoning directly from Figure~\ref{ForkTransition}: 
By the time the blue $\omega_{12}$ transition is just getting started, the red $\omega_{23}$ transition is well along,
and has connected to a partner in P.M.2 through polarizer 2.  
Since the polarizer only transmits a vector potential aligned with its axis of polarization, the $z$ axes
of both the cascade atom and atom $\alpha$ will be well aligned with polarizer 2.  Now, as the blue oscillation just begins to build,
its axis, as part of the same $P$ wave function, will also be aligned with polarizer 2, at rotation angle $\theta_2$. 

The way that these glass-plate-stack polarizers work is that they only pass the component of propagating vector potential along their axis of polarization,
the orthogonal component being reflected out of the direction of propagation.  Thus, the fraction of $\theta_2$ polarized vector potential
that can pass through a perfect $\theta_1$ oriented polarizer 1 is just $\cos{\!(\theta_2-\theta_1)}$.
Thus, the amplitude of the nascent $\omega_{12},\,\theta_2$ polarized  ``offer wave'' propagating outward from the cascade atom through
polarizer 1 will be proportional to $\cos{\!(\theta_2-\theta_1)}$.  By Feynman's {\it Grand Principle}, the probability amplitude
of an ``offer wave'' actually forming a transaction is proportional to the amplitude of the vector potential.  Since, in this example,
the $\omega_{23}$ transaction in P.M.2 has already formed, the slightly later formation of an $\omega_{12}$ transaction in P.M.1
will count as a coincidence.  The probability of coincidence counts will therefore be proportional to the square of the probably amplitude
which, for this case, is  $\cos^2{\!(\theta_2-\theta_1)}$ which, when corrected as noted, gives the solid line in Figure~\ref{FreedClausResult} and is zero when $\phi = 90^\circ$. 
Once again, the bi-directional non-local nature of the quantum handshake predicts the observed outcome of this historic experiment.

\section{Conclusions}

The development of our understanding of quantum systems began with a physical insight of deBroglie: 
Momentum was the wave vector of a propagating wave of some sort.  Schr\"odinger is well known for developing
a sophisticated mathematical structure around that central idea, only a~shadow of which remains in current practice. 
What is less well known is that Schr\"odinger also developed a deep understanding of the {\it physical meaning}
of the mathematical quantities in his formalism.  That~physical understanding enabled him to {\it see} the mechanism
responsible for the otherwise mysterious quantum behavior.  Meanwhile, Heisenberg, dismissing visual pictures of quantum processes,  had developed a matrix formulation that dealt with only the probabilities of transitions---what he called ``measurables.''  It looked for a while as if we had two competing quantum theories, until Schr\"odinger and Dirac showed that they gave the same answers. 
However, the stark contrast between the two approaches was highlighted by the ongoing disagreement in which
Bohr and Heisenberg maintained that the transitions were events with no internal structure, and therefore
there was nothing  left to be understood, while Einstein and Schr\"odinger believed that the statistical
formulation was only a stopgap and that a deeper understanding was possible and was urgently needed.  This argument still rages on.

Popular accounts of these ongoing arguments, unfortunately, usually focus
on the 1930 Solvay Conference confrontation between Bohr and Einstein that was centered around Einstein's clock paradox,
a clever attempted refutation of the uncertainty principle \cite{Jammer_1974}.
Einstein is generally considered to have lost to Bohr because he was ``stuck in classical thinking.'' 
However, as detailed in {\textbf {\textit {The Quantum Handshake}}} \cite{Cramer_2016},  Einstein's effort was doomed from the start and was also beside the point.  
The~uncertainty principle is simply a Fourier-algebra property of {\it any system described by waves}.
Both parties to the Solvay argument lacked any real clarity as to how to handle the intrinsic wave nature of matter. 
In the introduction, we quoted Einstein's deepest concern with statistical QM: \\
\indent{\it There must be a deeper structure to the quantum transition.}

Back in 1926, the field was faced with a choice:  Schr\"odinger's wave function in three-dimensional space, or
Heisenberg and Born's matrices, in as many dimensions as you like.
The choice was put forth clearly by Hendrik Lorentz \cite{Lorentz_1926} in a letter to Schr\"odinger in May, 1926:
\begin{quote}{\it
"If I had to choose now between your wave mechanics and the matrix mechanics, I would give the preference to the former because of its greater intuitive clarity, so long as one only has to deal with the three coordinates x,y,z.  If, however, there are more degrees of freedom, then I cannot interpret the waves and vibrations physically, and I must therefore decide in favor of matrix mechanics.  However, your way of thinking has the advantage for this case too that it brings us closer to the real solution of the equations; the eigenvalue problem is the same in principle for a higher dimensional q-space as it is for a three-dimensional space.

"There is another point in addition where your methods seem to me to be superior.  Experiment acquaints us with situations in which an atom persists in one of its stationary states for a certain time, and we often have to deal with quite definite transitions from one such state to another.  Therefore, we need to be able to represent these stationary states, every individual one of them, and to investigate them theoretically.  Now a matrix is the summary of all possible transitions and it cannot at all be analyzed into pieces.  In your theory, on the other hand, in each of the states corresponding to the various eigenvalues, E plays its own role.''}
\end{quote}

Thus, the real choice was between the intuitive clarity of Schr\"odinger's wave function and the ability of 
Heisenberg--Born matrix mechanics to handle more degrees of freedom.  That ability was immediately put to the test
when Heisenberg \cite{Heisenberg_1926} worked out the energy levels of the helium atom, in which two electrons shared the same orbital state and their correlations could not be captured by wave functions with only three spatial degrees of freedom.  That amazing success set the field on the path of eschewing Schr\"odinger's views and moving into multi-dimensional Hilbert space,
which was further ossified by Dirac and von~Neumann.  Schr\"odinger's equation
had been demoted to a bare matrix equation, engendering none of the {\it intuitive clarity}, the ability to
{\it interpret the waves and vibrations physically} so treasured by Lorentz. 

The matrix formulation of statistical QM, as now universally taught in physics classes, saves us the ``tedious'' process
of analyzing the details of the transaction process.  That's the good news.  The bad news is that
{\it it actively prevents us from learning anything about the transaction process, even if we want~to!}

What has been left out is, as Einstein said, any ``{\it description of the individual system}''.

Thus, it was left to the more practical-minded electrical engineers and applied scientists to resurrect, each in their own way,
Schr\"odinger's way of thinking because they needed a ``{\it description of the individual system}'' to make progress. 
Electrons in conductors were paired into standing waves, which could carry current
when the propagation vector of one partner was increased and that of the other partner decreased. 
Energy gaps resulted from the interaction of the electron wave functions with the periodic crystal lattice. 
Those same electron wave functions can ``tunnel'' through an energy gap in which they decay exponentially with distance.
The electromagnetic interaction of the collective wave functions in superconducting wires
leads to a new formulation of the laws of electromagnetism without the need for Maxwell's equations \cite{Mead_2000}. 
The field of Quantum Optics was born.  Conservation of momentum became the matching of wavelengths of
waves such that interaction can proceed.  When one such wave is the wave function of an electron in the conduction band
and the other is the wave function of a hole in the valence band of a semiconductor, matching of the wavelengths
of electron, hole, and photon leads to light emission near the band-gap energy.  

When that emission intensity 
is sufficient, the radiation becomes coherent---a semiconductor laser.  These insights, and many more like them, have made
possible our modern electronic technology, which has transformed the entire world around us. 

Each of them requires that, as Lorentz put it: {\it we...~represent these stationary states,
every individual one of them, and to investigate them theoretically.}   

Each of them also requires that we
analyze the transaction involved very much the way we have done in this paper.

What we have presented is a detailed analysis of the most elementary form of quantum transition, indicating that
the simplest properties of solutions of Schr\"odinger's equation for single-electron atomic states, the conservation of energy, and 
a symmetry property of relativistic laws of electromagnetic propagation, together with Feynman's insight that all paths should be counted,
give a unique form to the photon transaction between two atoms.  

We have extended this approach to experiments involving
three atoms.  The reason we can treat situations with more than one electron using a wave-mechanics Schr\"odinger equation
that only works for one electron is that the non-local bi-directional electromagnetic coupling between wave functions can be
factored into a retarded wave propagating forward in time and an advanced wave propagating backward in time, the vector potential
of each partner in a photon transaction being incorporated in the opposite partner's one-electron Schr\"odinger equation.

These calculations are, of course,
not general proofs that in every system the offer/confirmation exchange always triggers the formation of a transaction. 
They do, however, represent demonstrations of that behavior in tractable cases and constitute prototypes of more general transactional behavior.
They further demonstrate that the transaction model is implicit in and consistent with the  Schr\"odinger wave mechanics formalism, and they
demonstrate how transactions, as a space-time standing waves connecting emitter to absorber, can form. 
 
We see that the missing ingredients in previous failed attempts by others to derive wave function collapse
from the standard quantum formalism were:
\begin{enumerate} 
\item{ Advanced waves were not explicitly used as part of the process.}
\item{ The ``focusing'' property of the advanced-retarded radiation pattern had not been identified.}
\end{enumerate} 

Although many complications are avoided by the simplicity of these two-atom  and three-atom systems, they clearly illustrate that
{\it there is internal structure to quantum transitions} and that {\it this structure is amenable to physical understanding}. 
Each of them is an example of Einstein's  ``{\it description of the individual system}''. 
Through the Transactional Interpretation, the standard quantum formalism is seen as an ingenious shorthand
for arriving at probabilities without wading through the underlying details that Schr\"odinger described as ``tedious''.

Although the internal mechanism detailed above is of the simplest form, it describes the most mysterious behavior
of quantum systems coupled at a distance, as detailed in \cite{Cramer_2016}.  All of these behaviors can be exhibited by
single-electron quantum systems coupled electromagnetically.  The only thing ``mysterious'' about our development
is our unorthodox use of the advanced-retarded electromagnetic solution to conserve energy and speed up the transition. 
Therefore, we have learned some interesting things by analyzing these simple transactions!

This experience brings us face to face with the obvious question:  {\bf What if Einstein was right?}

{\it If there is internal structure to these simple quantum transitions, there must also be internal structure
to the more complex questions involving more than one electron, which cannot be so simply factored!}\\ 
In this case, we should find a way to look for it.
That would require that we effectively time-travel back to 1926 and {\it grok} the questions those incredibly talented scientists
were grappling with at the time.  

To face into the {\it conceptual} details of questions involving an overlapping multi-electron system is a daunting task that has defeated
every attempt thus far.  We strongly suspect that the success achieved by the matrix approach---adding three more
space dimensions and one spin dimension for each additional electron---came at the cost of being
``lost in multi-dimensional Hilbert space.''  Heisenberg's triumph with the helium atom led into a rather short tunnel that narrowed
rapidly in the second row of the periodic table.

Quantum chemists work with complex quantum systems that share many electrons in close proximity, and thus
must represent many overlapping degrees of freedom.  Their primary goal is to find the ground state of such systems.
Lorentz's hope---that the intuitive insights of Schr\"odinger's wave function in three dimensions would
{\it bring us closer to the real solution} in systems with more than one electron---actually helped in the early days of quantum chemistry: 
Linus~Pauling visualized chemical bonds that way, and made a lot of progress with that approach. 
It is quite clear that the covalent bond has a wave function in three dimensions, even if we don't yet have
a fully ``quantum'' way of handling it in three dimensions.   
The Hohenberg--Kohn theorems \cite{Hohenberg_1964} demonstrate that the
ground-state of a many-electron system is uniquely determined by its electron density,
which depends on only three spatial coordinates. 
Thus, the chemists have a three-dimensional wave function for many electrons! 
They use various approaches to minimize the total energy, which then gives the best estimate of the true ground~state. 

These approaches have evolved into {\bf Density Functional Theory} (DFT), and are responsible for amazingly successful
analyses of an enormous range of complex chemical problems. 
The original Thomas--Fermi--Hohenberg--Kohn idea was to make the Schr\"odinger equation just about the 3d density. 
The practical implementations do not come close to the original motivation because half-integer spin, Pauli exclusion, and 3N dimensions
are still hiding there. 
DFT, as it stands today, is a practical tool for generating numbers rather than a fundamental way of thinking. 
Although it seems unlikely at present that a more intuitive view of the multi-electron wave function will emerge from DFT,
the right discovery of how to adapt 3D thinking to the properties of electron pairs could be a major first step in that direction.

When we look at even the simplest two-electron problems, we see that our present understanding uses
totally ad hoc rules to eliminate configurations that are otherwise sensible:  The most outrageous
of these is the {\bf Pauli Exclusion Principle}, most commonly stated as:
{\it Two electrons can only occupy the same orbital state if their spins are anti-parallel.}

It is the reason we have the periodic table, chemical compounds, solid materials, and electrical conductors.
It is just a rule, with no underlying {\it physical} understanding.  
We have only mathematics to cover our ignorance of {\it why it is true physically.}
~[The matrix formulation of QM
has a much fancier mathematical way of enforcing this rule. The associated quantum field theory axiom is {\it not} a physical understanding.]

There is no shame in this---John Archibald Wheeler said it well:
\begin{quote}{\it
"We live on an island surrounded by a sea of ignorance.

As our island of knowledge grows, so does the shore of our ignorance."}
\end{quote}

The founders made amazing steps forward in 1926.

Any forward step in science {\it always} opens new questions that we could not express previously.

However, we need to make it absolutely clear {\it what it is that we do not yet understand:}
\begin{itemize}
\item{We do not yet understand the mechanism that gives the 3D
 wave function its ``identity'', which causes it to be normalized.}
\item{We do not yet have a {\it physical} picture of how the electron's wave function can be endowed with half-integer ``spin'', why it requires a full 720$^{\circ}$  (twice around) rotation to bring the electron's wave function back to the same state, why both matter and antimatter electron states exist, and why the two have opposite parity.}
\item{We do not yet have a {\it physical} understanding of how two electron wave functions interact to enforce Pauli's Exclusion Principle.}
\end{itemize}

However, our analysis {\it has} allowed us to understand {\it conceptually} several things that have been hidden under the statistics:
We saw that the Bose--Einstein property of photons can be understood
as arising from the symmetry of electromagnetic coupling together with the movement of electron charge
density of a superposed state.  There was nothing ``particle-like'' about the electromagnetic coupling.
Indeed, the two-way space-time symmetry of the photon transaction cannot really be viewed
as the one-way symmetry of the flight of a ``photon particle.''  Thus, looking at the mechanism of the
"wave-function collapse'' gives us a deeper view of the ``boson'' behavior of the photon: 
The rate of growth of oscillation of the superposed state is, by Equation~\eqref{dEdtav}, proportional to the oscillating
electromagnetic field.  When the oscillating currents of all the atoms are in phase, the amplitudes
of their source contributions add, and any new atom is correspondingly more likely to synchronize
its contribution at that same phase.  \\Thus, the ``magic'' bosonic properties of photons, including the
quantization of energy $\hbar\omega$ and tendency to fall into the same state, are simply properties
of single-electron systems coupled electromagnetically:  Their two-way space-time symmetry is
in no way “particle-like.”  It seems as though there is, after all, a fundamental conceptual difference
between ``matter'' and ``coupling''.

Perhaps, it is the stubborn determination of theoretical physicists to make everything into particles residing in a multi-dimensional Hilbert space
that has delayed for so long
Lorentz's {\it greater intuitive clarity}---a deeper {\it conceptual, physical understanding} of simple quantum systems.  

Thanks to modern quantum optics, we are experimentally standing on the shoulders of giants:
We can now routinely realize radio techniques, such as phase-locked loops, at optical frequencies. 
The old argument that ``everything is just counter-clicks'' just doesn't cut it in the modern world!

Given the amazing repertoire of these increasingly sophisticated experiments with coherent optical-frequency quantum systems,  many 
of the ``mysterious'' quantum behaviors seem more and more physically transparent when viewed as arising from the transactional symmetry of the interaction, rather than from the historic ``photon-as-particle'' view.  The bottom line is that Schr\"odinger wave mechanics can easily deal with issues of quantum entanglement and nonlocality in atomic systems coupled by matched advanced/retarded 4-potentials.  It remains to be seen whether this wave-based approach can be extended to systems involving the emission/detection of quark-composites or leptons.

Our Caltech colleague Richard Feynman  left a legacy of many priceless quotations; a great one is:
\begin{quote}
"However, the {\it Real Glory} of {\it Science}\\ is that we can {\it Find} a {\it Way} of {\it Thinking}\\ such that the {\it Law} is {\it Evident!}"
\end{quote}

What he was describing is {\bf Conceptual Physics}. 

From our new technological vantage point, it is possible to develop Quantum Science in this direction,
and make it accessible to beginning students. 

 We urge new generations of talented researchers to take this one on.
 
Be Fearless---as they were in 1926!\\

\vskip20pt
\authorcontributions{Both authors contributed to this work in roughly equal amounts.  It started as a short internal report written by JGC discussing Mathematica calculations based on CAM's book.  Following this, CAM made major contributions in expansion to the present length and  produced most of the discussion of equations and formalism.}

\funding{This research received no external funding.}

\acknowledgments{The authors are grateful to Jamil Tahir-Kheli for sharing his unmatched knowledge
and understanding of physics history, and his deeply insightful critique of the approach developed here,
during many thoughtful discussions down through the years. We thank Ruth Kastner and Gerald Miller for helpful comments and for rubbing our noses in the probabilistic approach to QM.  We are particularly grateful to  Nick Herbert for asking about transition time in our calculations, which led us to important new insights.  David Feinstein, Glenn Keller, Ed Kelm, and Lloyd Watts caught
many bugs and offered helpful suggestions.  We thank Jordan Maclay for asking us to write this paper for a special issue of {\em Symmetry} and for making useful comments and suggestons.  Finally, we thank our four anonymous reviewers, whose penetrating comments and questions led to major improvements in this paper. }

\conflictsofinterest{The authors declare that there are no conflicts of interest.}


\abbreviations{The following abbreviations are used in this manuscript:\\

\noindent
\begin{tabular}{@{}ll}
DFT & density functional theory\\
EPR experiment & Einstein, Podolsky, and Rosen experiment demonstrating nonlocality\\
NCT & neoclassical theory, i.e., Schr\"odinger's wave mechanics plus Maxwell's equations\\
QM & quantum mechanics\\
TI & the Transactional Interpretation of quantum mechanics \cite{Cramer_2016}\\
WFE & Wheeler--Feynman electrodynamics\\
\end{tabular}} 

\reftitle{References}



\end{document}